\documentclass[sigconf]{acmart}
\usepackage{subcaption}

\AtBeginDocument{%
  }

\copyrightyear{2026}
\acmYear{2026}
\setcopyright{cc}
\setcctype{by-nc-nd}
\acmConference[CHI '26]{Proceedings of the 2026 CHI Conference on Human Factors in Computing Systems}{April 13--17, 2026}{Barcelona, Spain}
\acmBooktitle{Proceedings of the 2026 CHI Conference on Human Factors in Computing Systems (CHI '26), April 13--17, 2026, Barcelona, Spain}
\acmDOI{10.1145/3772318.3791332}
\acmISBN{979-8-4007-2278-3/2026/04}

\newcommand{\hl}[1]{\textcolor{black}{#1}}
\newcommand{\hx}[1]{\textcolor{black}{#1}}

\begin{document}

\title{Improving Data Quality via Pre-Task Participant Screening in Crowdsourced GUI Experiments}

\author{Takaya Miyama}
\affiliation{%
  \institution{Meiji University}
  \city{Tokyo}
  \country{Japan}}
\email{cs252040@meiji.ac.jp}

\author{Satoshi Nakamura}
\affiliation{%
  \institution{Meiji University}
  \city{Tokyo}
  \country{Japan}}
\email{nkmr@meiji.ac.jp}

\author{Shota Yamanaka}
\affiliation{%
  \institution{LY Corporation}
  \city{Tokyo}
  \country{Japan}
}
\email{syamanak@lycorp.co.jp}

\begin{abstract}
\hl{
In crowdsourced user experiments that collect performance data from graphical user interface (GUI) interactions, some participants ignore instructions or act carelessly, threatening the validity of performance models.
We investigate a pre-task screening method that requires simple GUI operations analogous to the main task and uses the resulting error as a continuous quality signal.
Our pre-task is a brief image-resizing task in which workers match an on-screen card to a physical card; workers whose resizing error exceeds a threshold are excluded from the main experiment.
The main task is a standardized pointing experiment with well-established models of movement time and error rate.
Across mouse- and smartphone-based crowdsourced experiments, we show that reducing the proportion of workers exhibiting unexpected behavior and tightening the pre-task threshold systematically improve the goodness of fit and predictive accuracy of GUI performance models, demonstrating that brief pre-task screening can enhance data quality.
}
\end{abstract}



\maketitle
\section{Introduction}
Crowdsourced experiments are conducted over the Internet, enabling participation regardless of time or place.
Accordingly, researchers can execute studies quickly while recruiting numerous and diverse participants, and more readily detect statistical significance from the resulting data ~\cite{reips2002standards}.
However, because experimenters cannot observe participants' circumstances in crowdsourced settings, some participants may disregard instructions or act carelessly, which degrades data quality ~\cite{Oppenheimer09imc}.
Thus, data-quality management is a critical challenge in crowdsourced experiments ~\cite{Seaborn25}.

\hl{In HCI, graphical user interface (GUI)-based user experiments have long been conducted in controlled laboratory settings for purposes such as (i) validating and comparing predictive models of interaction performance (e.g., Fitts' law~\cite{mackenzie1992fitts,soukoreff2004towards}), (ii) evaluating the effectiveness of new interaction techniques such as the Bubble Cursor~\cite{grossman2005bubble,yamanaka2023varying}, and (iii) comparing performance across different user groups (e.g., younger vs.\ older adults or people with/without motor impairments~\cite{sharif2020reliability,Findlater13age}). In such studies, the primary aim is to obtain precise, reliable measurements of how users typically perform under standardized task conditions for model fitting and controlled comparison, rather than to capture the full diversity of uncontrolled real-world behaviors.}

\hl{However, laboratory-based GUI experiments are typically limited to relatively small numbers of participants~\cite{komarov2013crowdsourcing,findlater2017differences}. Crowdsourcing has therefore become attractive for this class of GUI experiments because it allows researchers to collect much larger and more demographically diverse samples for estimating central tendencies of user performance and model parameters.
At the same time, however, crowdsourced settings may increase the risk that some participants will ignore instructions or act inattentively, which threatens the validity of performance measurements and model predictions.
For example, there are reports that crowdsourced participants tend to perform pointing tasks less accurately than laboratory participants, with more than twice the error rate observed~\cite{findlater2017differences}.
Thus, the characteristics of recruited participants may influence outcomes and raise concerns about degraded data quality.
This motivates the need for mechanisms that can detect and filter out such non-compliant participants when the research goal is (e.g.) precise model evaluation under controlled task conditions.}

This work thus aims to improve data quality in crowdsourced GUI experiments by ensuring that only participants conforming to requirements in a GUI-based pre-task proceed to the main experiment.
\hx{While our experiments focus on target pointing, we expect the same screening principle to transfer to other downstream GUI-interaction tasks that rely on quantitative performance measures (e.g., goal crossing, path steering, dragging, typing, and object selection; see Section~\ref{sec:main_task}).}
\hx{Although our approach shares the general goal of identifying low-quality workers with existing techniques such as gold-standard tasks~\cite{Kazai11gold} and attention checks~\cite{Oppenheimer09imc}, it differs in that our pre-task is designed to yield a task-relevant continuous error value that directly reflects operational carefulness in GUI interaction (even though gold tasks and attention checks can also be summarized using continuous measures such as accuracy or response time). These continuous scores can then be thresholded with varying strictness, analogous to choosing different statistical outlier criteria, providing a tunable \emph{minimum-quality} requirement for participation in the main experiment.}
\hl{We instantiate this idea using an image-resizing task as the pre-task and a standard pointing experiment as the main task. Resizing a card image to a prescribed physical size is a simple GUI manipulation whose error sensitively reflects how carefully participants follow instructions, whereas pointing tasks are a canonical testbed in HCI with well-established procedures and performance models (e.g., Fitts' law). This combination allows us to quantify how pre-task-based screening influences the reliability and predictive validity of interaction-performance models in a representative GUI-experiment setting.}

We hypothesize that participants who follow instructions in the pre-task are more likely to act conscientiously in the main task, and we examine whether users who will perform the main task as intended (judged by the lawful regularity of their performance) can be identified using only pre-task data.
To this end, we simulate experimental outcomes under varying levels of inclusion of participants \hl{screened as nonconforming to pre-task requirements}, and we assess the utility of this method.
Concretely, we vary (i) the proportion of nonconforming participants in the overall pool and (ii) the pre-task threshold \(T\) of resizing error used to judge nonconformity.
Then, we compute the goodness of fit of models that estimate GUI performance (movement time and error rate) to evaluate whether screening improves model fit.

\hx{Because researchers must choose a screening strictness \(T\) without knowing how much nonconforming behavior will occur in a crowdsourced pool, we use a simulation framework that systematically varies \(N\), \(T\), and \(X\) (the nonconforming proportion) and quantifies how these choices affect model fit (\(R^2\)) and, in smartphone settings, predictive accuracy for unseen task conditions.}

\hl{Importantly, the goal of this screening is \textit{not} to select ``top-performing'' participants. Instead, it serves as a minimum-quality check that removes workers who do not follow instructions or who behave inattentively, in the same spirit as common exclusion rules based on $\pm$3SD or IQR. Thus, the screening threshold controls how lenient or strict this filtering is, rather than privileging high-skill participants.
We discuss potential biases of this design choice, for example, against participants with motor or visual impairments, in Section~\ref{sec:limitations}.}

Across three experiments involving mouse and smartphone interaction, we show that reducing the proportion of nonconforming participants and tightening the nonconformity threshold both improve the goodness of fit of GUI performance models and the prediction accuracy for unseen task conditions.
Because the information used to judge nonconformity does not rely on outcomes from the main GUI experiment, these results indicate that a brief pre-task can improve the data quality of the main experiment.
This enables researchers to screen workers before they proceed to the main experiment, thereby mitigating the risk of drawing incorrect conclusions (e.g., that a given model estimates GUI performance poorly).

Our contributions are threefold:
\begin{itemize}
    \item We propose a new screening method in which a brief pre-task requiring simple GUI interactions (finished in nine seconds on average) is used to determine which workers proceed to the main experiment.
    \item We show that changing the nonconformity threshold in the pre-task (namely, error of image resizing from the prescribed size) and the proportion of such workers in the sample systematically affects the goodness of fit and prediction accuracy of models that estimate GUI performance (movement time and error rate) in the main experiment.
    \item We demonstrate that these effects are robust across mouse interaction (Experiment~1) and two smartphone-interaction settings with different experimental configurations (Experiments~2 and~3).
\end{itemize}

\section{Related Work}
\hl{This work lies at the intersection of two research areas: (1) crowd\-sourc\-ing-based experimentation, which enables large-scale data collection but raises concerns about data quality, and (2) GUI interaction research, particularly studies on pointing performance. Because our method introduces a GUI-based pre-task into a crowdsourced experiment in order to screen participants, both strands of prior work are needed to motivate and situate our approach. Accordingly, Section~2.1 reviews prior work on crowdsourcing experiments and data-quality management, whereas Section~2.2 summarizes key concepts from pointing-task research (including movement time $MT$, error rate $ER$, and performance models such as Fitts' law) that underpin our experimental design and analyses.}

\subsection{Data Quality in Crowdsourced Experiments}
To address concerns about data quality in crowdsourced experiments, prior work has compared outcomes between crowdsourced and laboratory settings.
In psychology and sociology topics, results are often similar across both environments, although some studies report differences ~\cite{crump2013evaluating,horton2011online}.
There are also efforts to make online environments more laboratory-like; for example, Li et al.~\cite{li2020controlling} proposed a task that matches the size of a physical card (e.g., a credit card) to an on-screen card image to compute a display's pixel density and thereby control the physical size of visual stimuli.

Research has also examined inattentive participants, with one report finding that 45.9\% of crowdsourced participants exhibited inattentive behaviors ~\cite{bruhlmann2020quality}.
Curran~\cite{curran2016methods} proposed effective methods for detecting inattentive outcomes in response times and open-ended text responses.
Oppenheimer et al.~\cite{Oppenheimer09imc} proposed the Instructional Manipulation Check (IMC) to assess whether participants read instructions.
They showed that results differ between participants who pass the IMC and those who do not.
Despite these concerns, one review reports that 55\% of psychology studies conducting online experiments did not evaluate data quality ~\cite{gottfried2024practices}.

For GUI experiments as well, crowdsourced and laboratory results have been compared.
Komarov et al.~\cite{komarov2013crowdsourcing} validated the Bubble Cursor ~\cite{Grossman05}, reporting that its movement-time advantage over a standard cursor observed in the laboratory also replicated in crowdsourcing.
Schwab et al.~\cite{schwab2019evaluating} conducted panning and zooming tasks and reported that Fitts' law holds in both PC and mobile environments.
Findlater et al.~\cite{findlater2017differences} found that, in mouse and touch pointing tasks, crowdsourced participants exhibited shorter movement times and higher error rates than laboratory participants, suggesting a tendency to prioritize speed over accuracy when instructed to act ``as fast and accurately as possible.''

Taken together, diverse evaluations and countermeasures have been explored for data quality in crowdsourced experiments, but the applicability of existing quality-improvement methods to task results of GUI-based experiments (e.g., error rates in pointing tasks) remains unexplored.
Our work investigates whether a pre-task tailored to GUI-specific operations can improve data quality upon these efforts.

\subsection{Pointing and Other GUI-Related Tasks}
\hl{When users interact with GUI elements like icons and hyperlinks, their performance, such as how long it takes to reach a target and whether the selection succeeds or fails, has been studied extensively in HCI.
A central line of work models these behaviors using movement time ($MT$) and error rate ($ER$), which capture how task difficulty and user strategy jointly shape interaction outcomes.}
A representative model for pointing is Fitts' law, which predicts $MT$ for the first click in a trial based on the target distance $A$ and target width $W$ ~\cite{fitts1954information,mackenzie1992fitts}.
There is also research on $ER$, defined as the percentage of trials in which clicks fall outside the target region ~\cite{meyer1988optimality,wobbrock2008error}, showing that $ER$ increases as actions become faster and decreases as actions become more cautious.
There are prediction models for $ER$ as well for rectangular and circular targets ~\cite{bi2016predicting,yamanaka2020rethinking}.

Although participants are typically instructed to perform the task as fast and accurately as possible, individuals might unintentionally shift their biases toward either speed or accuracy ~\cite{sharif2020reliability,yamanaka2022test}.
Moreover, by running tasks under bias-emphasized conditions, it is possible to evaluate pointing behavior as a function of emphasis.
For example, movement time $MT$ decreases while error rate $ER$ increases under speed emphasis ~\cite{yamanaka2023varying}.
We adopt this approach (emphasizing either speed or accuracy in pointing tasks) to examine whether participants follow the given instructions.

\hl{Beyond target pointing, many other GUI interaction tasks have been studied in HCI, including goal crossing~\cite{yamanaka2020necessary}, path steering~\cite{Accot97}, dragging~\cite{mackenzie1991comparison}, typing~\cite{banovic2019limits}, and object selection~\cite{yamanaka2022effectiveness}. These tasks share the property that user performance can be quantified (e.g., via time or accuracy), and many rely on regularities similar to those captured by Fitts' law or its extensions. Because our screening approach evaluates how carefully participants perform simple GUI manipulations before the main experiment, we expect the same idea to be applicable to a broader class of GUI experiments beyond pointing, especially when the research goal is to obtain reliable performance measurements under controlled task conditions.}

\section{Approach}
\subsection{Concept}
\begin{figure}[ht]
    \centering
    \includegraphics[width=1.0\linewidth]{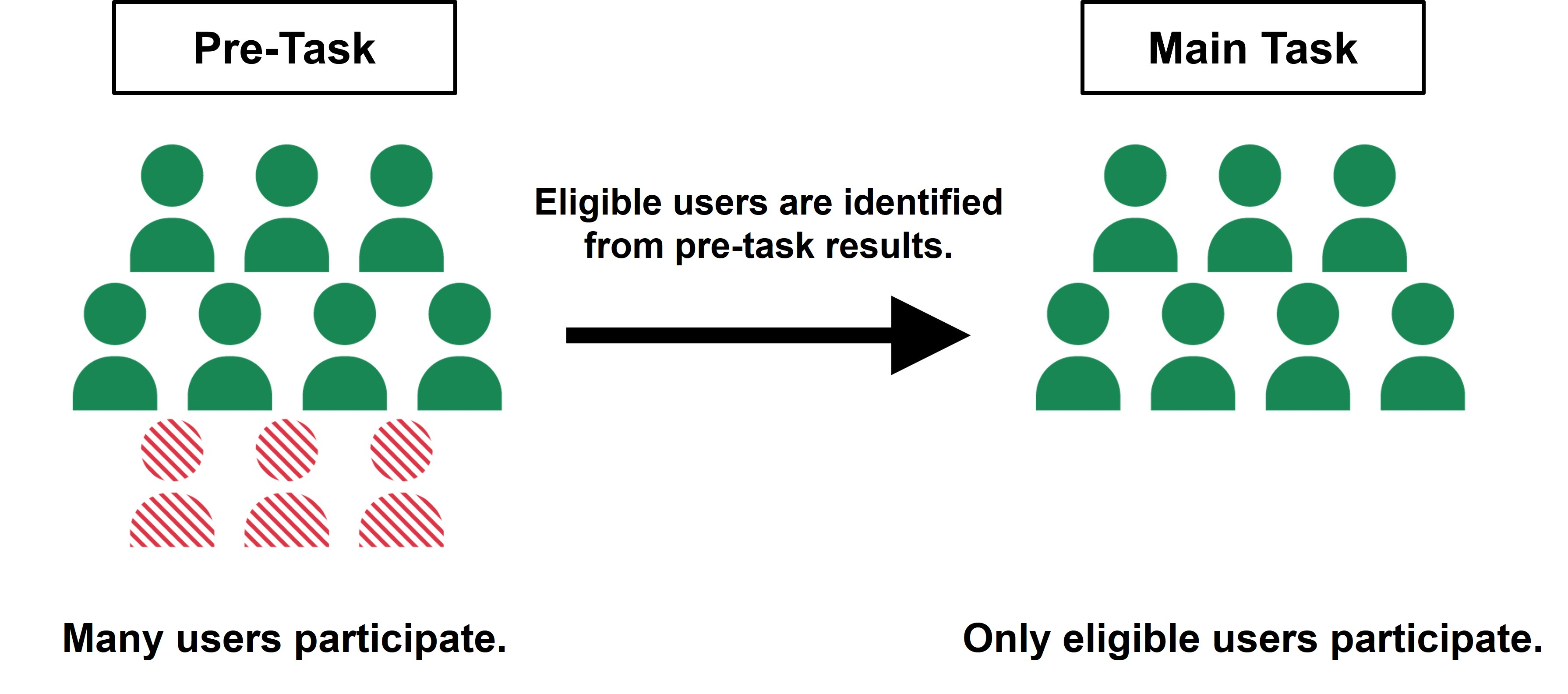}
    \vspace{-18pt}
    \caption{Overview of the proposed approach}
    \label{fig:approach_concept}
    \Description{Workflow overview: a brief pre-task screens participants before the main GUI experiment. Only workers who pass the pre-task proceed to the pointing task, aiming to improve data quality and model evaluation reliability.}
\end{figure}

Our approach is to improve data quality in crowdsourced GUI experiments by screening participants with a pre-task before the main experiment and assigning the intended main task only to conforming workers.
Concretely, a pre-task requires operations related to those in the main task and identifies suitable workers based on their performance in this pre-task.
In this way, the main experiment can be administered only to the screened workers.
This increases the likelihood that most participants will perform interactions as intended, thereby improving data quality.
An overview of the proposed approach is shown in Figure~\ref{fig:approach_concept}.

\subsection{Pre-Task for Participant Screening}
\label{sec:approachPreTask}
\begin{figure}[ht]
    \vspace{-5pt}
    \centering
    \includegraphics[width=0.69\linewidth]{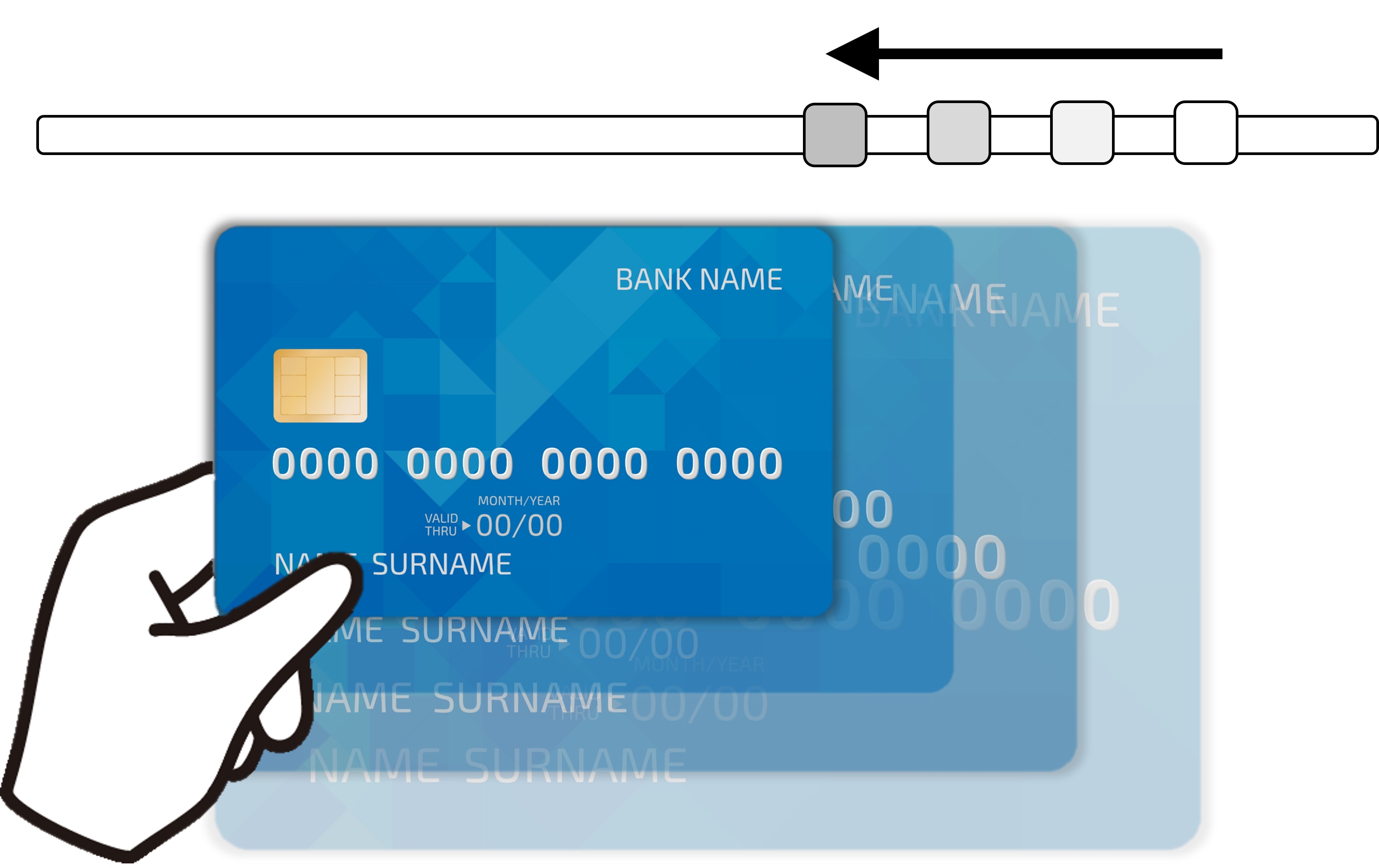}
    \caption{The size-adjustment task by Li et al.~\cite{li2020controlling}}
    \label{fig:approach_sizeTask}
    \Description{Illustration of the size-adjustment pre-task: participants place a physical ID-1 card on the display and resize an on-screen card image to match its size using a slider or handle.}
\end{figure}

\hl{In this study, we assume that researchers have a primary GUI experiment that they ultimately wish to analyze, and that this experiment is administered only after screening participants using a brief pre-task.
We refer to this primary GUI experiment as the ``main task.''}
Because our main task in this study is target pointing, a pre-task including operations related to pointing can be used to screen conforming participants.
We focus on the size-adjustment task proposed by Li et al.~\cite{li2020controlling}, in which participants adjust an on-screen card image to match the size of a physical card (e.g., a credit card) placed on the screen, see Figure~\ref{fig:approach_sizeTask}.
The accuracy of the operation can be evaluated by measuring the error between these sizes.
We assume that this is suitable as a pre-task when the main task requires accuracy as in pointing.

Although the size-adjustment task is typically administered prior to an experiment to control the physical size of visual stimuli~\cite{li2020controlling}, in our previous experiments we found some participants exhibiting \hl{unexpected behavior} (e.g., submitting without resizing at all, or completing the resizing in an extremely short time).
Therefore, by adopting the size-adjustment task as a pre-task, \hl{we expect to identify such participants in advance and restrict the main task to participants conforming to pre-task requirements.}

\subsection{Evaluation and Research Questions}
We evaluate the proposed approach by running crowdsourced experiments in which the size-adjustment task is administered as a pre-task followed by a main task.
\hl{As mentioned, this pre-task is not for detecting a top-performing group; rather, it serves as a minimum-quality criterion that excludes workers who do not follow instructions or who behave inattentively in the pre-task.
In this sense, the threshold \(T\) of the error between physical and on-screen cards to detect nonconforming workers plays a role similar to common statistical exclusion rules (e.g., $\pm$3SD or IQR-based filtering), allowing researchers to choose more lenient or more conservative settings depending on their study design without biasing the sample toward only high-skill users.}
Participants whose error is below a threshold $T$ pass, whereas those at or above the threshold fail.
Building on this, we simulate experimental outcomes when participants who fail the size-adjustment task (the non-passing group) are mixed into the sample.
Specifically, we vary the proportion of the non-passing group in the overall pool and the error threshold, compute the goodness of fit of models that estimate GUI operation performance, and evaluate whether screening improves model fit.

We address the following research questions (RQs):
\begin{description}
  \item[RQ1] Can a simple GUI pre-task (image resizing) effectively screen out nonconforming participants?
  \item[RQ2] How do the proportion of the non-passing group and the error threshold affect the goodness of fit and predictive accuracy of performance models (movement time and error rate) for the main pointing task?
  \item[RQ3] Is the proposed screening method consistently effective across device types (PC/smartphone) and across different error-handling policies (with or without re-aiming when missing a target)?
\end{description}

Our expectation is that model fit will improve as the proportion of the non-passing group decreases and as the threshold is tightened, and that nonconforming participants can be effectively screened regardless of device.
We conducted the following three experiments to evaluate these expectations:
\begin{description}
  \item[Experiment~1] PC mouse-based experiment; in case of an error, participants re-aim until the target is successfully selected.
  \item[Experiment~2] iPhone-based experiment; in case of an error, participants re-aim until the target is successfully selected.
  \item[Experiment~3] iPhone-based experiment; even if an error occurs, the next trial begins immediately.
\end{description}
We first conducted a preliminary check in Experiment~1 to confirm that screening functions as intended, and then decided to perform a more rigorous evaluation in Experiments~2 and~3.
We targeted the iPhone environment in Experiments~2 and~3 because, by restricting the device to iPhone, we can control the physical size of on-screen stimuli in \hl{millimeters (mm)} even in crowdsourced experiments and thus evaluate models more rigorously ~\cite{Usuba22iss,Yamanaka24iss}.
We also prepared multiple error-handling policies because under a ``re-aim until success'' policy, inattentive participants may try to perform as quickly and accurately as possible to finish a task in a short time, making them difficult to distinguish from conscientious participants.
Previous studies have used both with~\cite{Grossman05,Forlines08,Casiez11Surfpad,Yamanaka22chiBias} and without re-aiming when errors occur~\cite{iso2012,Wobbrock11dim,wobbrock2008error}.

All experiments were accessible through a web system we developed, and participants were recruited via Yahoo! Crowdsourcing ~\cite{YCrowd} \hl{ without demographic stratification (e.g., no age balancing).
Workers voluntarily chose to participate in the tasks from our listings, and we did not pre-screen them by demographic or performance attributes. Thus, our sample reflects a natural, unstratified subset of the platform's worker population. We chose Yahoo! Crowdsourcing because our previous GUI experiments on this platform revealed a non-trivial number of inattentive workers, making it a suitable environment for evaluating the effectiveness of the proposed screening method. Demographic information (age and gender), as reported to the platform, was recorded and is summarized in the supplementary materials.}

This study was approved by the research ethics committee at our institution including the legal department (serving as the equivalent of an \hl{Institutional Review Board (IRB)}) as well as by the crowdsourcing platform.
We obtained informed consent from all workers, who agreed that we may record, analyze, and publish data to the extent necessary for research.

\section{Experiment 1: Mouse-Based Experiment on PC}
\subsection{Task and Design}
\subsubsection{Overview}
In Experiment~1, we conducted a mouse-based study in a PC environment to preliminarily verify whether screening via the size-adjustment pre-task works as intended.
Upon accessing the experimental system, participants first viewed an explanation, then completed the size-adjustment task twice, one practice block of the pointing task, and four main (data-collection) blocks of the pointing task.
In the practice block, participants were instructed to act ``as fast and accurately as possible'' to familiarize themselves with the pointing interaction. 
Then, in the main blocks, we asked them to act either ``as fast as possible'' or ``as accurately as possible'' to test if they faithfully follow our instructions.

\begin{figure*}[ht]
    \centering
    \begin{minipage}[t]{0.49\linewidth}
        \centering
        \includegraphics[width=0.85\linewidth]{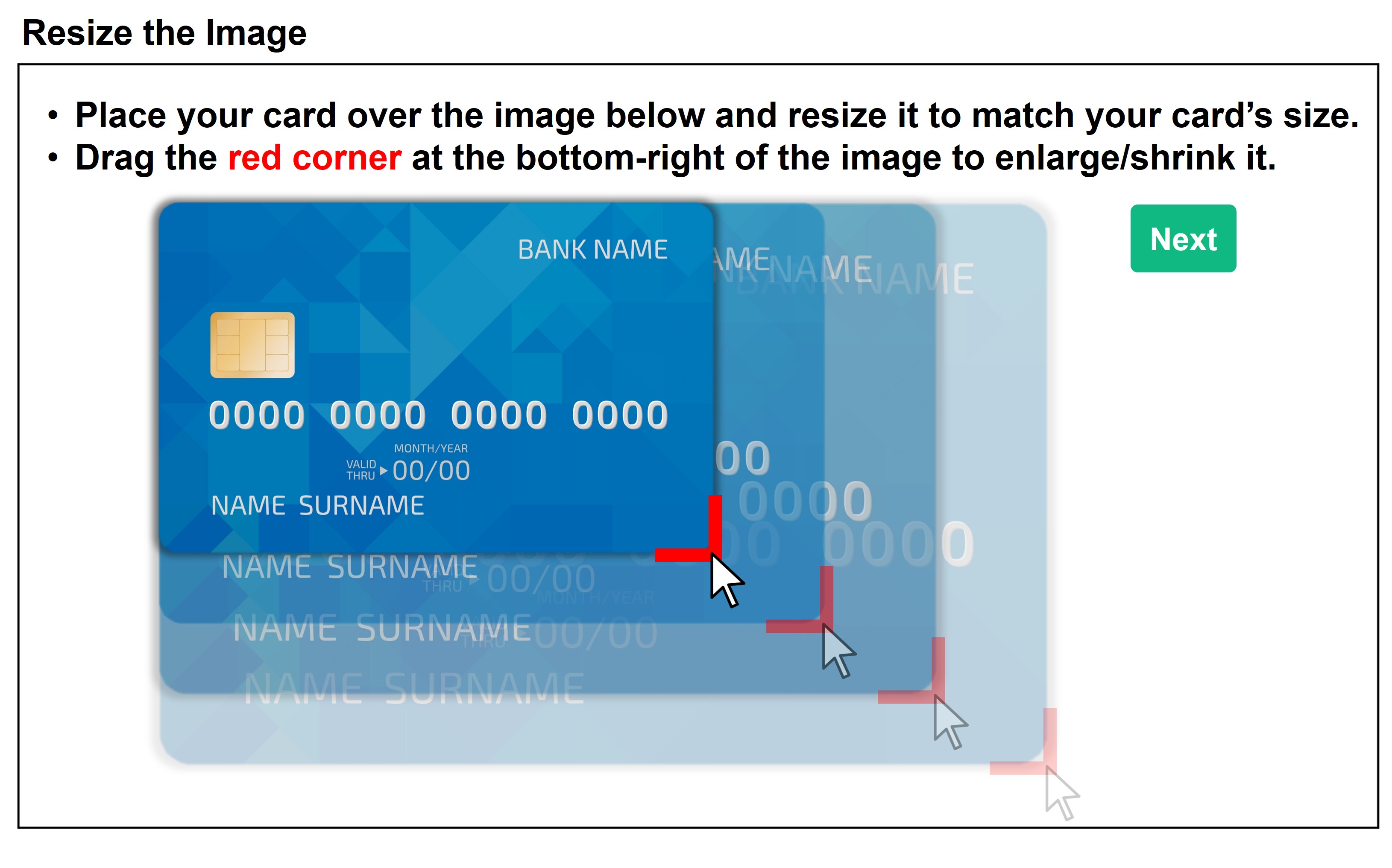}
        \vspace{-10pt}
        \caption{Size-adjustment task in Experiment~1}
        \label{fig:exp1_sizeTask}
        \Description{Screenshot of the PC size-adjustment screen in Experiment 1. The on-screen card image is resized by dragging its corner to match a physical card placed on the monitor.}
    \end{minipage}
    \hfill
    \begin{minipage}[t]{0.49\linewidth}
        \centering
        \includegraphics[width=0.85\linewidth]{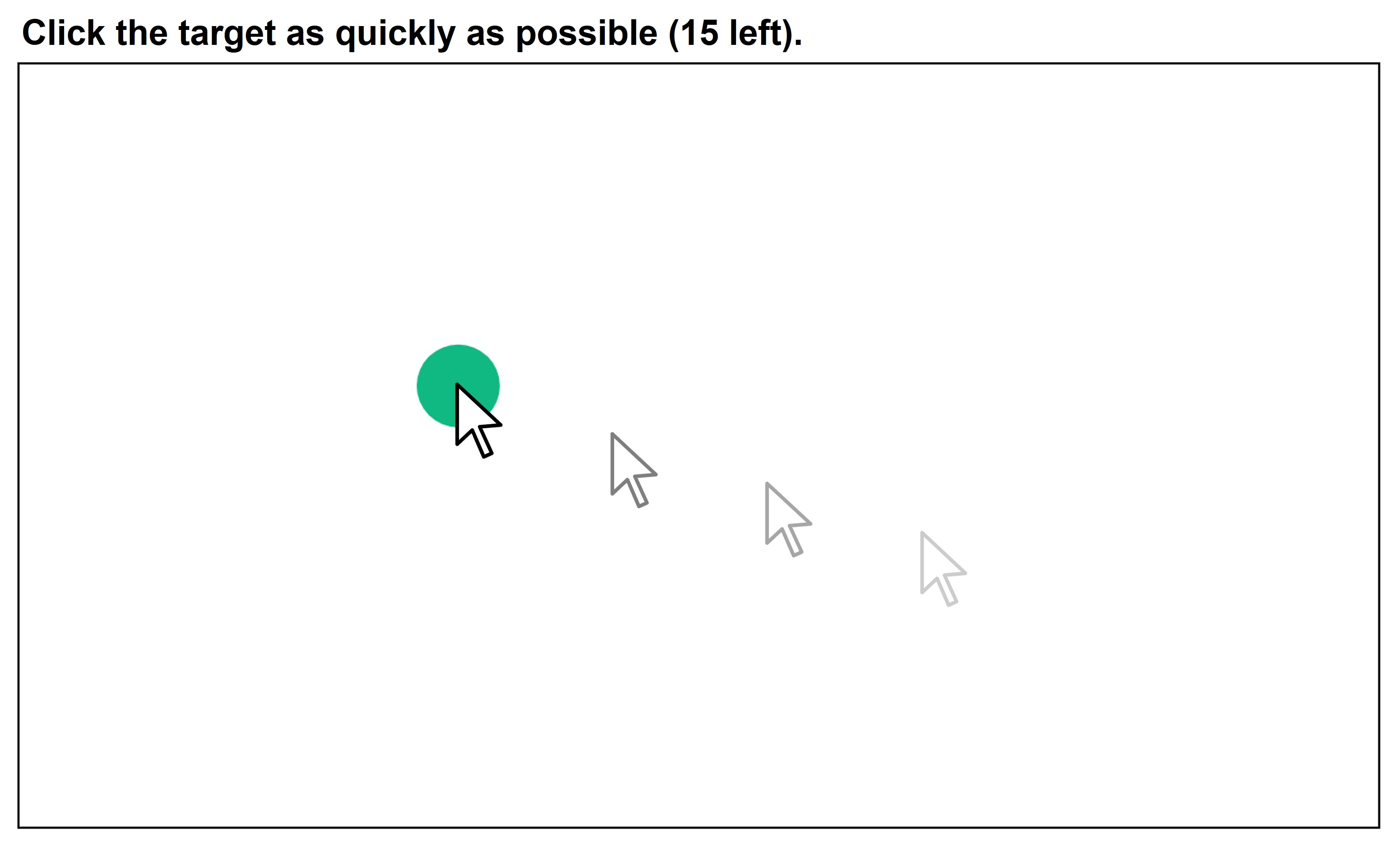}
        \vspace{-10pt}
        \caption{Pointing task in Experiment~1}
        \label{fig:exp1_fittsTask}
        \Description{PC pointing task in Experiment 1 with circular targets in a 1280×720 px area. Participants click targets; on a miss they re-aim until a successful selection.}
    \end{minipage}
\end{figure*}

\subsubsection{Size-Adjustment Task}
As described in Section~\ref{sec:approachPreTask}, the size-adjustment task required participants to place a physical card on the display and adjust an on-screen card image to match its size.
Because pixel density varies across participants' displays in a PC-based crowdsourcing setting, a single size-adjustment trial is insufficient to judge whether a participant adjusted accurately.
Therefore, we administered the task twice and assessed each participant's accuracy based on the discrepancy between the two adjustments.

To avoid cases in which the initial on-screen card size happened to be close to the physical card, thereby eliminating the need to adjust, we set the initial size to be either extremely small or large.
Specifically, the initial width was set to 100~{px} or 900~{px}, and these two conditions were assigned once each across the two repetitions in random order. We allowed the use of physical cards conforming to the ISO/IEC 7810 ID-1 standard (53.98~{mm} $\times$ 85.60~{mm})\hl{~\cite{ISO7810}}.
Representative examples widely used in Japan include credit cards, transportation IC cards such as Suica and PASMO, health insurance cards, and driver's licenses.

The size adjustment was performed on the screen shown in Figure~\ref{fig:exp1_sizeTask} by dragging the lower-right corner of the card image.
This behavior closely follows the \texttt{virtual-chinrest} plugin\footnote{\url{https://www.jspsych.org/v7/plugins/virtual-chinrest/}} of jsPsych~\cite{de2015jspsych}, developed based on Li et al.'s work~\cite{li2020controlling}.
Although size adjustment is typically used to control the physical size of visual stimuli, across Experiments~1–3 we did not control target sizes in the pointing task based on the size-adjustment results.
This is because some participants would operate inaccurately during the size adjustment; if we had computed pixel density (pixels per mm) from their results and used it to scale target sizes in the main task, we might have displayed excessively large or small stimuli.
Thus, the size-adjustment task served solely as a pre-task to identify conforming users.

\subsubsection{Pointing Task: Design and Procedure}
Participants clicked circular targets on the screen according to the given instruction.
After clicking a button labeled \textit{Start}, targets appeared within a $1280\times720$~{px} region (Figure~\ref{fig:exp1_fittsTask}).
When a target was clicked, the next target appeared, and trials proceeded in this manner.
If a click missed the target, participants continued to \textit{re-aim} until they succeeded.
Upon a successful click, the next trial began.

The target distance $A$ was fixed at 510~{px}, and the target diameter $W$ took values of 8, 38, and 78~{px}.
Target positions were random apart from satisfying the distance $A$ constraint. In the main blocks, participants performed the task under two instructions, following prior work ~\cite{Zhai04speed,Yamanaka24ijhcimerit}:
\begin{itemize}
  \item \textbf{As fast as possible:} Click the target as quickly as you can after it appears, but do not act more carelessly than necessary such as clicking without aiming.
  \item \textbf{As accurately as possible:} Avoid errors (clicks outside the target), but do not spend more time than necessary.
\end{itemize}

Each practice and main block consisted of 15 trials.
Among the 15 targets, each $W$ condition appeared five times in random order.
Participants completed four main blocks in total, with the two instructions assigned to two blocks each in random order.
Thus, each participant completed \( 3W \times 5 \text{ trials} \times 2 \text{ instructions} \times 2 \text{ blocks} = 60 \text{ trials} \) in the main blocks.
Each $(W \times \text{instruction})$ condition consisted of \( 5 \text{ trials} \times 2 \text{ blocks} = 10 \) trials.

\subsection{Participants}
A total of 500 participants completed the experiment.
Each received a reward of 200~JPY (approximately 1.3~USD).
The entire task completion time was 4~min~4~s on average, yielding an effective hourly payment of 2{,}951~JPY (approximately 19.1~USD).
We excluded 29 participants who reported using a trackpad rather than a mouse in the questionnaire and 16 participants with missing data, leaving 455 participants (354 male, 101 female) for analysis.

\subsection{Results}
\subsubsection{Size-Adjustment Task}
\begin{figure*}[ht]
    \centering
    \begin{minipage}[t]{0.49\linewidth}
        \centering
        \includegraphics[width=0.75\linewidth]{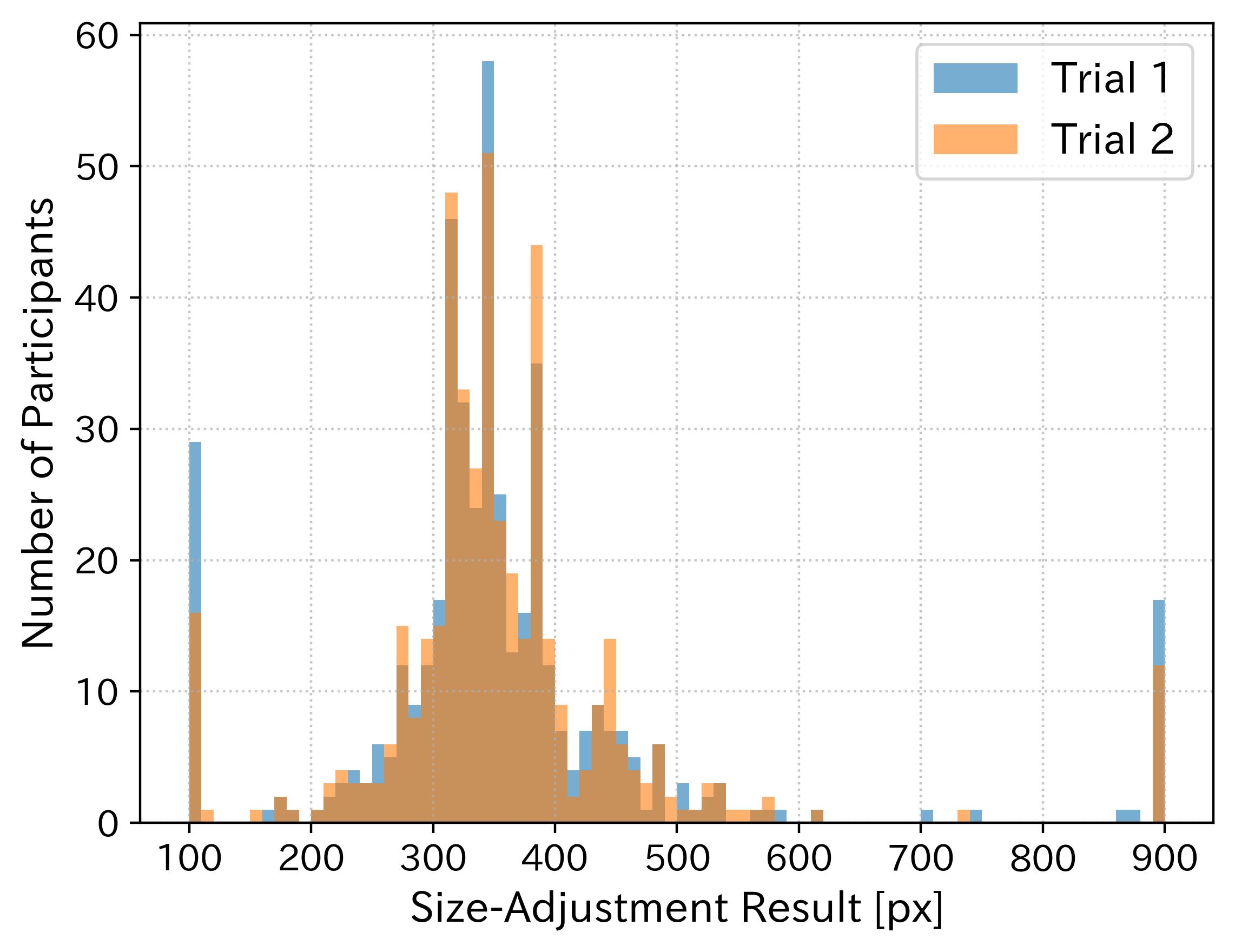}
        \vspace{-5pt}
        \caption{\hl{Distribution of the adjusted long-side length of the card image across the two size-adjustment trials in Experiment~1.}}
        \label{fig:exp1_size_width}
        \Description{Histogram of the adjusted long-side lengths (px) across two size adjustments in Experiment 1.}
    \end{minipage}
    \hfill
    \begin{minipage}[t]{0.49\linewidth}
        \centering
        \includegraphics[width=0.75\linewidth]{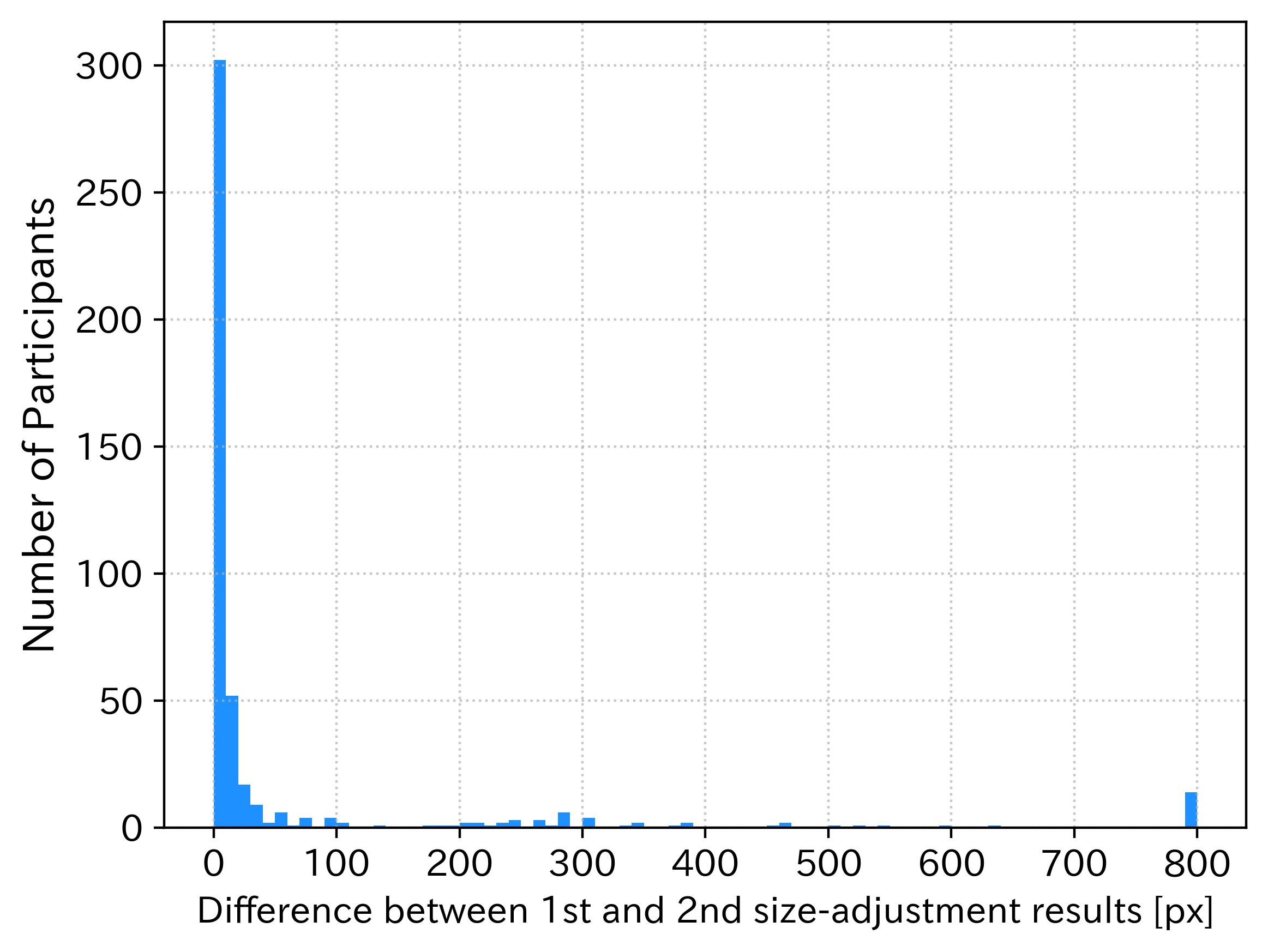}
        \vspace{-5pt}
        \caption{\hl{Distribution of absolute discrepancies between the two size-adjustment trials in Experiment~1.}}
        \label{fig:exp1_size_diff}
        \Description{Histogram of the absolute discrepancy (px) between the two size adjustments in Experiment 1.}
    \end{minipage}
    \hfill
    \begin{minipage}[t]{0.49\linewidth}
        \centering
        \includegraphics[width=0.75\linewidth]{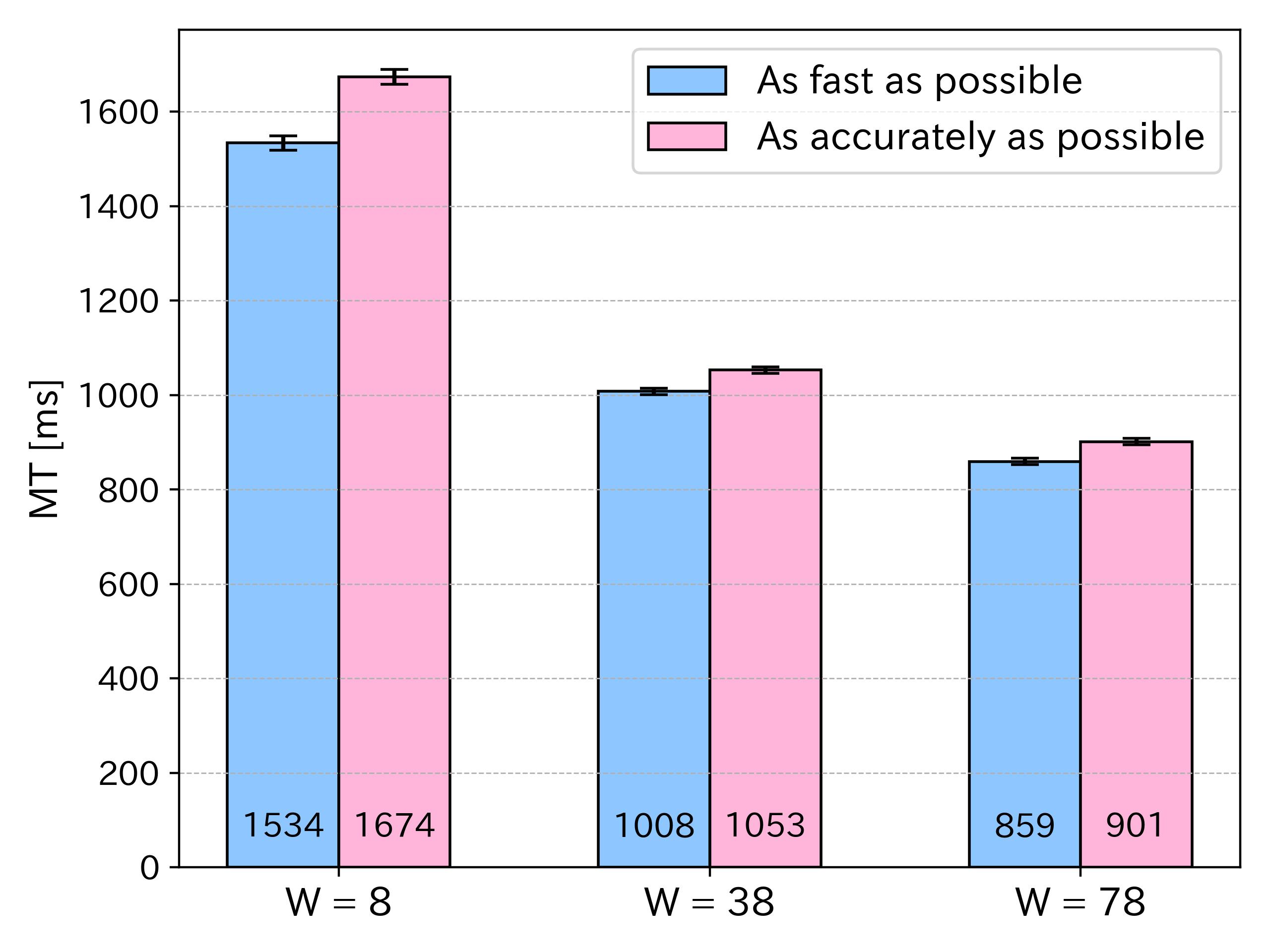}
        \vspace{-5pt}
        \caption{\hl{$MT$ results for each $W$ under the fast and accurate instruction conditions}}
        \label{fig:exp1_fitts_mtAll}
        \Description{Mean movement time (MT) with 95\% CIs across target widths and instruction conditions in Experiment 1.}
    \end{minipage}\hfill
    \begin{minipage}[t]{0.49\linewidth}
        \centering
        \includegraphics[width=0.75\linewidth]{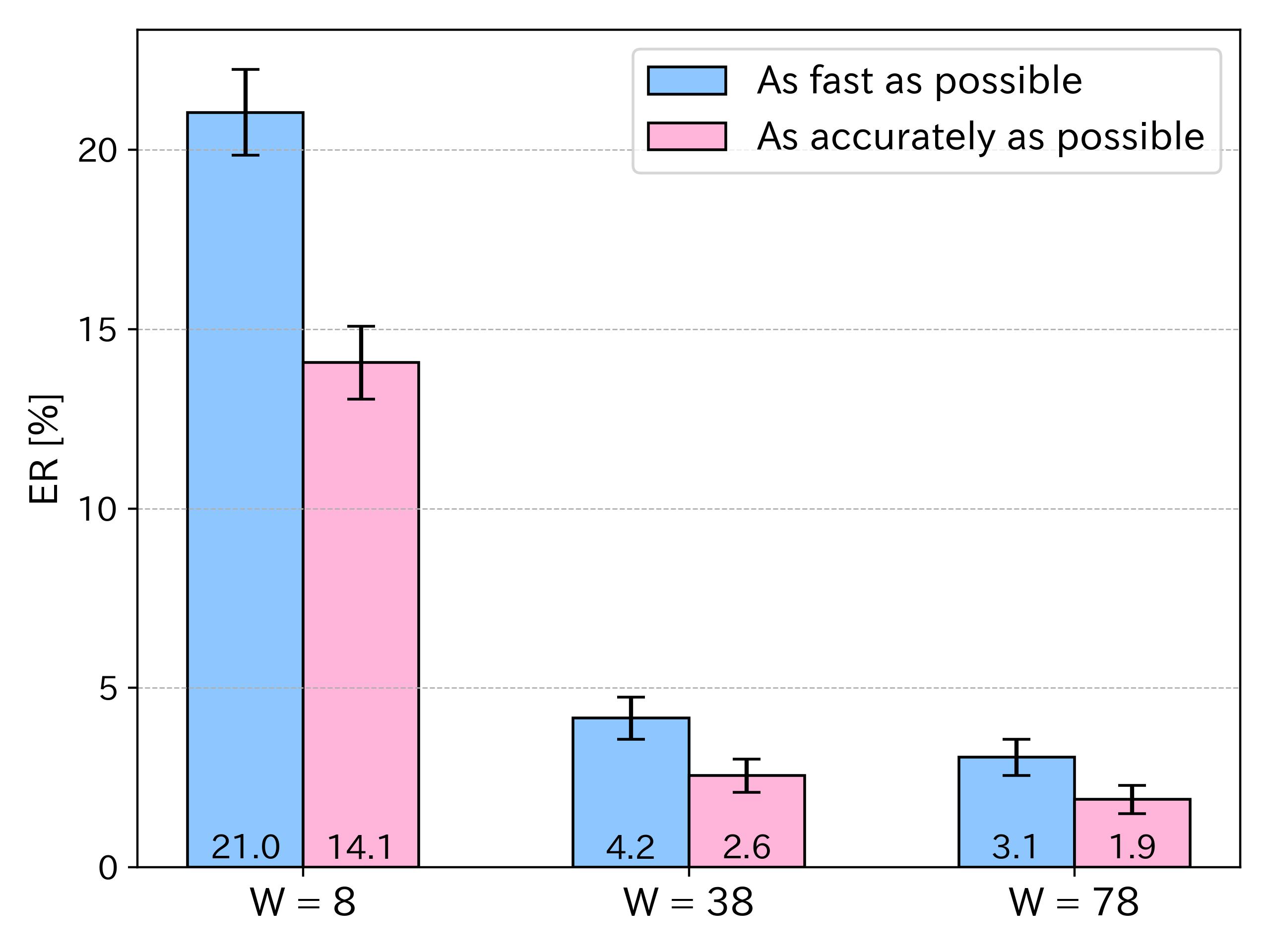}
        \vspace{-5pt}
        \caption{\hl{$ER$ results for each $W$ under the fast and accurate instruction conditions}}
        \label{fig:exp1_fitts_erAll}
        \Description{Mean error rate (ER) with 95\% CIs across target widths and instruction conditions in Experiment 1.}
    \end{minipage}
\end{figure*}

Figure~\ref{fig:exp1_size_width} shows the distribution of the long-side length (px) of the card image after the two adjustments.
Participants' adjusted sizes had a mean of 359.45~{px} (SD=144.17~{px}) in the first trial and 359.41~{px} (SD=121.20~{px}) in the second; 390 participants (86\%) had both adjustments between 200~{px} and 600~{px}.
\hl{Because each participant used a different display with its own pixel density, the absolute pixel length corresponding to the physical card naturally varied across participants. For any given participant, however, the two adjustment trials should yield similar pixel lengths if they follow the instructions and actually align the image with the physical card. Figure~\ref{fig:exp1_size_width} shows that many participants indeed produced consistent sizes within a plausible range (200--600~px)\footnotemark, while a noticeable subset left the image at the initial sizes (100 or 900~px) or produced clearly inconsistent pairs of sizes. These patterns indicate that some participants did not perform the required resizing operation at all or did so inattentively, highlighting the need to use the pre-task as a basis for screening.}

Figure~\ref{fig:exp1_size_diff} shows the distribution of absolute adjustment error across the two trials. Most participants had small errors, specifically, 354 (78\%) were below 20~{px}, but some had large errors, with 73 (16\%) at or above 50~{px}.
The drag-operation times were, on average, 8.03~{s} (SD=6.13~{s}) in the first trial and 6.27~{s} (SD=5.49~{s}) in the second.
However, 38 participants in the first and 25 in the second had 0~{s}, indicating no adjustment at all.

\footnotetext{If the long side of a credit card really corresponded to 200~px on the screen, the display would be approximately 37.3 inches assuming full HD resolution; if it were 600~px, it would be 12.4 inches. We set thresholds under the assumption that contemporary displays typically fall within this range, though displays outside this range may also be used. Therefore, in Experiments~2 and~3 we limited the devices to iPhones, for which we can obtain screen size in both mm and px, to evaluate the effectiveness of our screening more rigorously.}

\subsubsection{Pointing Task}\label{sec:exp1_result_fitts}
We first removed outliers.
For each participant and each \(\text{instruction} \times W\) condition (10 trials), we excluded trials with outlier click coordinates.
We rotated coordinates so that the direction from the previous target center to the current target center aligned with the positive $x$-axis; we then recorded the $x$-coordinate as positive if the click overshot the current target center and negative if it fell short~\cite{soukoreff2004towards,Wobbrock11dim}. Using the standard deviation $\sigma$ of these $x$-coordinates, we excluded trials whose $x$-coordinates deviated from the mean by more than $3\sigma$ ~\cite{soukoreff2004towards}.
For the first target in a block, the \textit{Start} button is treated as the \textit{previous target}.
Similarly, for the $10$ $MT$ values in each \(\text{instruction} \times W\) condition, we excluded outliers using the same $3\sigma$ rule.
Finally, using the mean $MT$ across all 60 trials per participant, we applied the $3\sigma$ rule and excluded that participant.
This procedure identified five participants (1.1\%) as outliers.
No trial-level outliers were detected for click coordinates or $MT$.

Figure~\ref{fig:exp1_fitts_mtAll} shows $MT$ and Figure~\ref{fig:exp1_fitts_erAll} shows $ER$ results.
Throughout the paper, error bars indicate 95\% confidence intervals.
\hl{
Because prior research has shown that ANOVA is robust against violations of the normality assumption \cite{Blanca17,Schmider10,Tsandilas24}, we consistently ran RM-ANOVAs (independent variables were instruction and $W$; dependent variables were $MT$ and $ER$).
Bonferroni correction was used to adjust the $p$-values in pairwise tests.
Analyses not directly related to our research questions (e.g., normality tests for $MT$ or complete results of omnibus tests across conditions) are provided in the supplementary materials.
}

\hl{
For $MT$, we found significant main effects of instruction ($F_{1,449}=173.4$, $p<0.001$, $\eta_g^2=0.026$) and $W$ ($F_{2,898}=4053$, $p<0.001$, $\eta_g^2=0.65$).
The interaction effect of $W \times$ instruction was significant ($F_{2,898}=62.00$, $p<0.001$, $\eta_g^2=0.010$).
For $ER$, we found significant main effects of instruction ($F_{1,449}=56.61$, $p<0.001$, $\eta_g^2=0.022$) and $W$ ($F_{2,898}=612.5$, $p<0.001$, $\eta_g^2=0.29$).
The interaction effect of $W \times$ instruction was significant ($F_{2,898}=34.41$, $p<0.001$, $\eta_g^2=0.014$).
}
Briefly, $MT$ and $ER$ decreased as the target size increased \hl{($p<0.001$ for all pairwise tests)}, and the participants changed their $MT$ and $ER$ according to the given instructions \hl{($p<0.001$ for both)}, consistent with prior work ~\cite{Yamanaka24ijhcimerit,Zhai04speed}.
Thus, on average, participants exhibited typical pointing behavior.

\subsection{Simulation}\label{sec:exp1_sim}
\subsubsection{Overview}
To assess the utility of our screening method, we simulated cases in which participants whose behavior is identified by the pre-task screening as nonconforming are mixed into the sample.
Assuming a researcher's goal is to evaluate the goodness of fit of existing performance models, we examine whether screening improves model fit.
To this end, we compute the fit of pointing-performance models as the proportion of nonconforming participants varies.

The GUI performance models we employ are given in Equations~(\ref{eq:exp1_fitts_mt})--(\ref{eq:exp1_fitts_er}).
Lowercase italics $a$–$f$ are regression coefficients estimated from the data.
Equation~(\ref{eq:exp1_fitts_mt}) is Fitts' law, which predicts movement time $MT$ from the target distance $A$ and diameter $W$ ~\cite{mackenzie1992fitts}.
Equation~(\ref{eq:exp1_fitts_sigma}) estimates spreads of click coordinates based on target size.
For each $W$, after rotating the task axis (x-axis) towards the current target as explained in Section~\ref{sec:exp1_result_fitts}, we regress the standard deviations $\sigma_x$ and $\sigma_y$ of click coordinates (with the target center as origin) along the $x$- and $y$-axes ~\cite{yamanaka2021utility,Yamanaka23HPO}.
Equation~(\ref{eq:exp1_fitts_er}) predicts the success probability $P(D)$ that a click $(x,y)$ falls within a circular target region $D$ of diameter $W$. Using $\sigma_x$ and $\sigma_y$ predicted by Equation~(\ref{eq:exp1_fitts_sigma}), we obtain $P(D)$ for each $W$ ~\cite{Yamanaka23HPO}.
\begin{equation}
    \label{eq:exp1_fitts_mt}
    MT=a+b\mathit{ID},\ \ \mathit{ID}=\log_2\left(\frac{A}{W}+1\right)
\end{equation}
\begin{equation}
    \label{eq:exp1_fitts_sigma}
    \sigma_x=c+dW,\ \ \sigma_y=e+fW 
\end{equation}
\begin{equation}
    \label{eq:exp1_fitts_er}
    P(D)=\iint_{D}\frac{1}{2\pi\sigma_x\sigma_y}\exp\left(-\frac{x^2}{2\sigma_x^2}-\frac{y^2}{2\sigma_y^2}\right)dxdy
\end{equation}
The error rate is predicted to be $1-P(D)$.
These models have shown high goodness of fit (typically $R^2$) in prior work ~\cite{soukoreff2004towards,Yamanaka23HPO}.
Hence, if data quality is sufficient, we should obtain similar results.
Conversely, if data quality is poor, for instance, if participants do not vary $MT$ or $ER$ with $W$, model fit will degrade.

\subsubsection{Procedure}
Our screening aims to split participants into a ``passing group'' and a ``non-passing group'' based on the pre-task, and to administer the main task only to the passing group.
After applying the outlier handling in Section~\ref{sec:exp1_result_fitts}, we classified conforming participants into the passing group as follows: those whose first and second adjusted sizes both fell between 200~{px} and 600~{px} (the proper range, see Figure~\ref{fig:exp1_size_width}) and whose absolute discrepancy between the two adjustments was smaller than a threshold $T$~px (see Figure~\ref{fig:exp1_size_diff}).
All others were assigned to the non-passing group.
We incorporated the range criterion to exclude participants whose two adjustments were both near extreme bounds of the slider (suggesting nonconformity) and those who might not have used a physical card yet happened to exhibit a small adjustment discrepancy.

In the simulation, for a total sample size $N$ and a non-passing proportion $X\%$, we randomly drew $N\times X\%$ participants from the non-passing group and $N\times(100-X)\%$ from the passing group.
We then computed model fit for Equations~(\ref{eq:exp1_fitts_mt}) and (\ref{eq:exp1_fitts_er}) on the pointing task data of the sampled set.
We repeated this sampling-and-fitting procedure 1{,}000 times and averaged the fit.

We varied $N$, $T$, and $X$ to analyze how the inclusion of nonconforming participants affects researchers' model-fit results.
Specifically, we set $N\in\{10,20,40,80\}$, $T$ from 5~{px} to 50~{px} in steps of 5~{px}, and $X$ from 0\% to 100\% in steps of 10\%.
We capped $N$ at 80 and $T$ at 50~{px} because larger $N$ requires more non-passing data, and overly lenient $T$ reduces the number of non-passing participants, making it difficult to secure sufficient data for analysis.
\hl{Because Experiment~1 serves as a preliminary PC-based check, in the main text we explain the results of $N=40$ as an example. The other sample sizes ($N=10,20,80$) showed similar quantitative trends and are thus summarized in the supplementary materials.}
Figure~\ref{fig:exp1_group_N} shows the numbers of passing and non-passing participants as $T$ varies.

\begin{figure}[ht]
    \centering
    \includegraphics[width=0.85\linewidth]{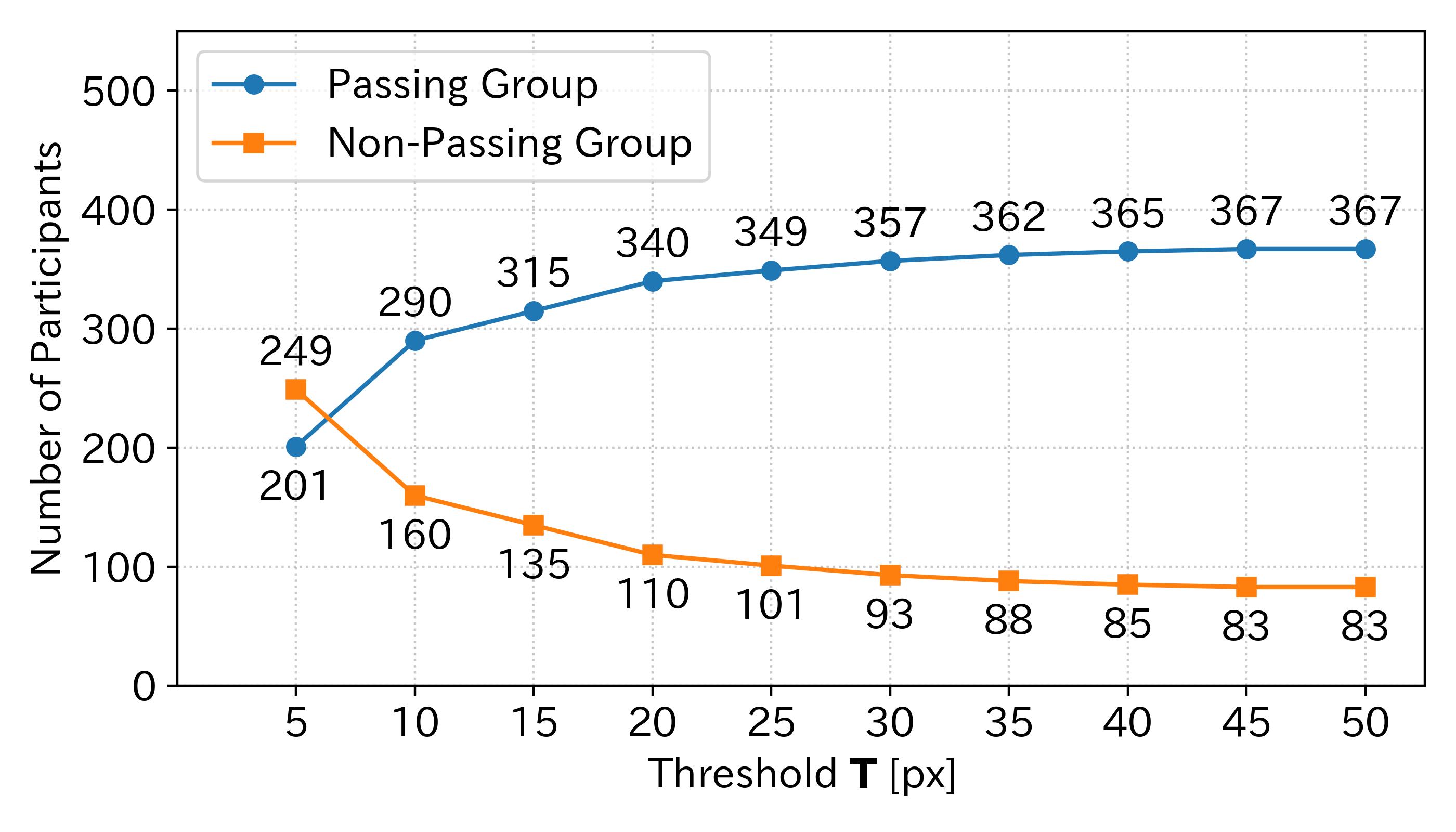}
    \caption{Numbers of passing vs.\ non-passing participants as a function of the threshold}
    \label{fig:exp1_group_N}
    \Description{Line chart showing counts of passing vs. non-passing participants as a function of threshold T (px) in Experiment 1. Group sizes vary systematically with T.}
\end{figure}

\subsection{Simulation Results}
\subsubsection{Movement Time ($MT$)}
\begin{figure*}[ht]
    \centering
    \begin{minipage}[t]{0.49\linewidth}
        \centering
        \includegraphics[width=\linewidth]{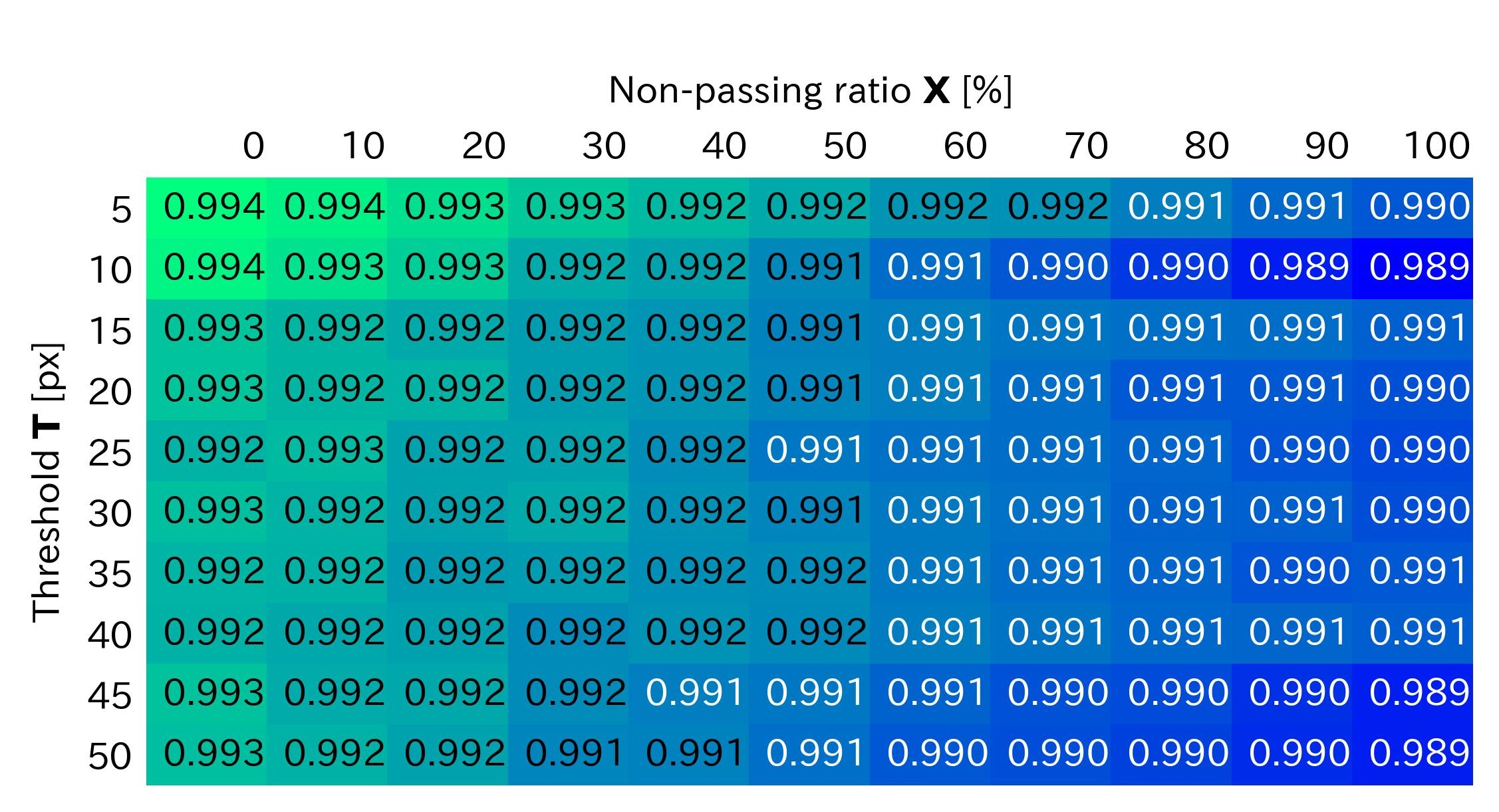}
        \vspace{-15pt}
        \caption{Experiment~1: Goodness of fit to Equation~(\ref{eq:exp1_fitts_mt}) for $MT$ (fast)}
        \label{fig:exp1_fitts_mt_speed_R2_sim}
        \Description{Heatmap (N=40) of R^2 for the MT model under the “fast” instruction in Experiment 1, across threshold T and non-passing proportion X. Fit remains very high with limited variation.}
    \end{minipage}
    \hfill
    \begin{minipage}[t]{0.49\linewidth}
        \centering
        \includegraphics[width=\linewidth]{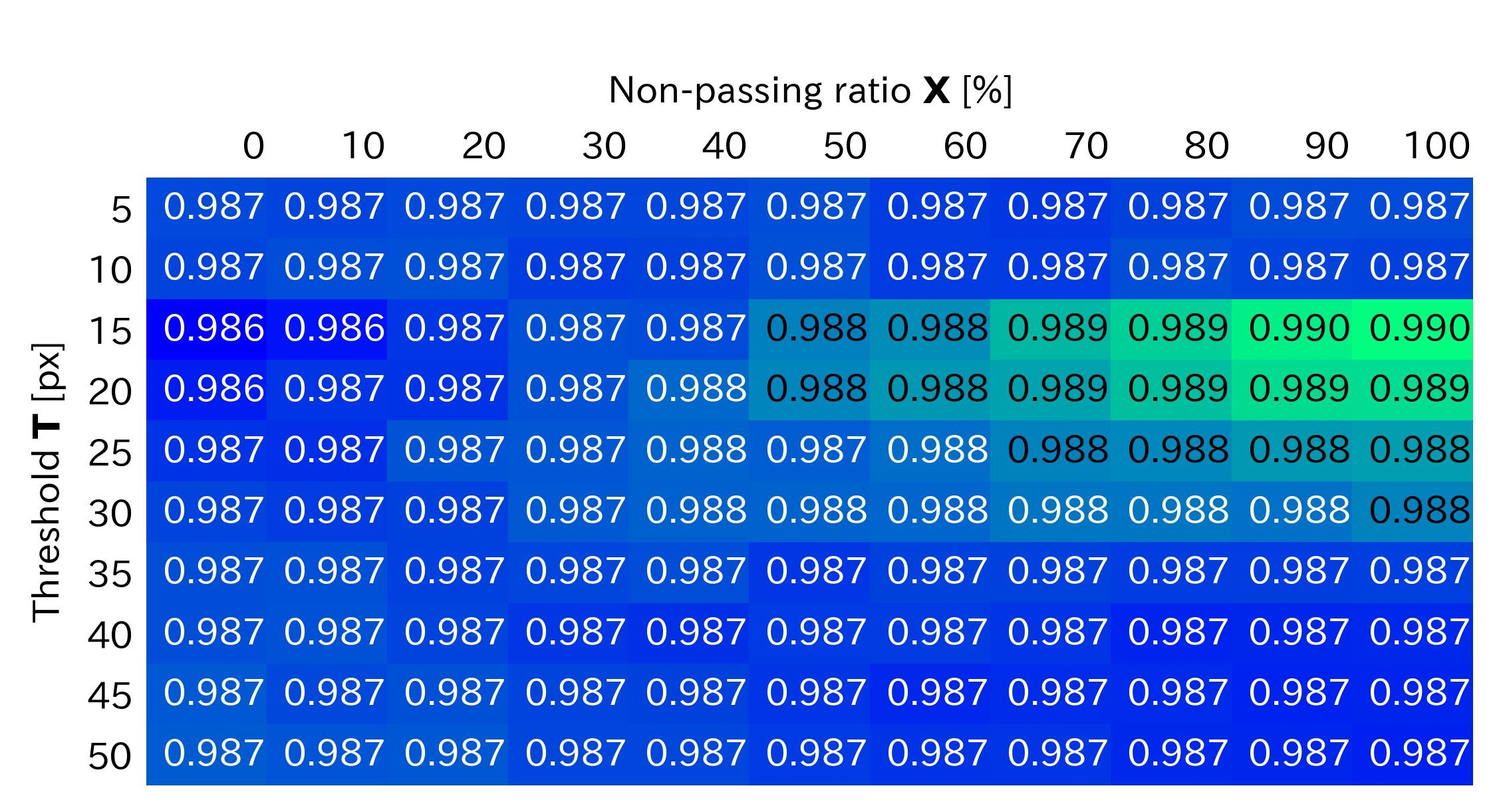}
        \vspace{-15pt}
        \caption{Experiment~1: Goodness of fit to Equation~(\ref{eq:exp1_fitts_mt}) for $MT$ (accurate)}
        \label{fig:exp1_fitts_mt_accuracy_R2_sim}
        \Description{Heatmap (N=40) of R^2 for the MT model under the “accurate” instruction in Experiment 1. R^2 stays above 0.98 with little sensitivity to T or X.}
    \end{minipage}
    \begin{minipage}[t]{0.49\linewidth}
        \centering
        \includegraphics[width=\linewidth]{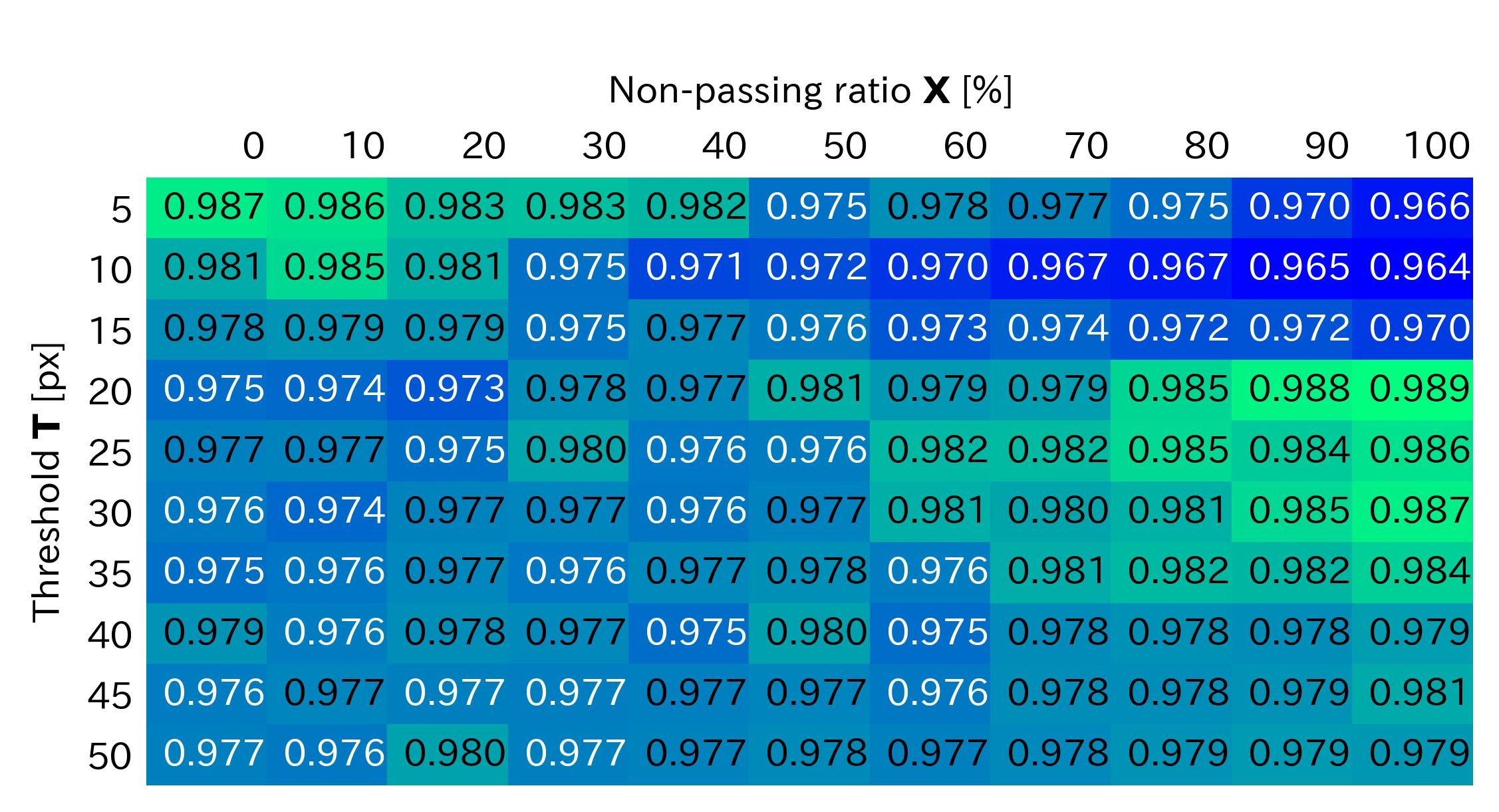}
        \vspace{-15pt}
        \caption{Experiment~1: Goodness of fit to Equation~(\ref{eq:exp1_fitts_er}) for $ER$ (fast)}
        \label{fig:exp1_fitts_er_speed_R2_sim}
        \Description{Heatmap (N=40) of R^2 for the ER model under the “fast” instruction in Experiment 1, varying T and X. No clear monotonic degradation with increasing X.}
    \end{minipage}
    \hfill
    \begin{minipage}[t]{0.49\linewidth}
        \centering
        \includegraphics[width=\linewidth]{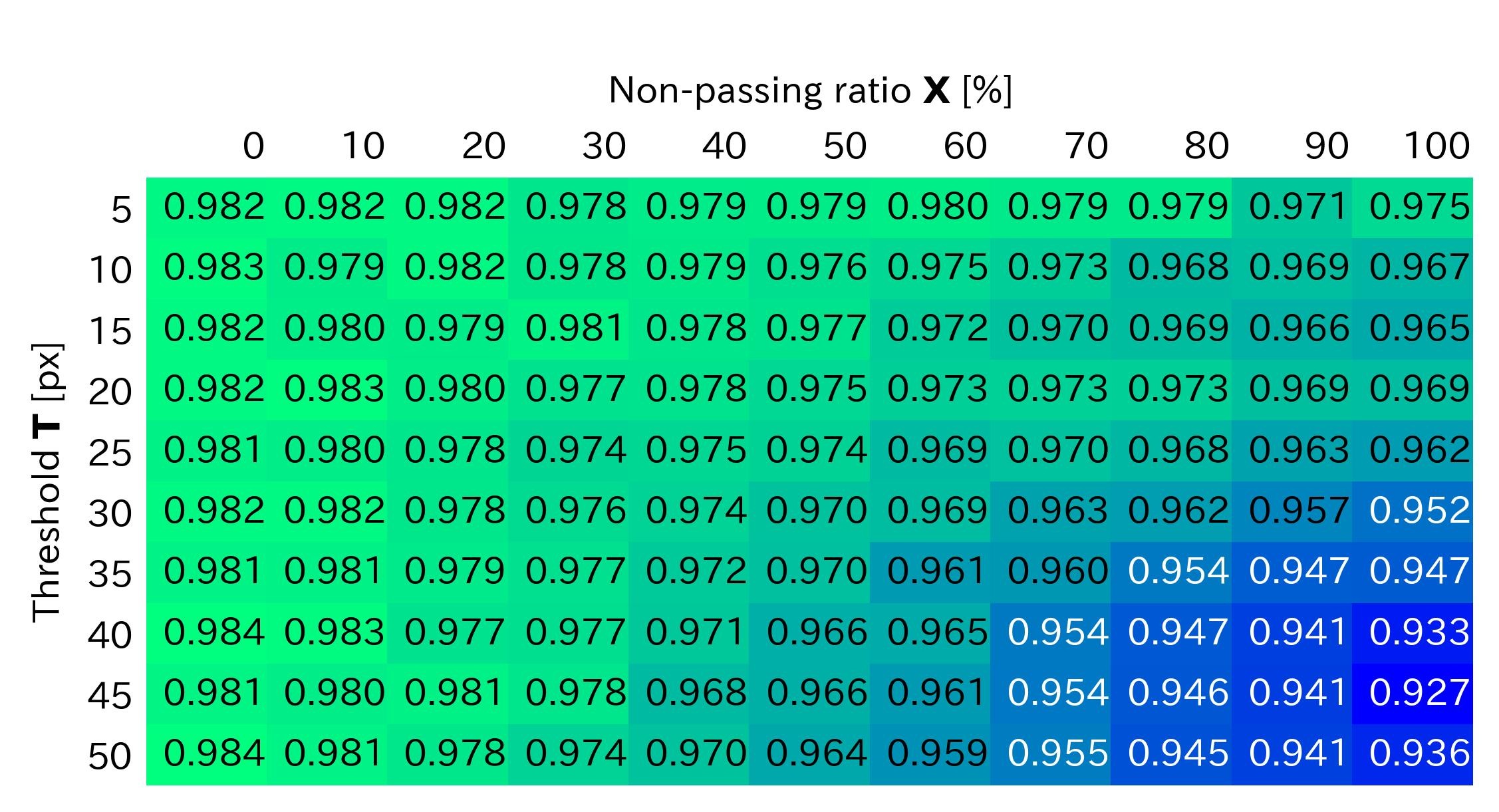}
        \vspace{-15pt}
        \caption{Experiment~1: Goodness of fit to Equation~(\ref{eq:exp1_fitts_er}) for $ER$ (accurate)}
        \label{fig:exp1_fitts_er_accuracy_R2_sim}
        \Description{Heatmap (N=40) of R^2 for the ER model under the “accurate” instruction in Experiment 1. Fit degrades from upper-left to lower-right as T loosens and X increases.}
    \end{minipage}
\end{figure*}

Figures~\ref{fig:exp1_fitts_mt_speed_R2_sim} and \ref{fig:exp1_fitts_mt_accuracy_R2_sim} show $R^2$ of Fitts' law for Equation~(\ref{eq:exp1_fitts_mt}) when varying $T$ and $X$ for movement time $MT$.
Figure~\ref{fig:exp1_fitts_mt_speed_R2_sim} corresponds to the ``fast'' instruction and Figure~\ref{fig:exp1_fitts_mt_accuracy_R2_sim} to the ``accurate'' instruction.
\hl{For $N=40$, $R^2$ remained above 0.98 for all $T$ and $X$ in both the fast and accurate conditions, with very little variation.}
A likely reason is that Fitts' law tends to achieve high fit even with relatively small samples and repetitions compared to $ER$ models ~\cite{yamanaka2021utility}.
In addition, participants with larger discrepancies in the pre-task are expected to act more carelessly in the main task, but Equation~(\ref{eq:exp1_fitts_mt}) models $MT$, so accuracy-based screening plays a smaller role.
Consequently, even a higher proportion of careless participants may have limited impact on $MT$-model fit.

Therefore, when validating an $MT$ model such as Equation~(\ref{eq:exp1_fitts_mt}) via crowdsourcing, the utility of our screening may be limited.
Note that we did not evaluate cross-validated prediction for unseen task conditions in Experiment~1, because the number of $W$ conditions (three) was too small to support an appropriate cross-validation design.

\subsubsection{Error Rate ($ER$)}\label{sec:exp1_sim_er}
Figures~\ref{fig:exp1_fitts_er_speed_R2_sim} and \ref{fig:exp1_fitts_er_accuracy_R2_sim} present the analogous analysis for error rate $ER$, showing $R^2$ for Equation~(\ref{eq:exp1_fitts_er}).
\hl{As with $MT$, the $R^2$ values for $N=40$ did not remarkably change with $T$ and $X$.}
Yet, in Figure~\ref{fig:exp1_fitts_er_accuracy_R2_sim}, $R^2$ tends to decrease from the upper-left to the lower-right.
This indicates that the model fit declines as the threshold $T$ is more lenient (large) and as the non-passing proportion is larger.
When $T$ is lenient, participants with very large pre-task discrepancies populate the non-passing group; as their proportion increases, model fit worsens.
Hence, when validating an $ER$ model such as Equation~(\ref{eq:exp1_fitts_er}) via crowdsourcing, screening participants and restricting the main experiment to conforming users can improve model fit.

The high prediction accuracy of Equation~(\ref{eq:exp1_fitts_er}) has been confirmed in previous studies~\cite{yamanaka2021utility,Yamanaka23HPO}, and thus this simulation result demonstrates that our screening method enables researchers to draw valid experimental conclusions on model evaluation.
Because our screening leverages accuracy in the size-adjustment pre-task, it is particularly beneficial when evaluating $ER$ models.
The trend that lower non-passing proportions yield higher $R^2$ further supports the usefulness of screening.

\hl{We also examined the other total sample sizes $N\in\{10,20,80\}$; larger $N$ tended to yield slightly higher $R^2$, but the qualitative pattern that stricter $T$ and smaller $X$ improve model fit was unchanged. Detailed heatmaps for these sample sizes are provided in the supplementary materials.}

By contrast, Figure~\ref{fig:exp1_fitts_er_speed_R2_sim} does not show the same trend.
Under the ``fast'' instruction, participants are forced to act quickly, and thus the difference in accuracy between the passing (presumably accurate) and non-passing (presumably careless) groups may be attenuated, which could explain the lack of a clear effect of $X$.

\subsection{Limitations}
Although Figure~\ref{fig:exp1_fitts_er_accuracy_R2_sim} suggests that screening is effective, the top-left cells (threshold $T=5$, non-passing proportion $X=0$) are not always the best-fitting across the heatmap.
Since $T=5$ is the strictest threshold and $X=0$ includes no non-passing participants, this cell represents the ideal sample for an experimenter.
A plausible reason behind this non-optimal result is that we used somewhat limited data size for precise model evaluation: only three $W$ conditions and $10$ trials per \(\text{instruction} \times W\) condition.

To more rigorously assess the utility of our screening, we should increase the number of $W$ levels and trials per condition in subsequent experiments.
Furthermore, our evaluation of the size-ad\-just\-ment task was based on the discrepancy between the two trials, but a participant who did not use the designated physical card might, by chance, exhibit a small discrepancy.
Thus, we could not guarantee that the adjusted size actually matched the physical card.
Addressing these limitations is necessary for a stricter validation of the size-adjustment task; we do so in Experiment~2.

\section{Experiment 2: iPhone-Based Experiment with Re-Aiming}
\hl{Regarding participant overlap across experiments, workers who took part in Experiment~1 were not prevented from later participating in Experiments~2 or~3, and we did not track whether such overlap occurred. By contrast, Experiments~2 and~3 were posted as separate tasks at different times, and we ensured that no worker participated in both.
Because Experiment~1 used a PC-based mouse interaction setting whereas Experiments~2 and~3 used smartphone-based touch interaction with different task structures, we consider any learning effects across experiments to be negligible.}

\begin{figure*}[ht]
    \vspace{10pt}
    \centering
    \includegraphics[width=0.65\linewidth]{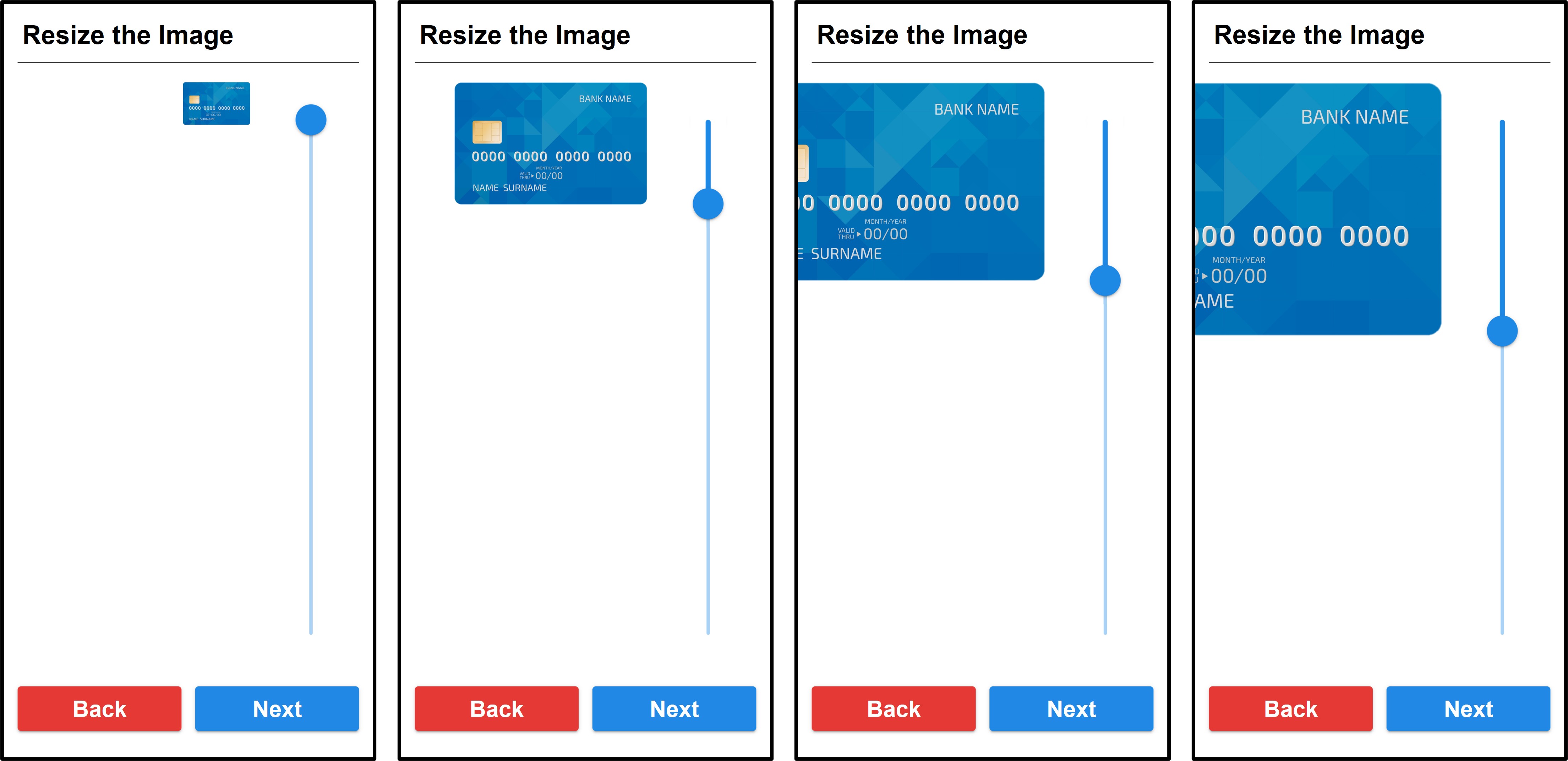}
    \caption{Size-adjustment screen in Experiment~2}
    \label{fig:exp2_sizeTask}
    \Description{iPhone size-adjustment screen in Experiment 2. Participants resize the on-screen card image (short side) to match a physical card placed on the device.}
\end{figure*}
\begin{figure*}[ht]
    \centering
    \includegraphics[width=0.85\linewidth]{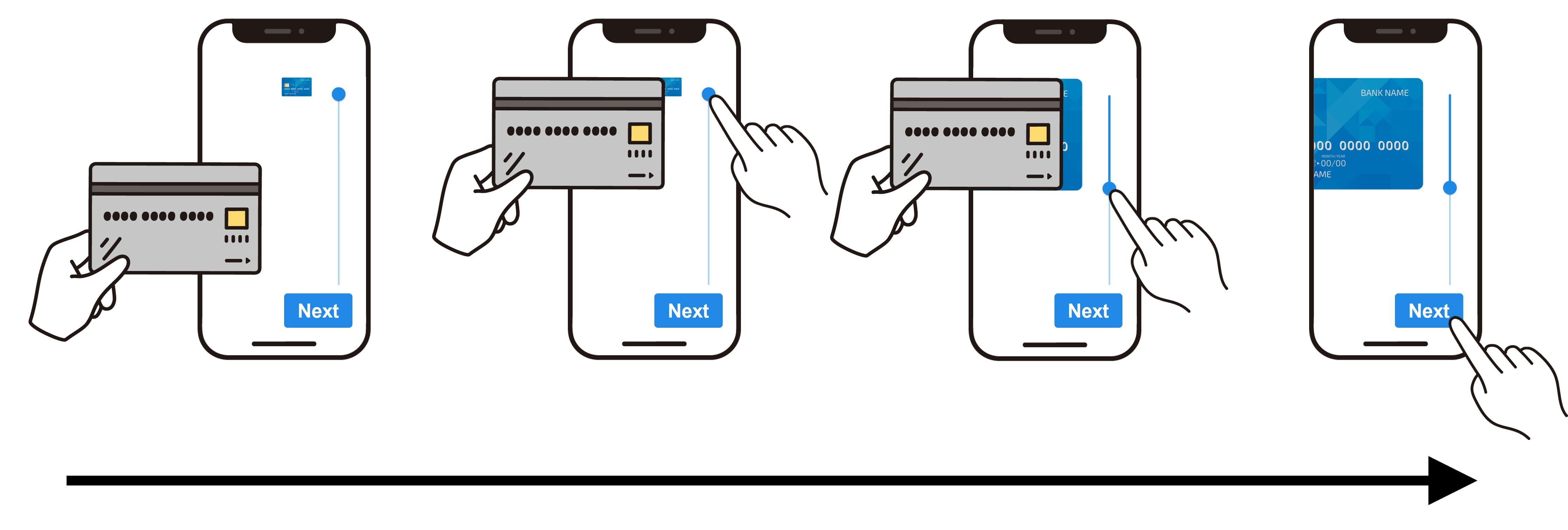}
    \vspace{-15pt}
    \caption{How participants operated the size-adjustment task}
    \label{fig:exp2_sizeTask2}
    \Description{Illustration of how participants operate the iPhone size-adjustment task in Experiment 2: placing a physical card on the screen and adjusting the on-screen card image.}
\end{figure*}

\subsection{Task and Design}
\subsubsection{Overview}
In Experiment~2, we examined more rigorously whether screening via the size-adjustment pre-task works as intended by restricting the devices to iPhones.
We can identify each device's display pixel density and accurately evaluate whether a participant's adjusted size matches the physical card.
Specifically, we used iPhone screen-resolution data\footnote{\url{https://www.ios-resolution.com/}} to infer each participant's device PPI from the resolution obtained by the experimental system.
We then converted the adjustment results from px to mm using that PPI, enabling strict measurement of error relative to the physical card.

We also tightened control in the pointing task.
First, targets were rendered in mm based on each participant's device PPI.
We further increased the number of $W$ conditions to nine and the number of trials per \(\text{instruction} \times W\) condition to 20, allowing a more rigorous evaluation of model fit.
This addressed the limitation in Experiment~1, where few $W$ levels and repetitions hindered thorough model evaluation.

The procedure largely followed Experiment~1: after instructions, participants completed one size-adjustment task, one practice block of the pointing task, and four main blocks.
Whereas Experiment~1 used two size adjustments, here a single adjustment sufficed because restricting to iPhones provided a correct reference for each participant.
The study targeted iPhone~7 and later models to properly render targets and texts.

\subsubsection{Size-Adjustment Task}
This pre-task was conducted on the screen shown in Figure~\ref{fig:exp2_sizeTask}; participants resized the image via a slider.
As in Figure~\ref{fig:exp2_sizeTask2}, participants placed a physical ISO/IEC~7810 ID-1 card on the iPhone and matched the shorter side of the on-screen card image to the card's shorter side.
We set the initial size of the shorter side of the image to 50~{px}, because this ensures that, on any iPhone model, it is smaller than the shorter side of the physical card (53.98~mm), thereby requiring slider dragging.\footnote{For example, iPhone 16 has 460 PPI and Scale Factor of 3, yielding 50~{px} as 8.28~{mm} on the screen.}
As in Experiment~1, we did not scale pointing task targets based on the pre-task's inferred pixel density.
Instead, we controlled target size using the PPI identified from screen resolution.

\begin{figure*}[ht]
    \centering
    \begin{minipage}[t]{0.49\linewidth}
        \centering
        \includegraphics[width=0.8\linewidth]{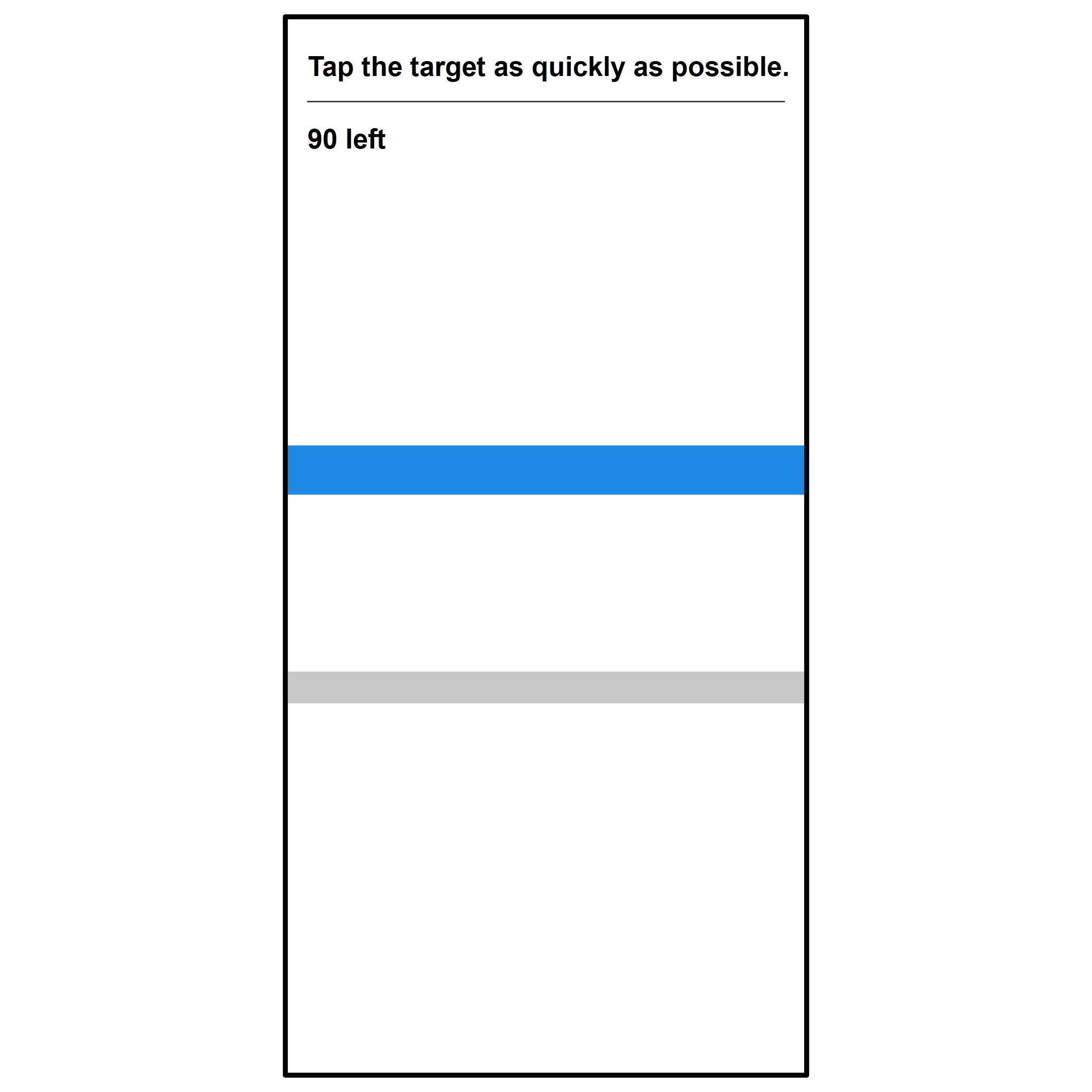}
        \caption{\hl{Smartphone pointing-task screen in Experiment~2. Two horizontally oriented rectangular targets span the full screen width, and their height is set to the target width $W$ in mm.}}
        \label{fig:exp2_fittsTask}
        \Description{iPhone pointing-task screen in Experiment 2 with two full-width horizontal targets of height W mm. Participants alternate tapping top and bottom targets.}
    \end{minipage}%
    \hfill
    \begin{minipage}[t]{0.49\linewidth}
        \centering
        \includegraphics[width=0.6\linewidth]{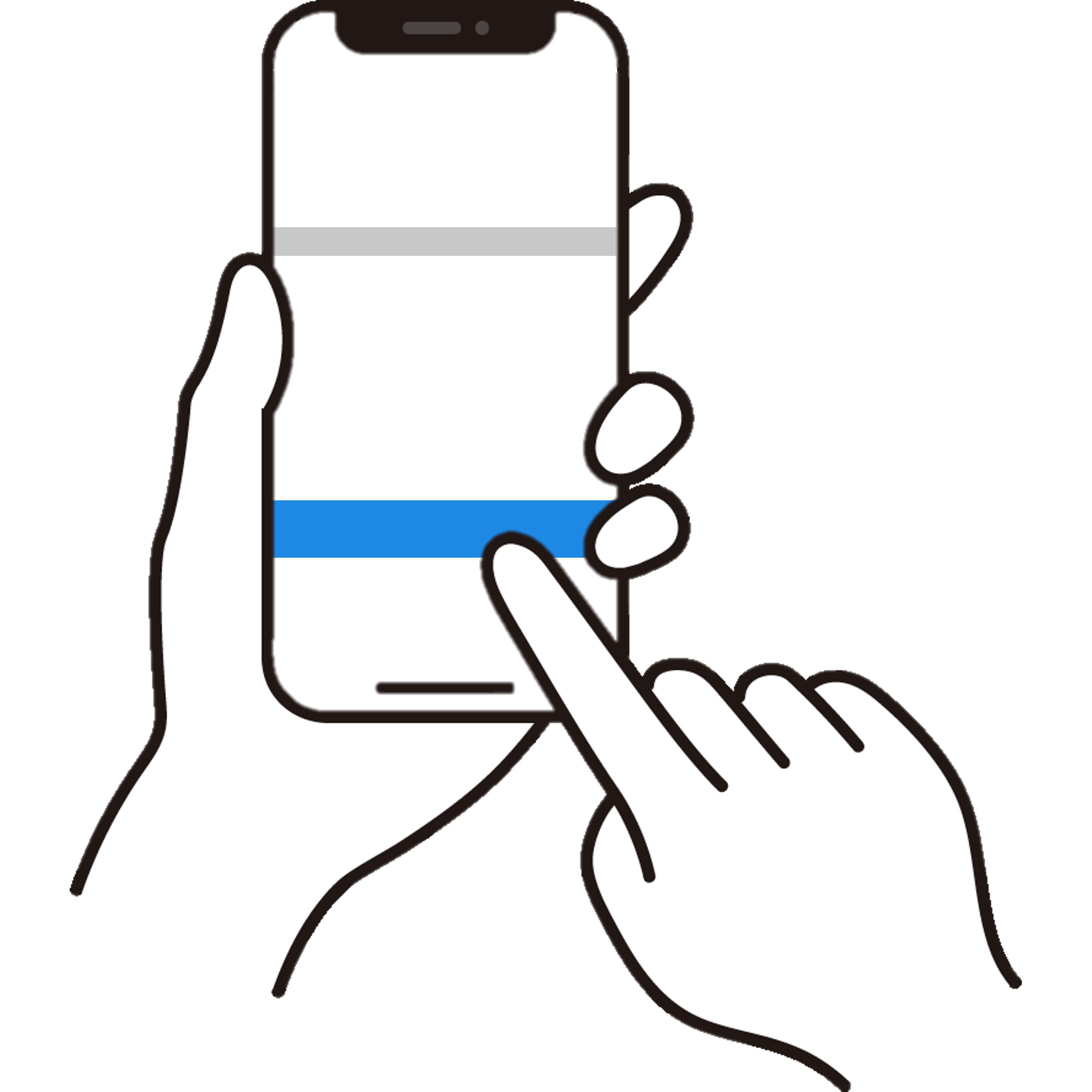}
        \caption{\hl{Instructed posture for the smartphone pointing task in Experiments~2. Participants were asked to hold the phone in the non-dominant hand and tap the targets with the index finger of the dominant hand.}}
        \label{fig:exp2_fittsTask2}
        \Description{Illustration of the tapping posture in Experiment 2: hold the phone with the non-dominant hand and tap with the dominant index finger.}
    \end{minipage}
\end{figure*}

\subsubsection{Pointing Task}
As shown in Figure~\ref{fig:exp2_fittsTask}, two horizontally long rectangular targets, which could have different heights, were displayed on the screen.
Each target spanned the full-screen width and was $W$~mm in height.
Participants tapped them alternately according to the instructions.
The top target initially served as the start button.
When participants tapped a target, its height changed to the next $W$ condition.
As in Experiment~1, if a tap missed, participants continued re-aiming until they succeeded.
Upon a successful tap, they could proceed to tap the opposite target.

The center-to-center distance of targets $A$ was 30~{mm} (one level). The target height $W$ took nine levels: 2, 2.8, 3.6, 4.4, 5.2, 6, 6.8, 7.6, and 8.4~{mm}.
This range covered very easy to difficult conditions, yielding error rates from approximately 0\% up to 29\% ~\cite{yamanaka2020rethinking,Yamanaka24iss}.
Target centers appeared at two fixed vertical positions and alternated top and bottom.
As in Experiment~1, the two instructions were ``as fast as possible'' and ``as accurately as possible,'' and the practice instruction was ``as fast and accurately as possible.''
Participants were instructed to hold the smartphone in the non-dominant hand and tap with the dominant index finger (Figure~\ref{fig:exp2_fittsTask2}) ~\cite{bi2016predicting,yamanaka2020rethinking}.

Each block comprised 90 pointing trials, with each of the nine $W$ levels appearing 10 times in random order.
In the main phase, participants completed four blocks; the two instructions appeared twice each in random order.
Thus, each participant completed \( 9W\times 10\ \text{trials} \times 2\ \text{instructions} \times 2\ \text{blocks} = 360\ \text{trials} \), yielding \(10\ \text{trials} \times 2\ \text{blocks} = 20\) trials per $(W \times \text{instruction})$ condition.

\subsection{Participants}
A total of 584 participants completed the experiment.
Each received 250~JPY (approximately 1.7~USD).
The mean task completion time was 6~min~43~s, yielding an effective hourly payment of 2{,}233~JPY (approximately 15.2~USD).
We excluded 8 participants with missing data and 42 for whom the iPhone PPI could not be identified, leaving 534 participants for analysis (297 male, 235 female, 2 other).

\subsection{Results}
\subsubsection{Size-Adjustment Task}
\begin{figure}[ht]
    \vspace{-10pt}
    \centering
    \includegraphics[width=0.85\linewidth]{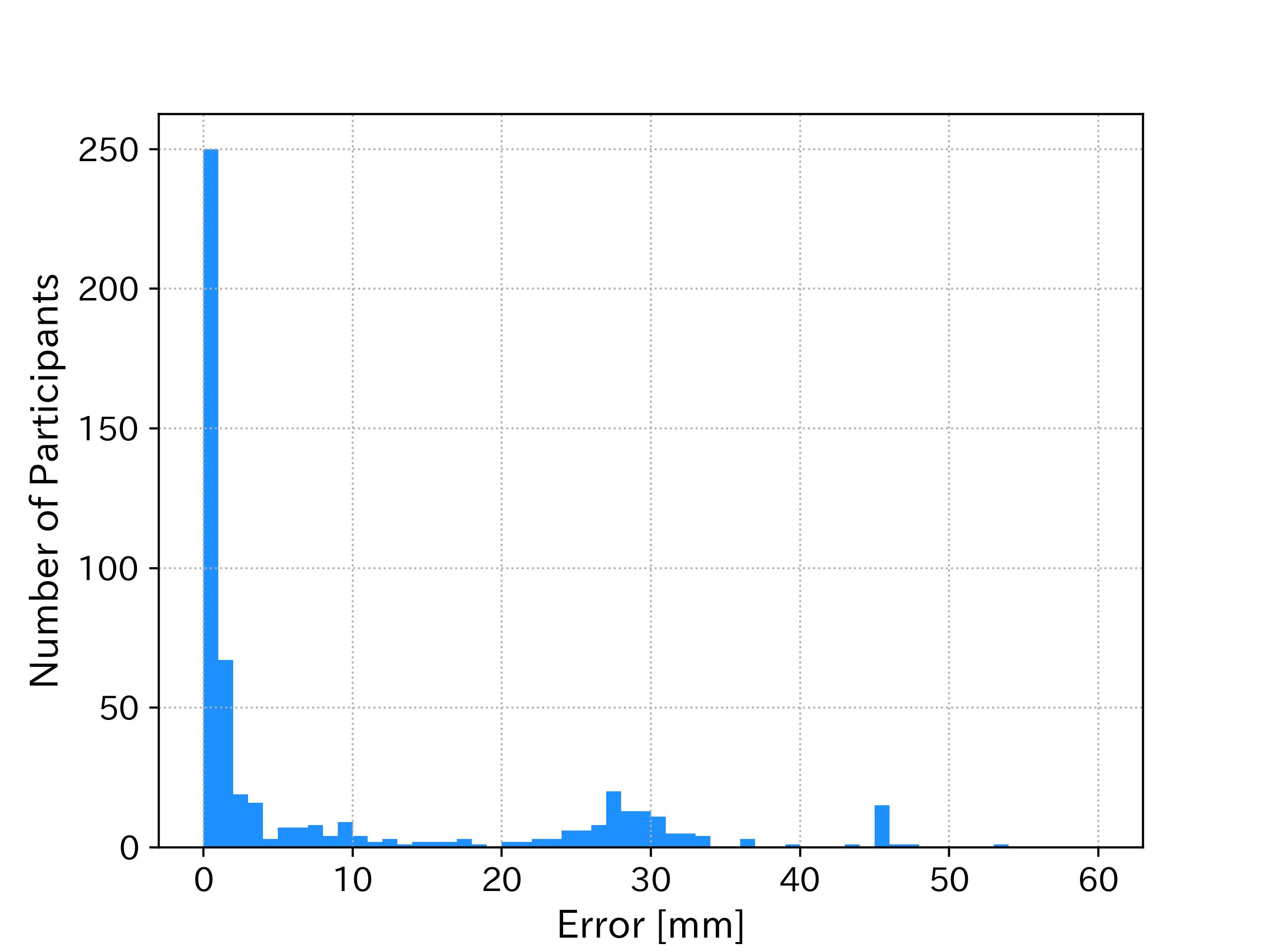}
    \vspace{-5pt}
    \caption{Distribution of size-adjustment errors}
    \label{fig:exp2_size_diff}
    \Description{Histogram of absolute size-adjustment errors (mm) in Experiment 2. Many participants show small errors; some exhibit large deviations.}
\end{figure}

Figure~\ref{fig:exp2_size_diff} shows the distribution of absolute errors relative to the correct size in the size-adjustment task.
Most participants had small errors, specifically, 317 (59\%) were below 2~{mm}, but some showed large errors, with 144 (27\%) at or above 10~{mm}.
The operation time for the adjustment averaged 8.86~{s} (SD=7.06~{s}); five participants recorded 0~{s}, indicating no adjustment at all.

\subsubsection{Pointing Task}\label{sec:exp2_result_fitts}
To remove outliers on tap coordinates, we measured $y$ as positive when the tap landed below the target center and negative when above, and excluded trials whose $y$-coordinates deviated from the mean by more than $3\sigma$.
Other criteria on trial- and participant-level $MT$ followed Experiment 1 (see Section~\ref{sec:exp1_result_fitts}).
This procedure excluded 1{,}042 trials for tap coordinates, 2{,}547 trials for $MT$, and four participants for $MT$.
These counts are not mutually exclusive, as some trials and participants met multiple exclusion criteria, and we analyzed the remaining 187{,}347 trials (97.5\%).

\begin{figure*}[ht]
    \centering
    \begin{minipage}[b]{0.49\linewidth}
        \centering
        \includegraphics[width=\linewidth]{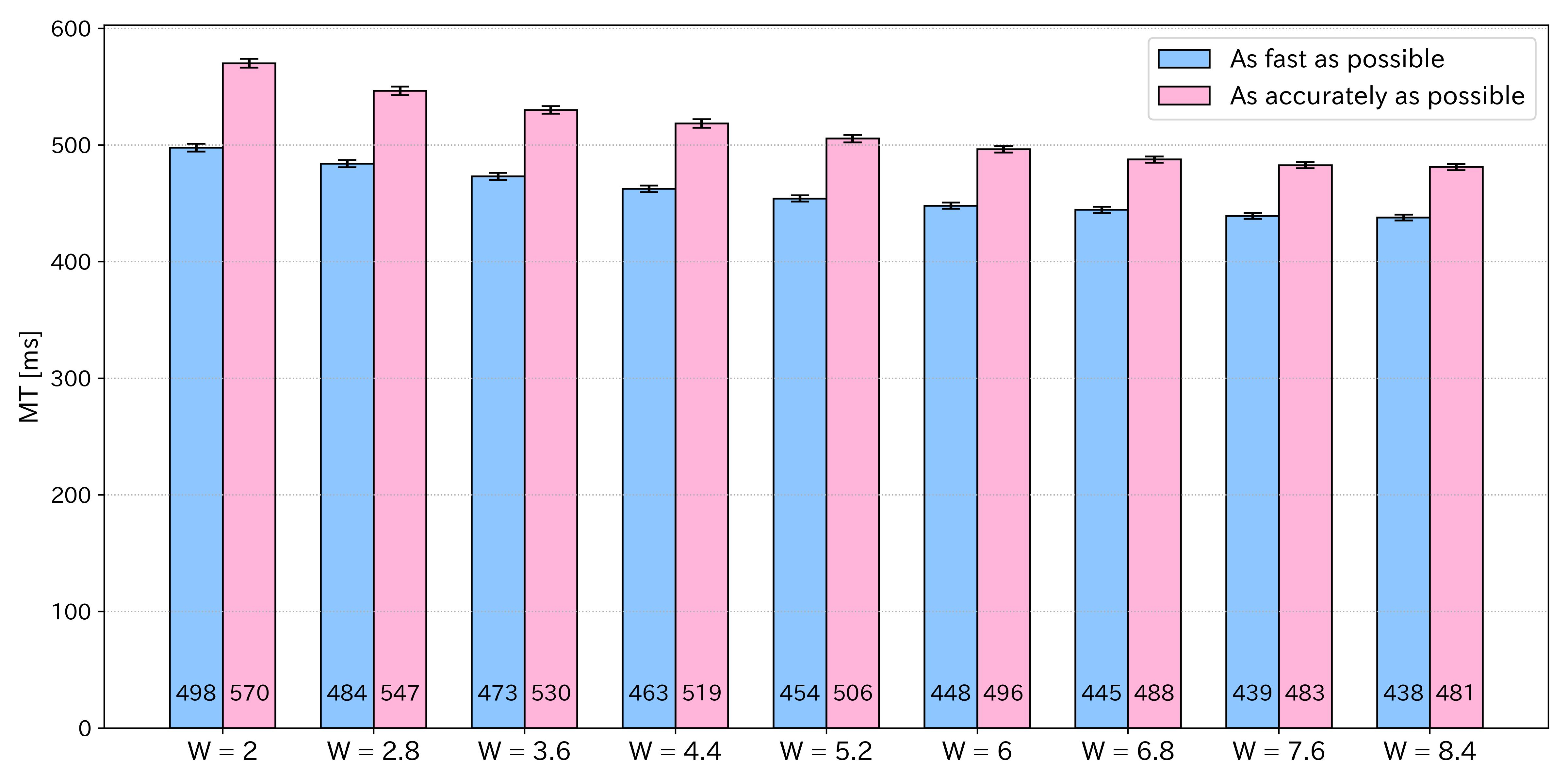}
        \vspace{-15pt}
        \caption{\hl{$MT$ results for each $W$ under the fast and accurate instruction conditions}}
        \label{fig:exp2_fitts_mtAll}
        \Description{Mean movement time (MT) with 95\% CIs by target height W and instruction in Experiment 2 (iPhone). MT varies with W and instruction.}
    \end{minipage}
    \hfill
    \begin{minipage}[b]{0.49\linewidth}
        \centering
        \includegraphics[width=\linewidth]{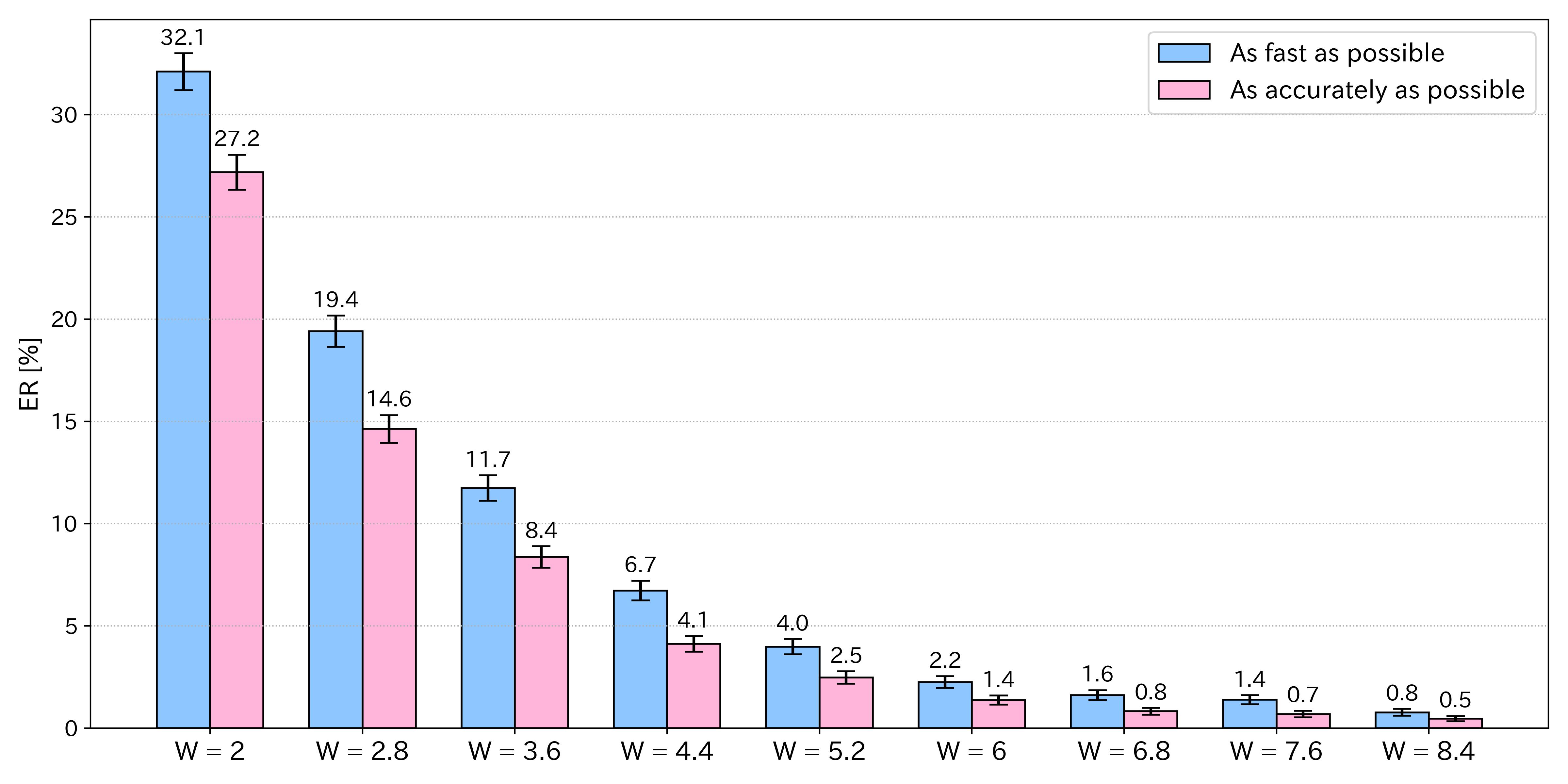}
        \vspace{-15pt}
        \caption{\hl{$ER$ results for each $W$ under the fast and accurate instruction conditions}}
        \label{fig:exp2_fitts_erAll}
        \Description{Mean error rate (ER) with 95\% CIs by target height W and instruction in Experiment 2 (iPhone). ER is more sensitive to W, reaching higher values at small W.}
    \end{minipage}
\end{figure*}

Figures~\ref{fig:exp2_fitts_mtAll} and \ref{fig:exp2_fitts_erAll} show the results of the pointing task in Experiment~2: $MT$ by $(W \times \text{instruction})$ and $ER$ by $(W \times \text{instruction})$, respectively.
\hl{
For $MT$, we found significant main effects of instruction ($F_{1,529}=256.3$, $p<0.001$, $\eta_g^2=0.042$) and $W$ ($F_{8,4232}=493.3$, $p<0.001$, $\eta_g^2=0.035$).
The interaction effect of $W \times$ instruction was significant ($F_{8,4232}=38.75$, $p<0.001$, $\eta_g^2=0.0014$).
For $ER$, we found significant main effects of instruction ($F_{1,529}=121.9$, $p<0.001$, $\eta_g^2=0.014$) and $W$ ($F_{8,4232}=1751$, $p<0.001$, $\eta_g^2=0.51$).
The interaction effect of $W \times$ instruction was significant ($F_{8,4232}=22.22$, $p<0.001$, $\eta_g^2=0.0084$).
When $W$ increased from 2.0 to 8.4~mm under the fast condition, $MT$ changed from 498 to 438~ms, i.e., change ratio was $(498-438)/498= 12\%$.
In contrast, $ER$ changed from 32.1 to 0.8\%, i.e., change ratio was $(32.1-0.8)/32.1= 98\%$.
A similar trend was found for the accurate condition.
This result indicates that $ER$ is more sensitive to $W$ than $MT$, which is also supported by the larger effect size ($\eta_g^2=0.035$ for $MT$ vs. $0.51$ for $ER$).
}
As intended, $ER$ reached roughly 30\% at the smallest $W$.

\subsection{Simulation}\label{sec:exp2_sim}
\subsubsection{Overview}
As in Experiment~1 (Section~\ref{sec:exp1_sim}), we simulated outcomes when participants whose behavior is identified by the pre-task screening as nonconforming are mixed into the sample.
In addition to Equation~(\ref{eq:exp1_fitts_mt}) to model $MT$ (Fitts' law), the GUI performance models used in Experiment~2 are Equations~(\ref{eq:exp2_fitts_sigma}) and (\ref{eq:exp2_fitts_er}), where $g$ and $h$ are regression coefficients estimated from the data.
Equation~(\ref{eq:exp2_fitts_sigma}) shows that the variance of tap $y$-coordinates is linearly related to the square of target size ~\cite{bi2016predicting,yamanaka2020rethinking}.
We can then predict the tap success probability $P(-W/2 \le Y \le W/2)$ for each $W$ by Equation~(\ref{eq:exp2_fitts_er}) using $\sigma_y$ from Equation~(\ref{eq:exp2_fitts_sigma}).
\begin{equation}
    \label{eq:exp2_fitts_sigma}
    \sigma_y^2=g+hW^2 
\end{equation}
\begin{equation}
    \label{eq:exp2_fitts_er}
    P\left(-\frac{W}{2}\le Y\le\frac{W}{2}\right)=\operatorname{erf}\left(\frac{W}{2\sqrt{2}\sigma_y}\right)
\end{equation}

\subsubsection{Procedure}
\begin{figure}[ht]
    \vspace{-5pt}
    \centering
    \includegraphics[width=0.85\linewidth]{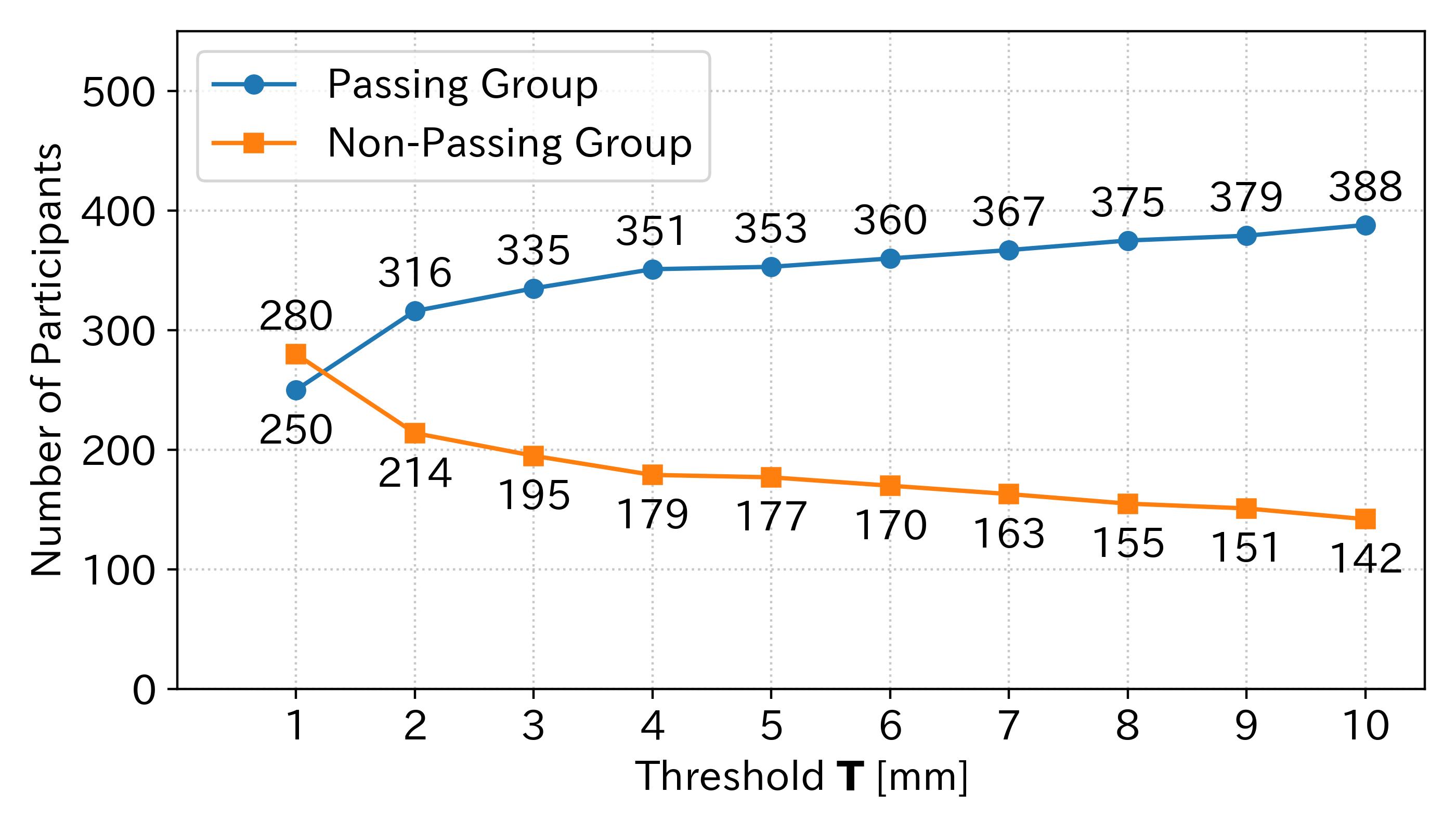}
    \caption{Numbers of passing vs.\ non-passing participants as a function of the threshold}
    \label{fig:exp2_group_N}
    \Description{Counts of passing vs. non-passing participants versus threshold T (mm) in Experiment 2. Passing counts decline with stricter thresholds.}
\end{figure}

In Experiment~2, participants whose adjustment error was smaller than a threshold $T$ (mm) were assigned to the passing group; all others were assigned to the non-passing group.
We varied $T$ from 1~{mm} to 10~{mm} in 1~{mm} steps.
The values of $N$ and $X$ were the same as in Experiment 1.
Figure~\ref{fig:exp2_group_N} shows the numbers of passing and non-passing participants as $T$ varies.
After applying the outlier handling in Section~\ref{sec:exp2_result_fitts}, we followed the same simulation procedure as in Experiment~1.

\subsection{Simulation Results}
\subsubsection{Movement Time ($MT$)}
\begin{figure*}[ht]
    \centering
    \includegraphics[width=0.49\linewidth]{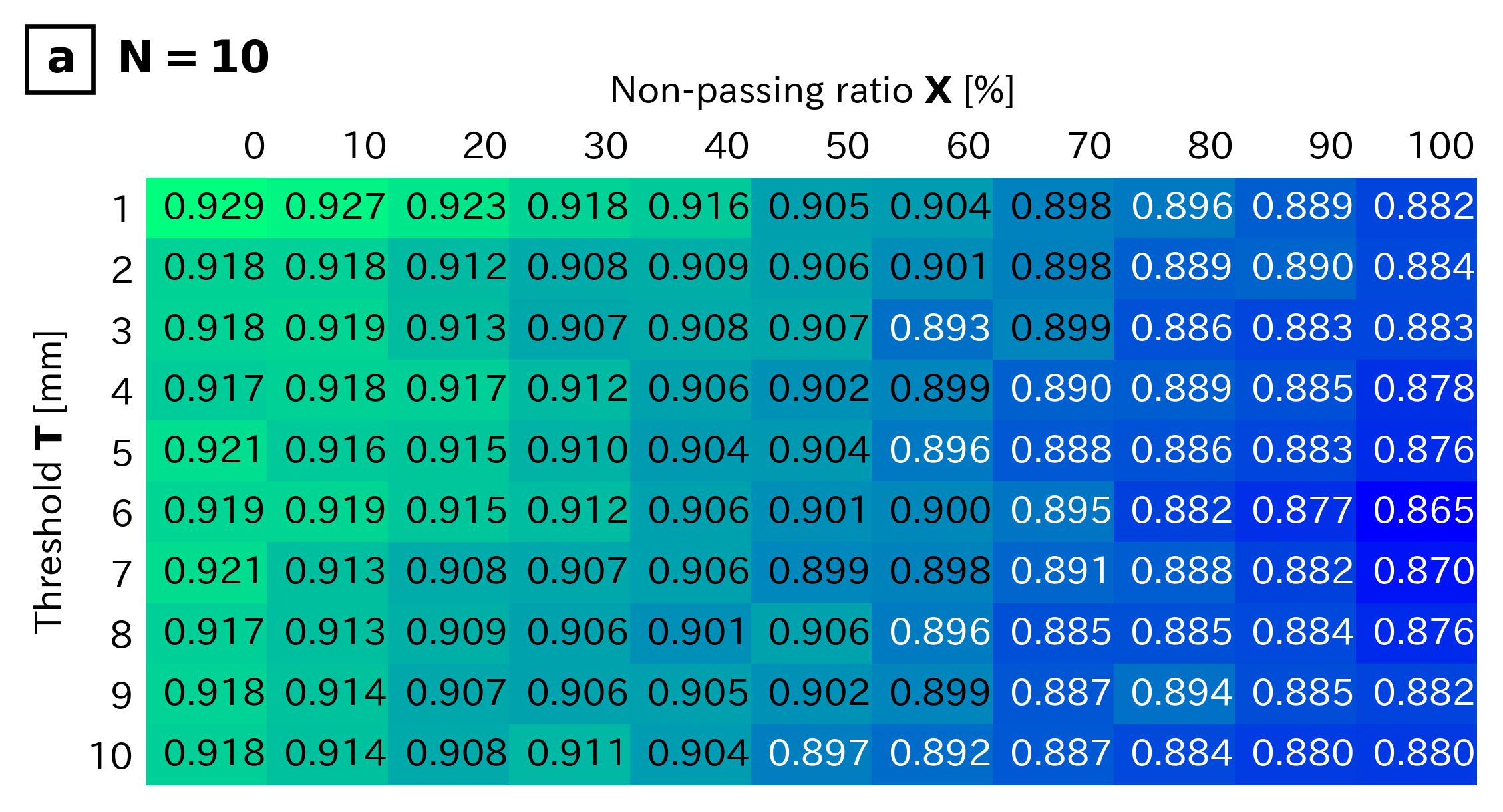}
    \includegraphics[width=0.49\linewidth]{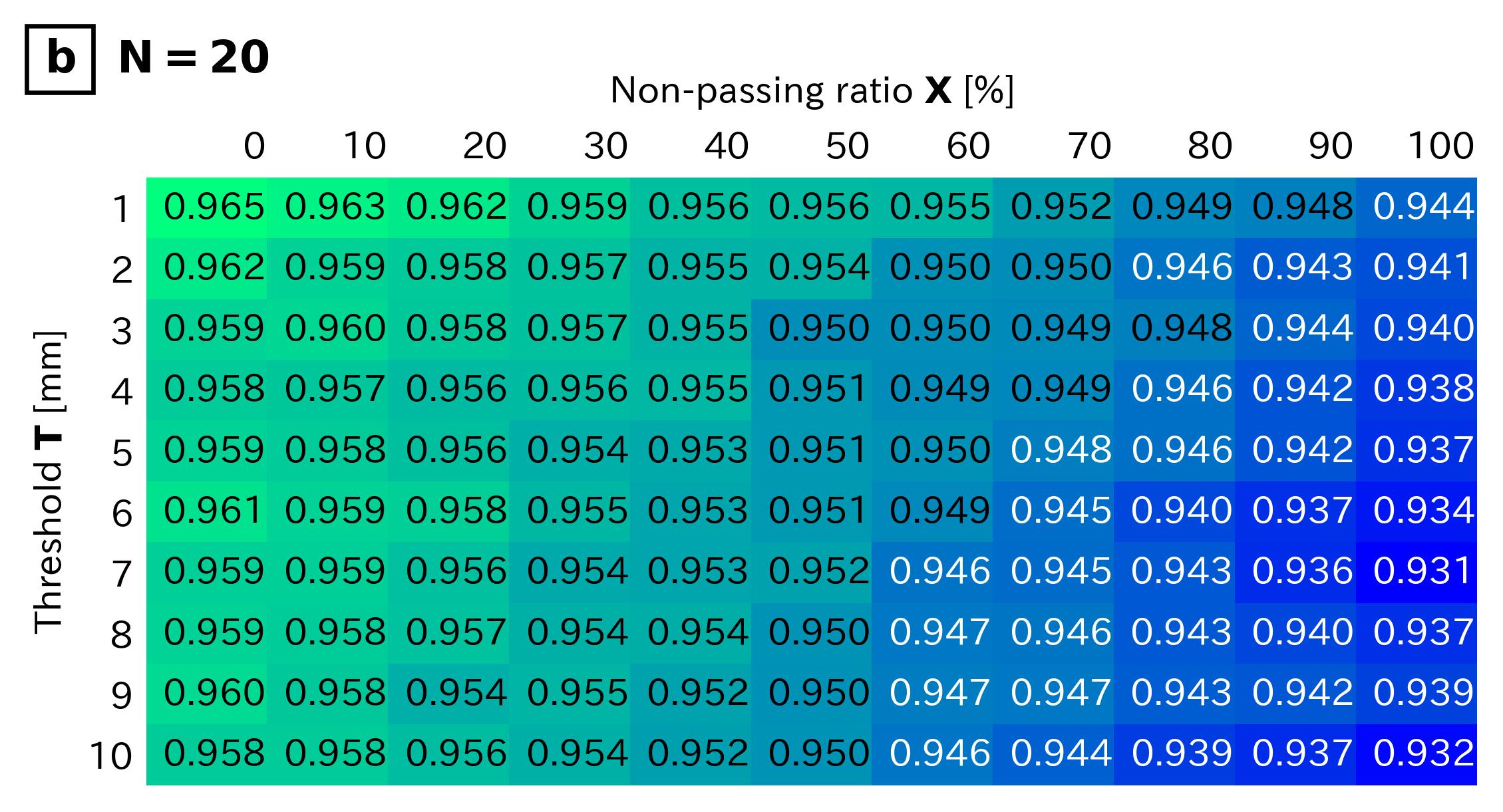}\\[1ex]
    \includegraphics[width=0.49\linewidth]{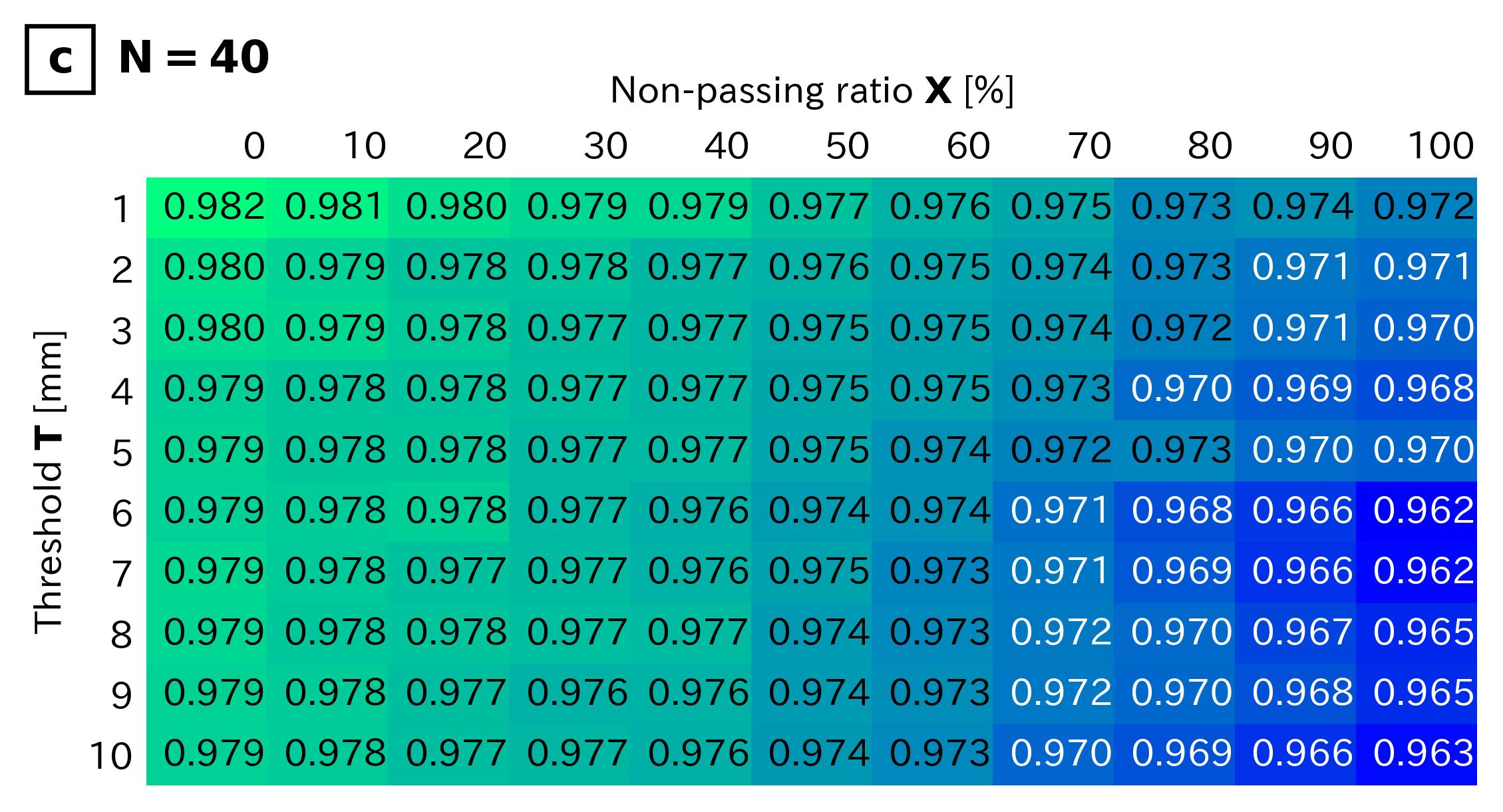}
    \includegraphics[width=0.49\linewidth]{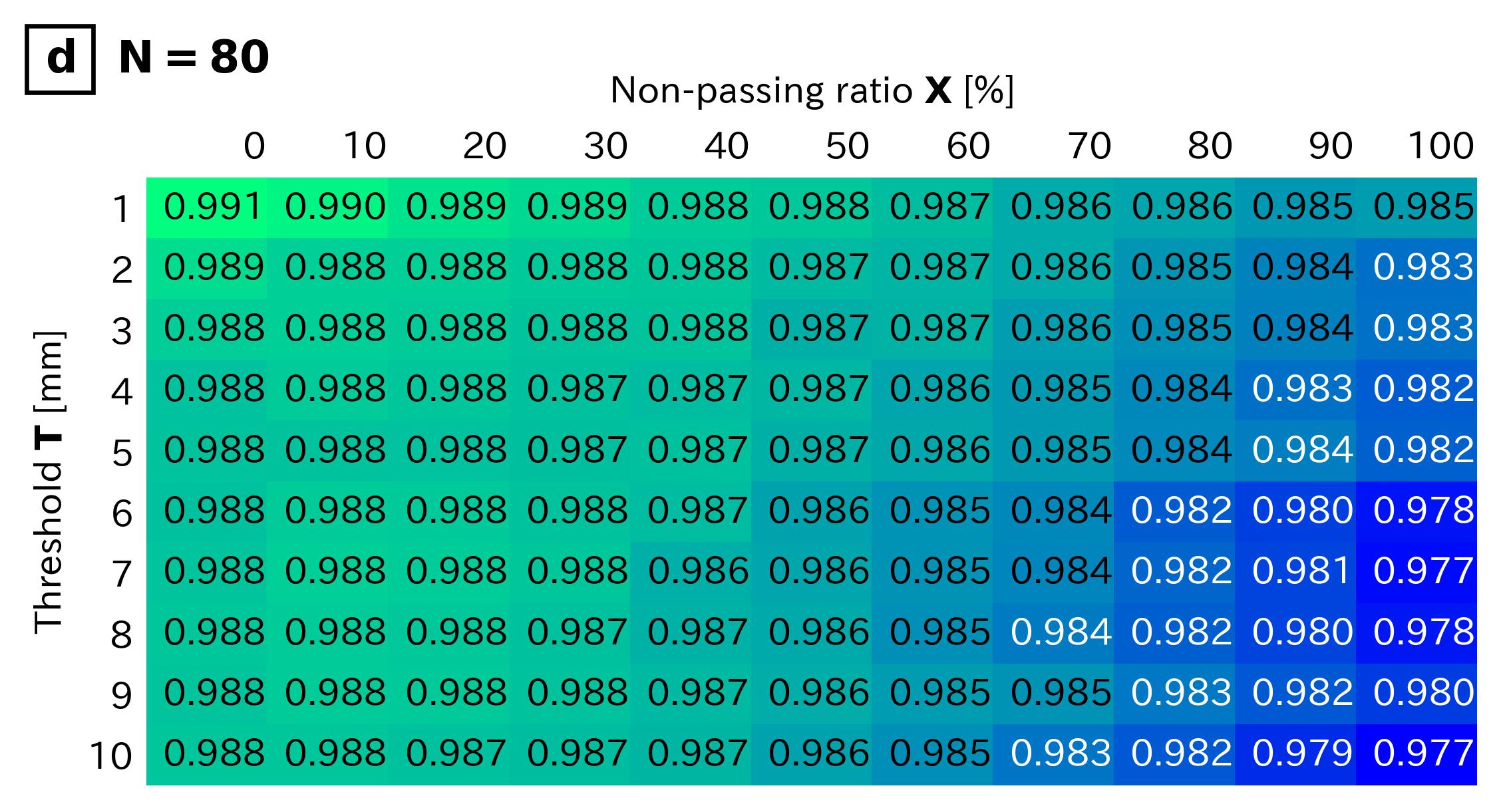}
    \caption{Experiment~2: Goodness of fit to Equation~(\ref{eq:exp1_fitts_mt}) for $MT$ (fast)}
    \label{fig:exp2_fitts_mt_speed_R2_sim}
    \Description{Four heatmaps (N=10,20,40,80) of R^2 for the MT model under “fast” instruction in Experiment 2, over T and X. Fit declines as T loosens and X increases; larger N partially mitigates the drop.}
\end{figure*}
\begin{figure*}[ht]
    \centering
    \includegraphics[width=0.49\linewidth]{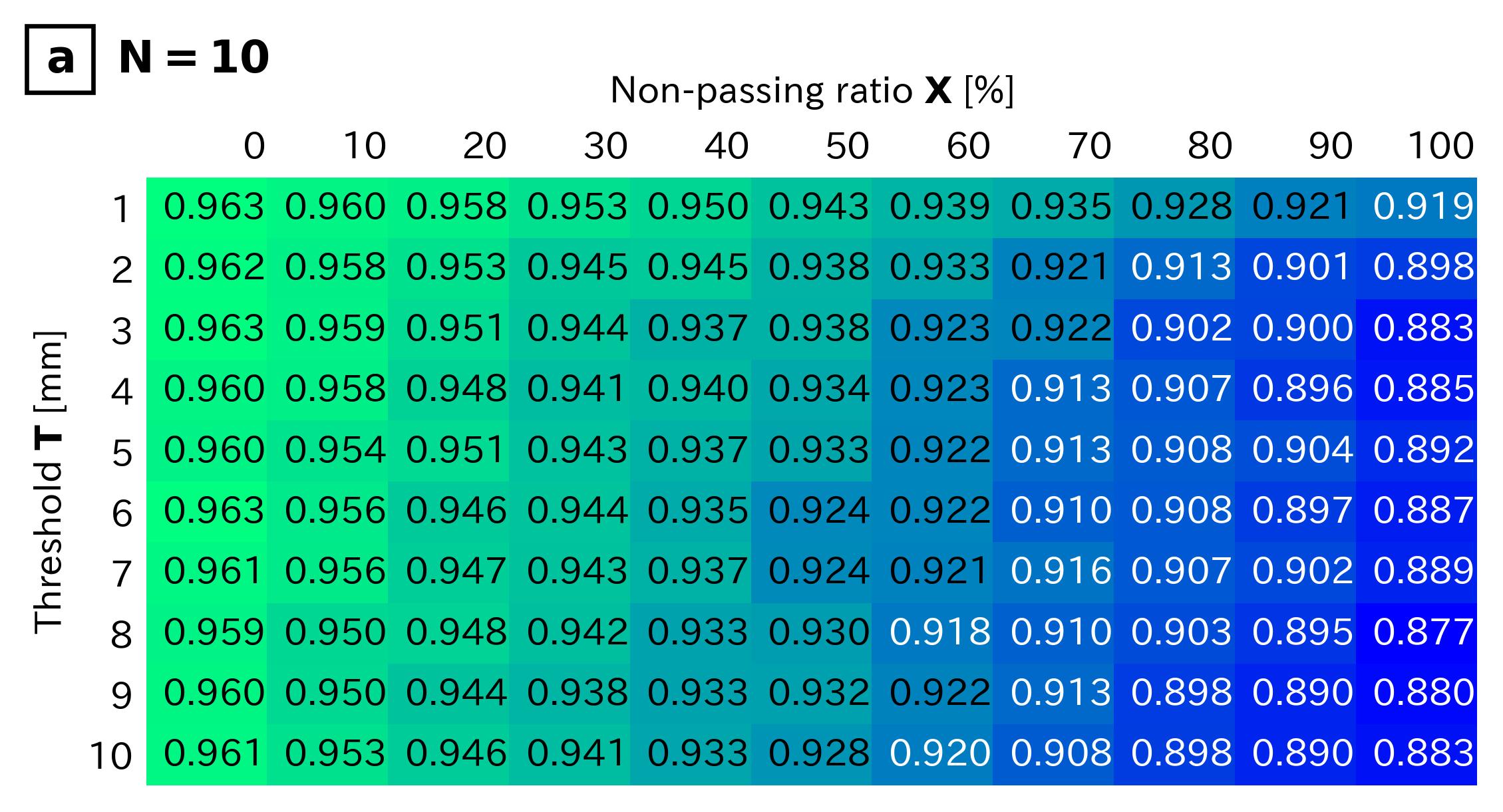}
    \includegraphics[width=0.49\linewidth]{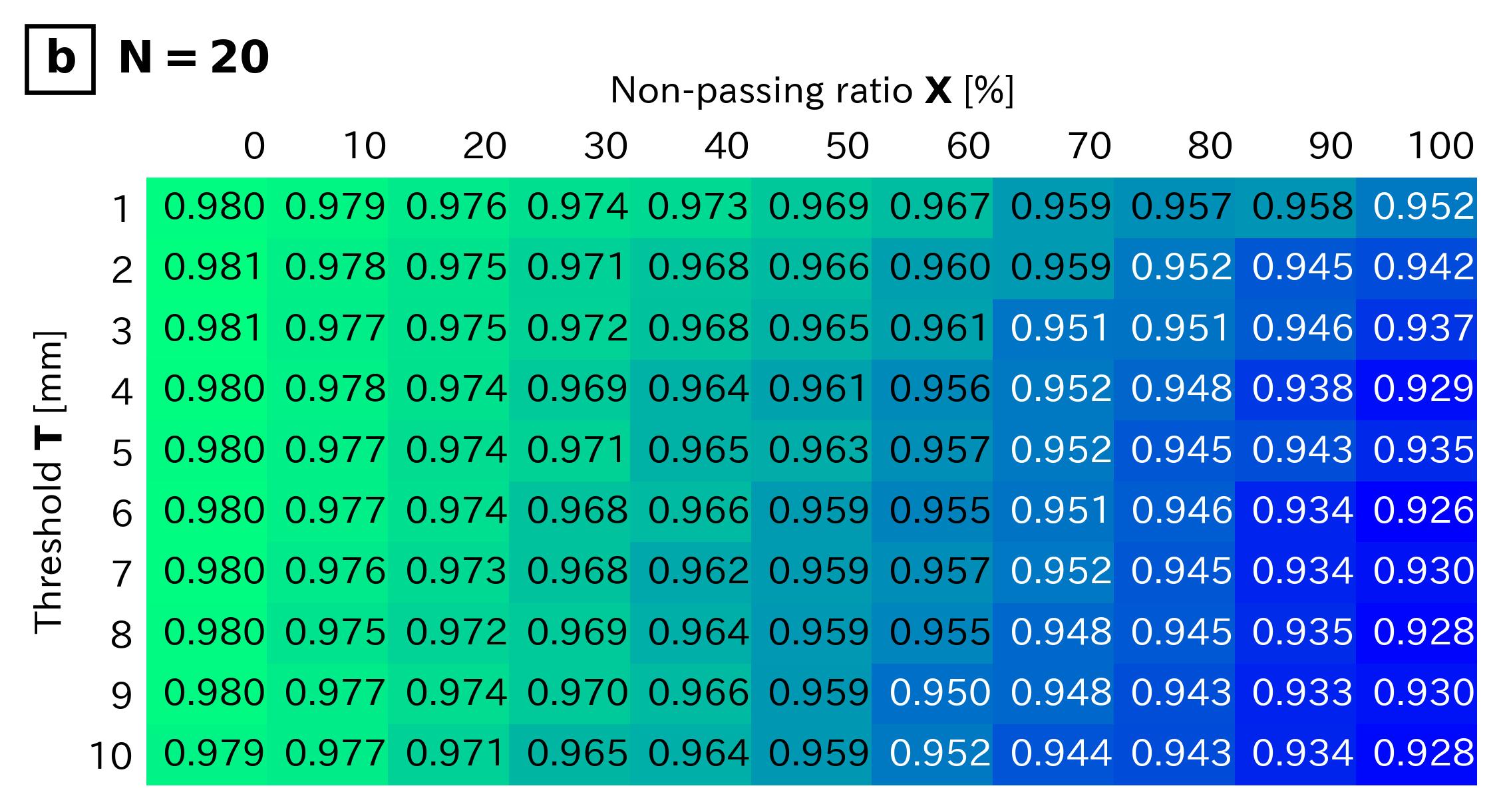}\\[1ex]
    \includegraphics[width=0.49\linewidth]{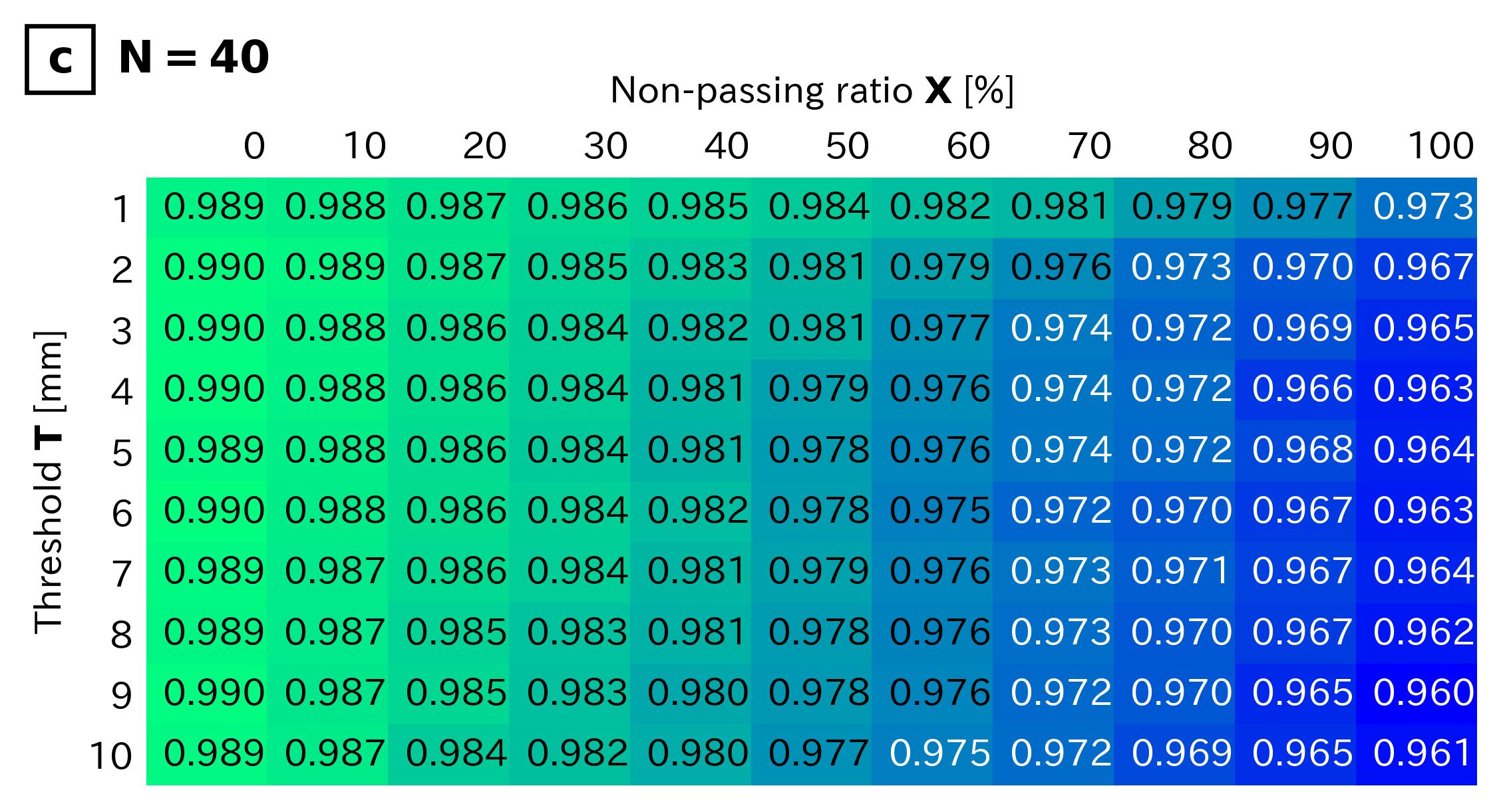}
    \includegraphics[width=0.49\linewidth]{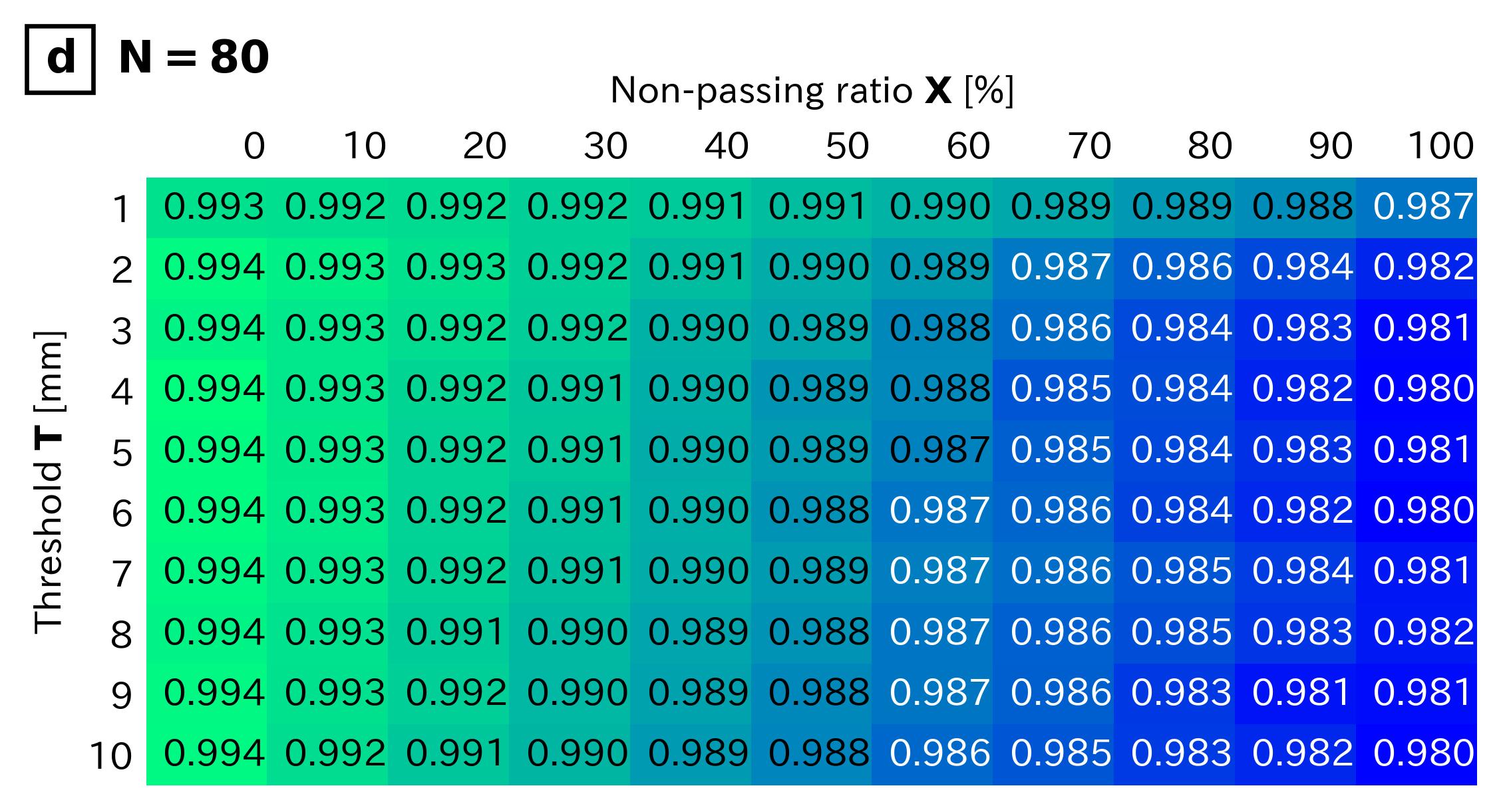}
    \caption{Experiment~2: Goodness of fit to Equation~(\ref{eq:exp1_fitts_mt}) for $MT$ (accurate)}
    \label{fig:exp2_fitts_mt_accuracy_R2_sim}
    \Description{Four heatmaps (N=10,20,40,80) of R^2 for the MT model under “accurate” instruction in Experiment 2. Similar degradation from upper-left to lower-right with larger X and looser T.}
\end{figure*}
\begin{figure*}[ht]
    \centering
    \includegraphics[width=0.49\linewidth]{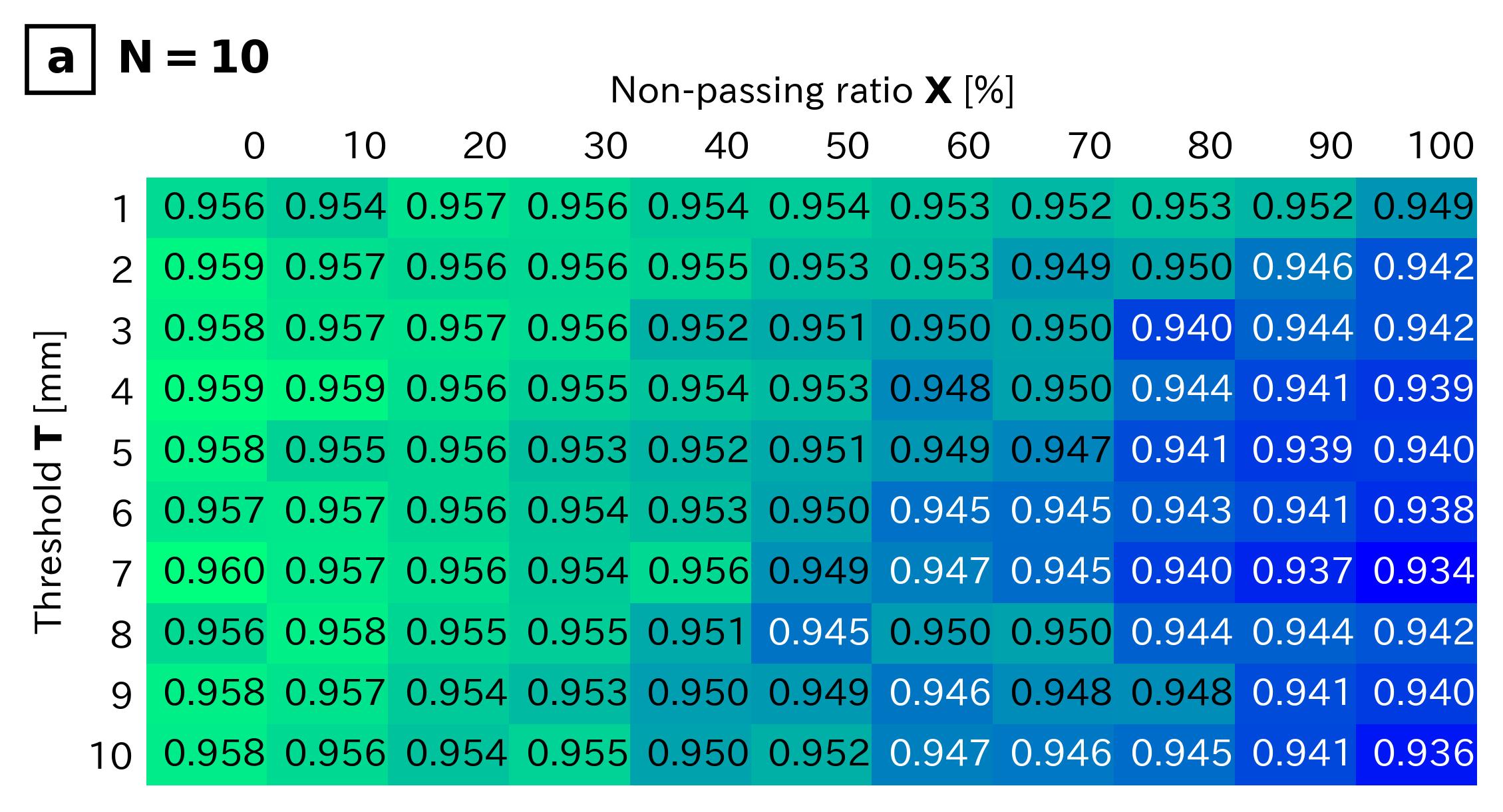}
    \includegraphics[width=0.49\linewidth]{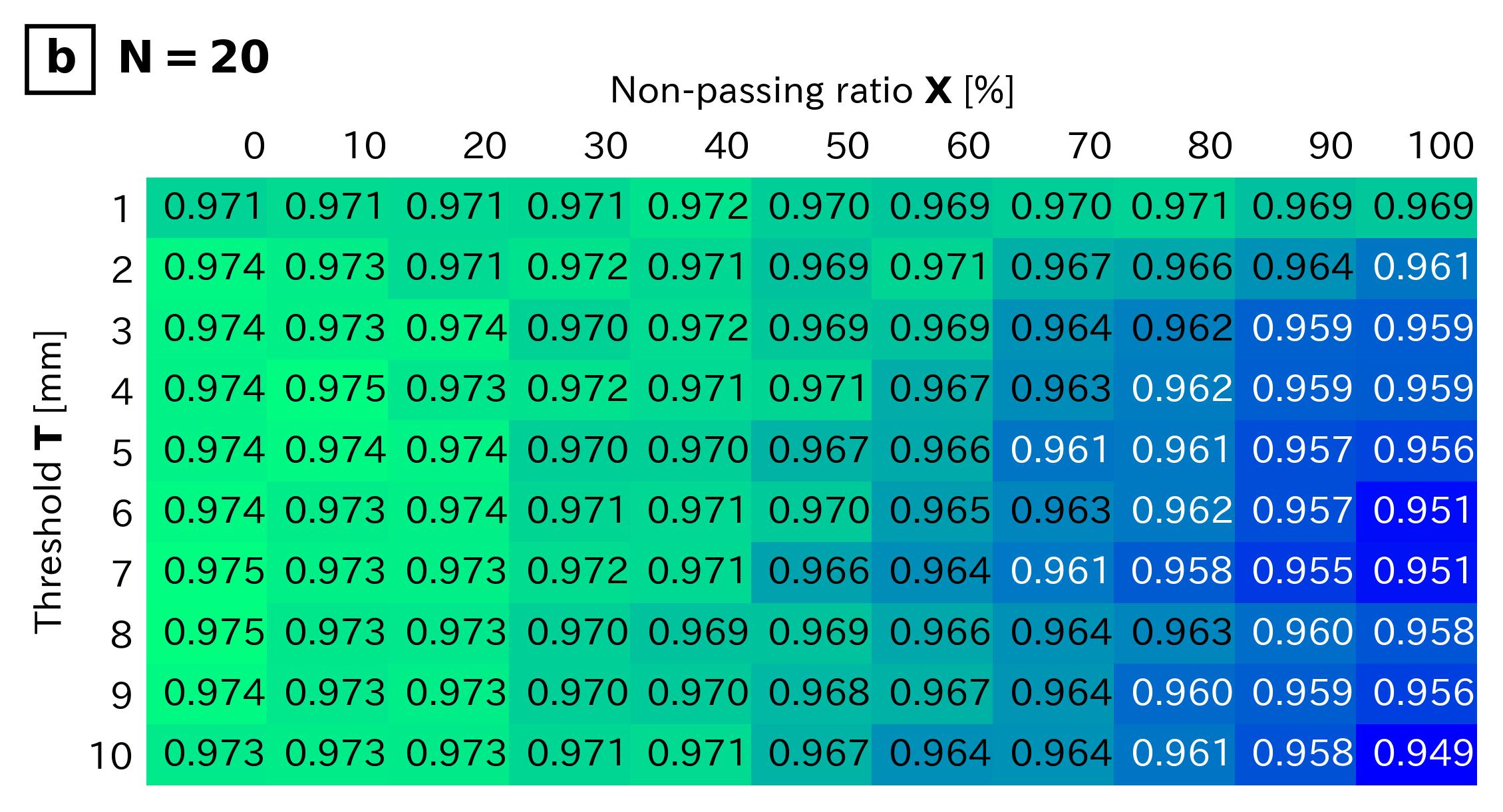}\\[1ex]
    \includegraphics[width=0.49\linewidth]{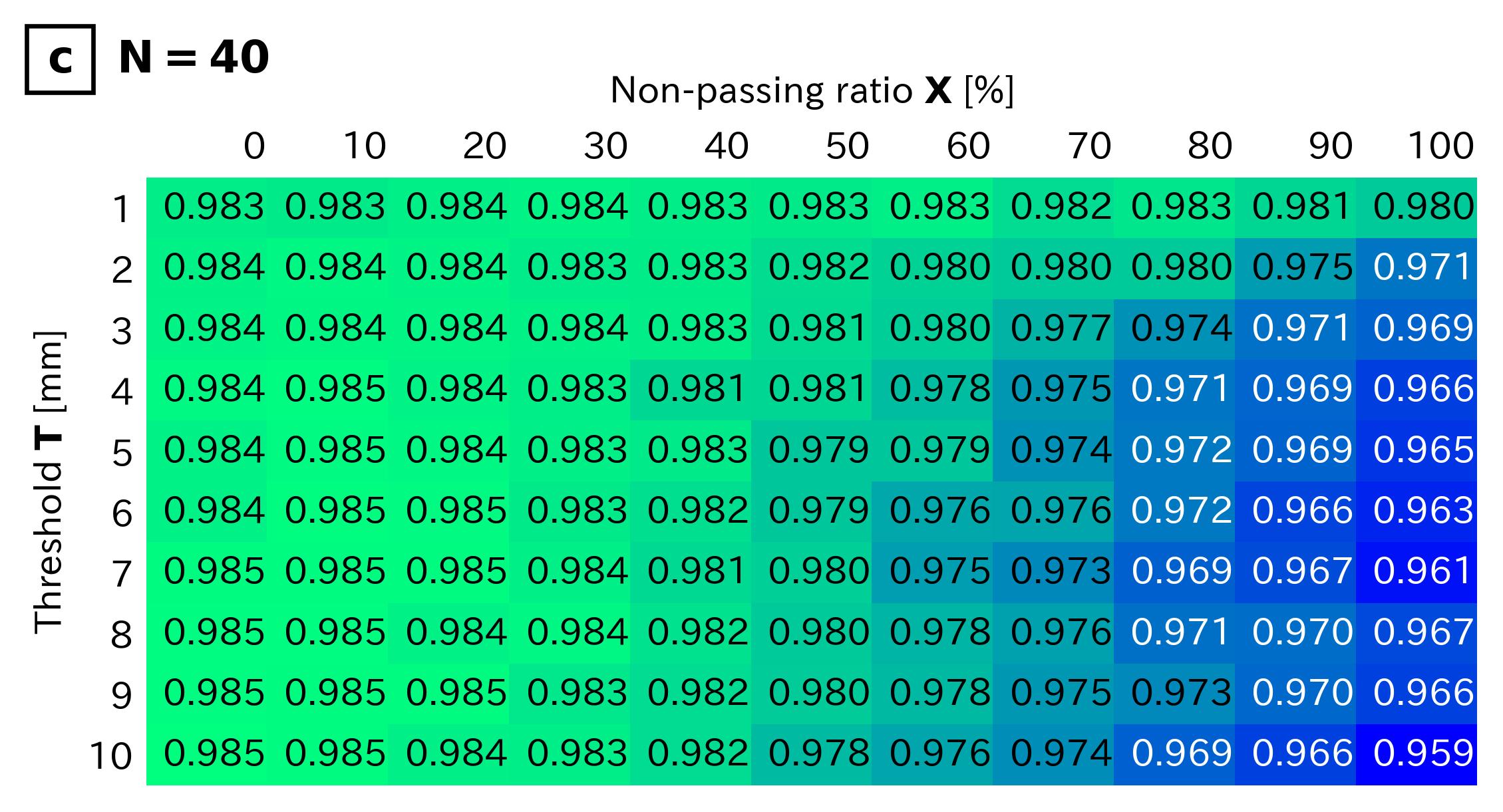}
    \includegraphics[width=0.49\linewidth]{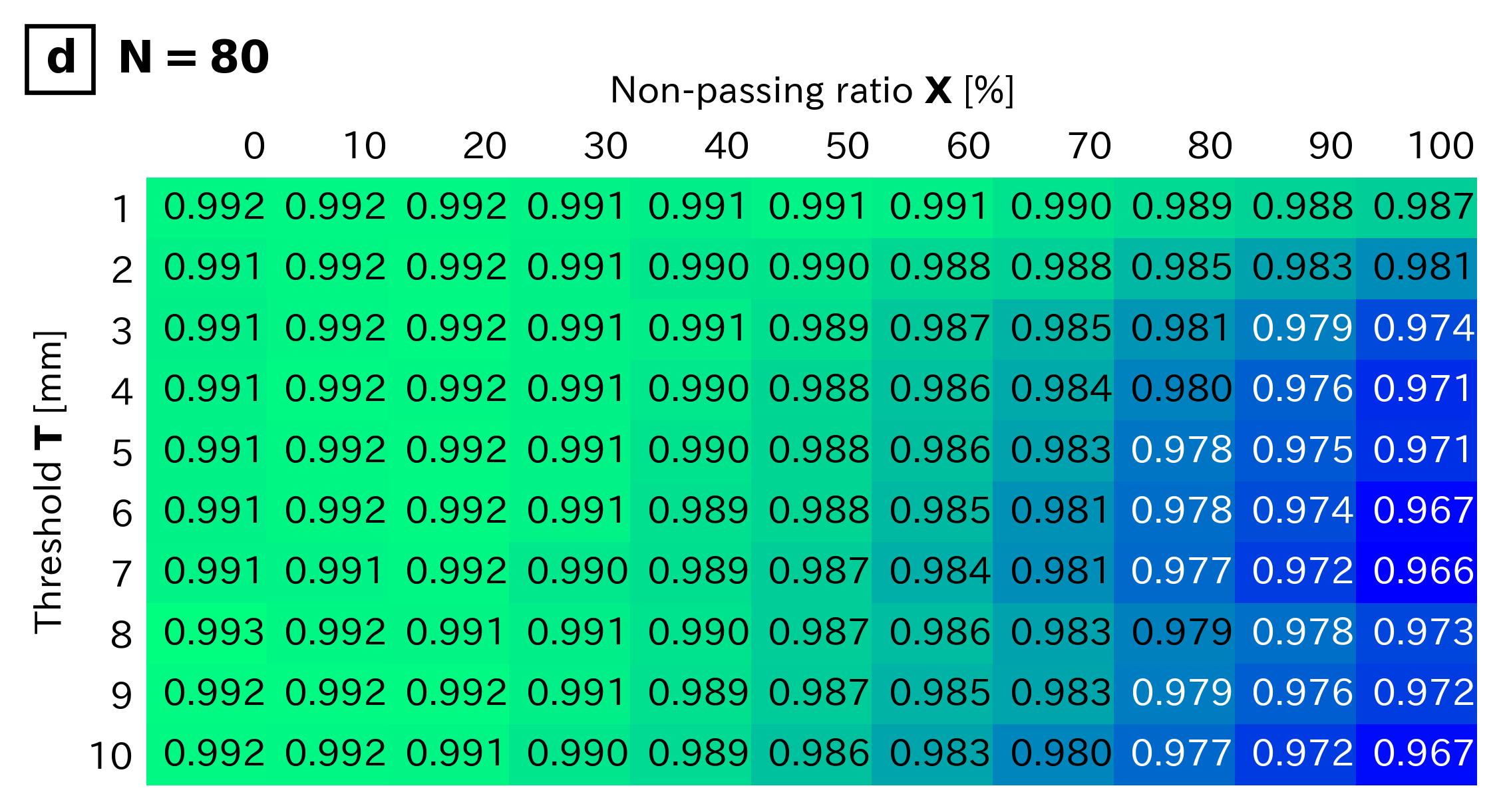}
    \caption{Experiment~2: Goodness of fit to Equation~(\ref{eq:exp2_fitts_er}) for $ER$ (fast)}
    \label{fig:exp2_fitts_er_speed_R2_sim}
    \Description{Four heatmaps (N=10,20,40,80) of R^2 for the ER model under “fast” instruction in Experiment 2. Fit decreases as the non-passing proportion X grows and T loosens.}
\end{figure*}
\begin{figure*}[ht]
    \centering
    \includegraphics[width=0.49\linewidth]{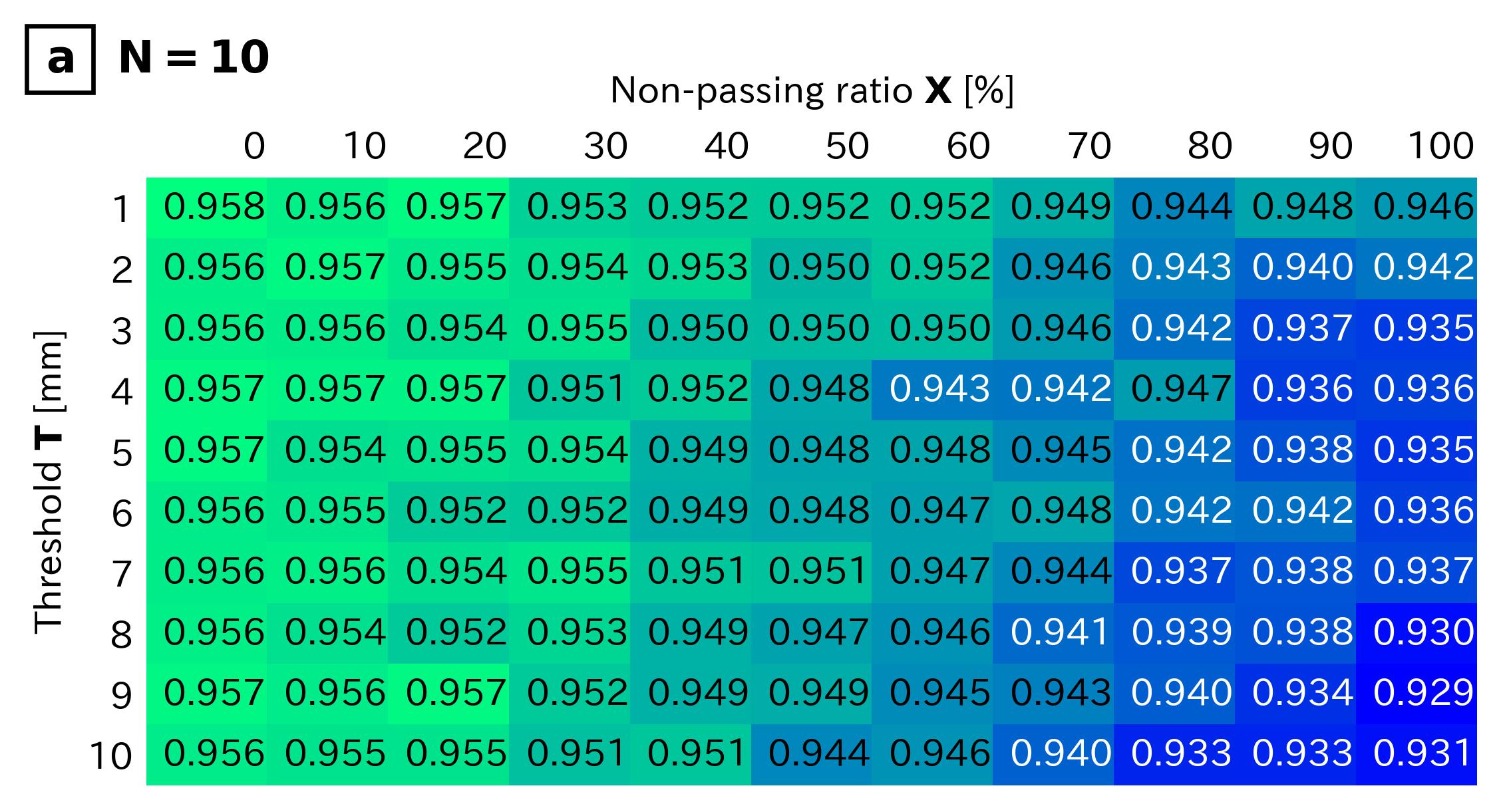}
    \includegraphics[width=0.49\linewidth]{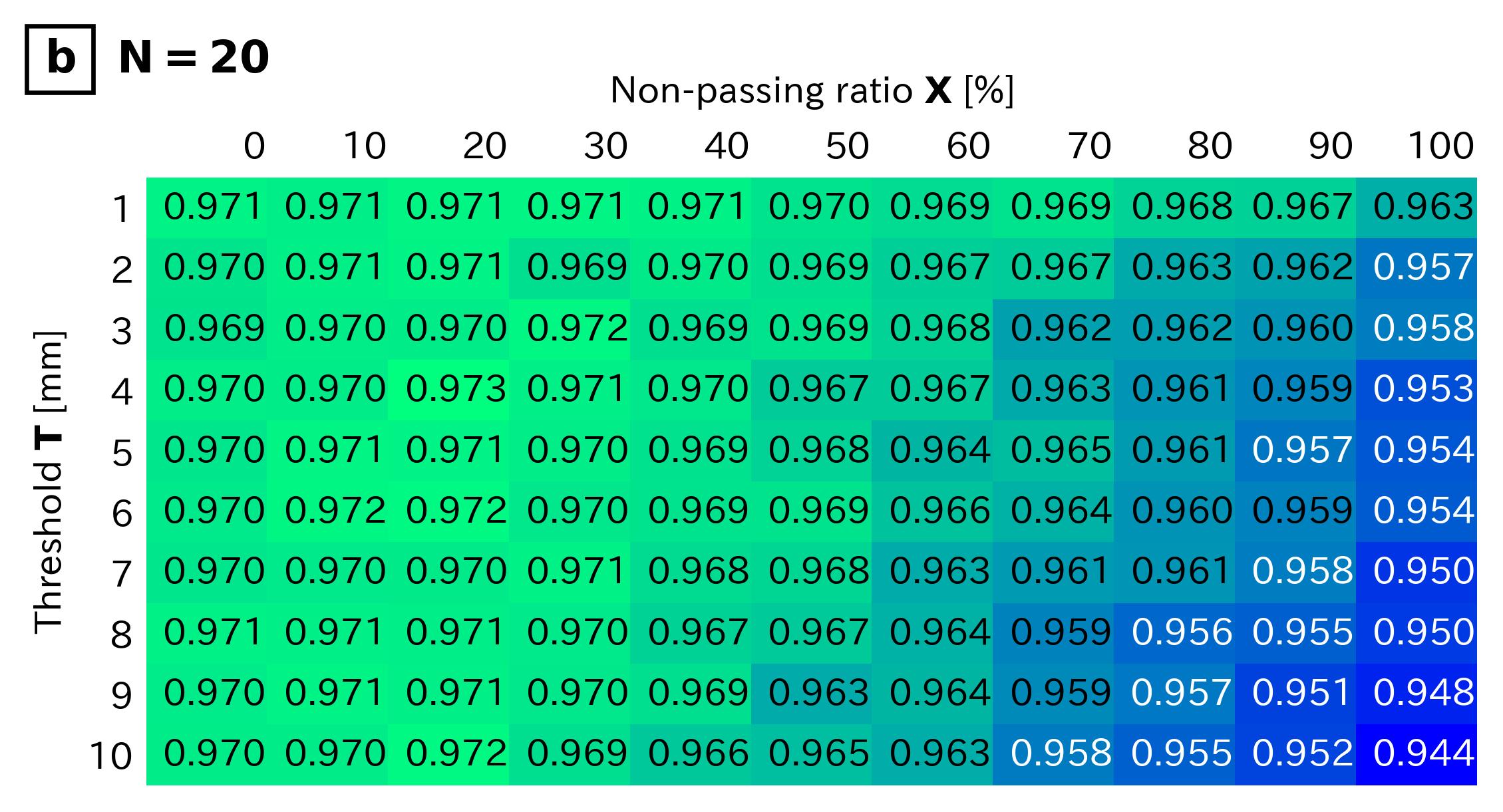}\\[1ex]
    \includegraphics[width=0.49\linewidth]{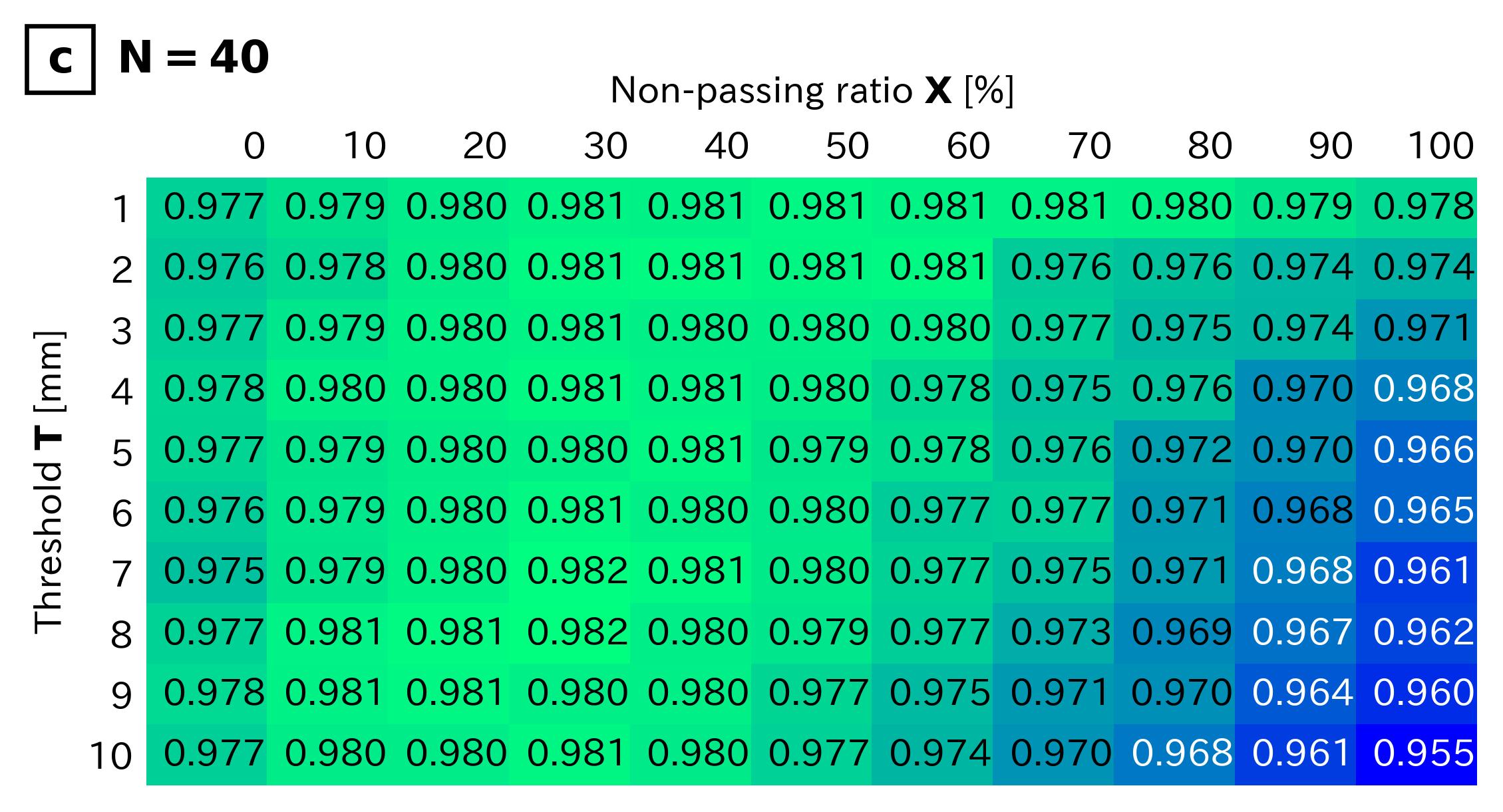}
    \includegraphics[width=0.49\linewidth]{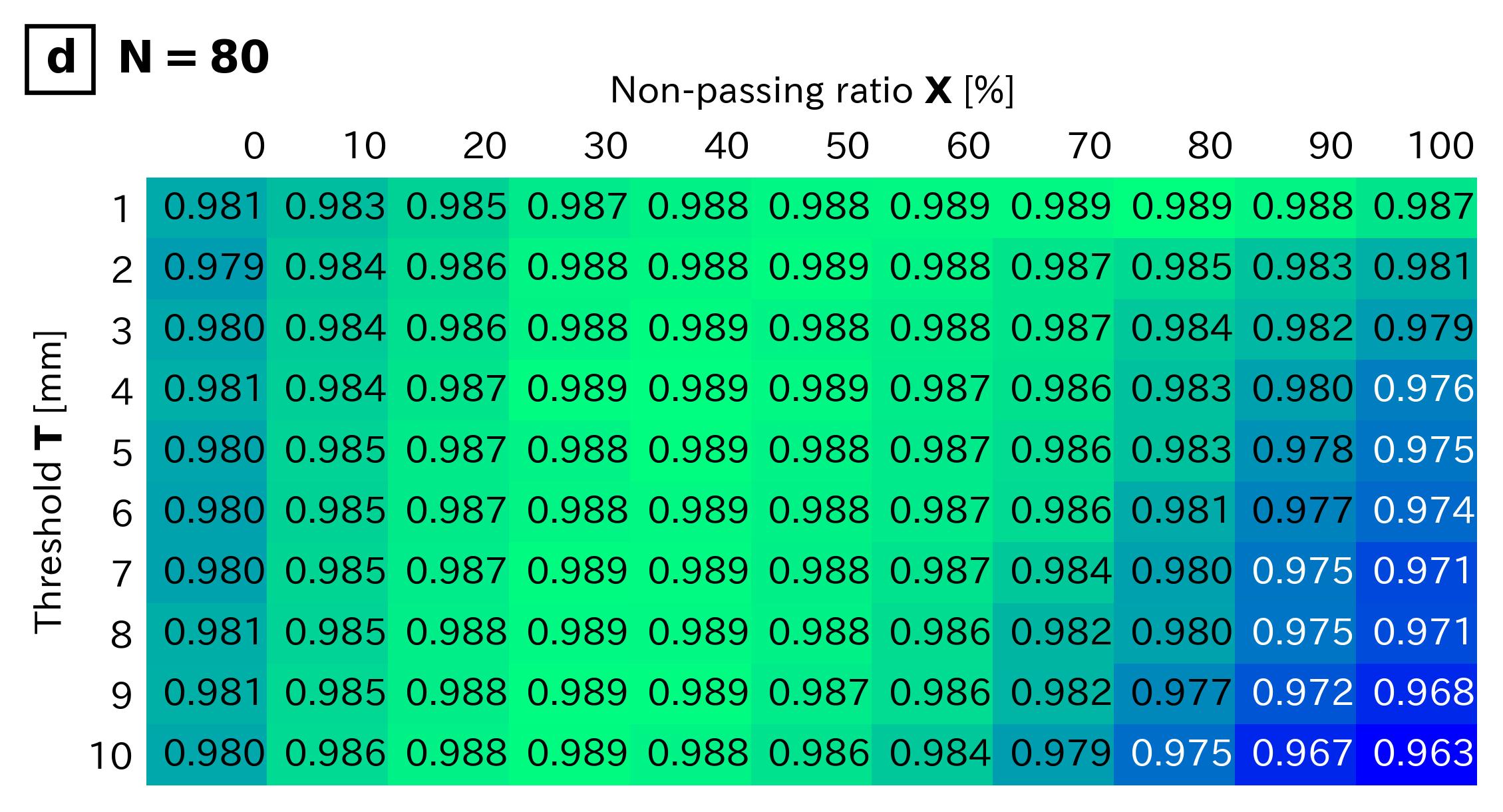}
    \caption{Experiment~2: Goodness of fit to Equation~(\ref{eq:exp2_fitts_er}) for $ER$ (accurate)}
    \label{fig:exp2_fitts_er_accuracy_R2_sim}
    \Description{Four heatmaps (N=10,20,40,80) of R^2 for the ER model under “accurate” instruction in Experiment 2. Stronger monotonic degradation with larger X and looser T; larger N improves stability.}
\end{figure*}

Figures~\ref{fig:exp2_fitts_mt_speed_R2_sim} and \ref{fig:exp2_fitts_mt_accuracy_R2_sim} show $R^2$ for Equation~(\ref{eq:exp1_fitts_mt}) when varying $T$ and $X$.
In both figures, $R^2$ tends to decrease from the upper-left to the lower-right of each heatmap: when the threshold is lenient and the non-passing proportion is large, model fit declines.
Thus, unlike Experiment~1, the non-passing proportion clearly affects the evaluation of Equation~(\ref{eq:exp1_fitts_mt}).

A likely reason for this difference is the device environment.
In Experiment~1 (PC with mouse), even an inattentive participant attempting to finish quickly still needed to move the pointer according to target distance, which may have produced relatively normal $MT$ values.
In Experiment~2 (smartphone), target positions were fixed and alternating, and thus, an inattentive participant could finish quickly by using, for example, two fingers or both hands to tap without moving a single finger.
Although we instructed participants to hold the phone in the non-dominant hand and tap with the dominant index finger (Figure~\ref{fig:exp2_fittsTask2}), inattentive participants may have ignored these instructions, leading to abnormal $MT$.
This may also explain the higher outlier rate compared to Experiment 1.
Hence, in crowdsourced smartphone settings where controlling participants' tapping techniques is difficult, our screening is particularly useful for validating $MT$ models like Equation~(\ref{eq:exp1_fitts_mt}).

As in Experiment 1 (see Section~\ref{sec:exp1_sim_er}), increasing $N$ improved $R^2$ in the upper-left regions of the heatmaps and mitigated the impact of larger $X$.
Nevertheless, $R^2$ still varied with $X$ in Figures~\ref{fig:exp2_fitts_mt_speed_R2_sim} and \ref{fig:exp2_fitts_mt_accuracy_R2_sim}.
Thus, when $N$ is not large enough to absorb the inclusion of nonconforming participants, our screening can suppress the impact of nonconforming participants and improve model fit.

\begin{figure*}[ht]
    \centering
    \begin{minipage}[b]{0.48\linewidth}
        \centering
        \includegraphics[width=\linewidth]{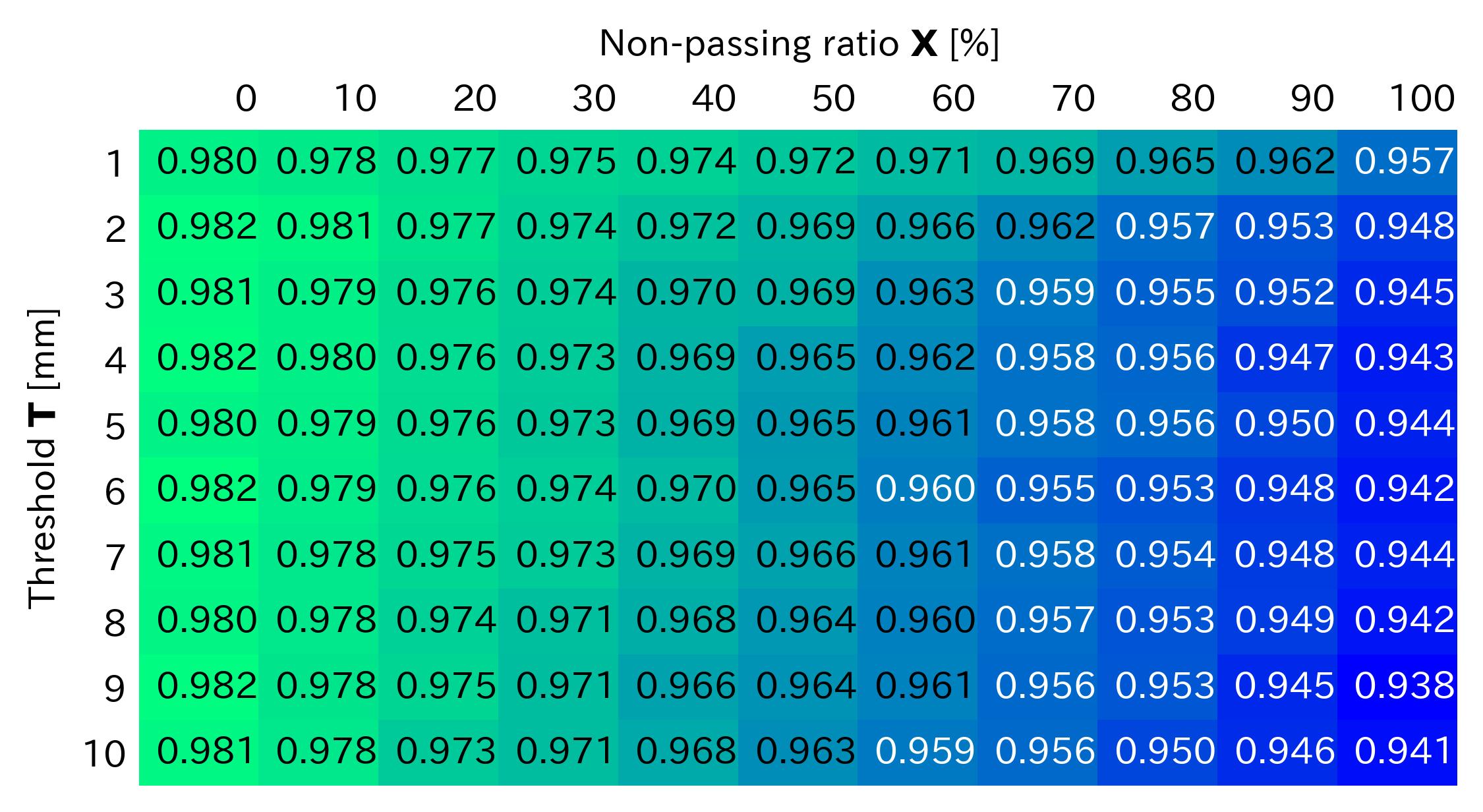}
        \vspace{-10pt}
        \caption{Cross-validation results of $R^2$ for $MT$ ($N=40$, accurate)}
        \label{fig:exp2_fitts_mt_accuracy_R2_loocv}
        \Description{Leave-one-W-out cross-validated R^2 heatmap for the MT model in Experiment 2 (example N=40, “accurate”). Predictive accuracy shows the same degradation as T loosens and X increases.}
    \end{minipage}
    \hfill
    \begin{minipage}[b]{0.48\linewidth}
        \centering
        \includegraphics[width=\linewidth]{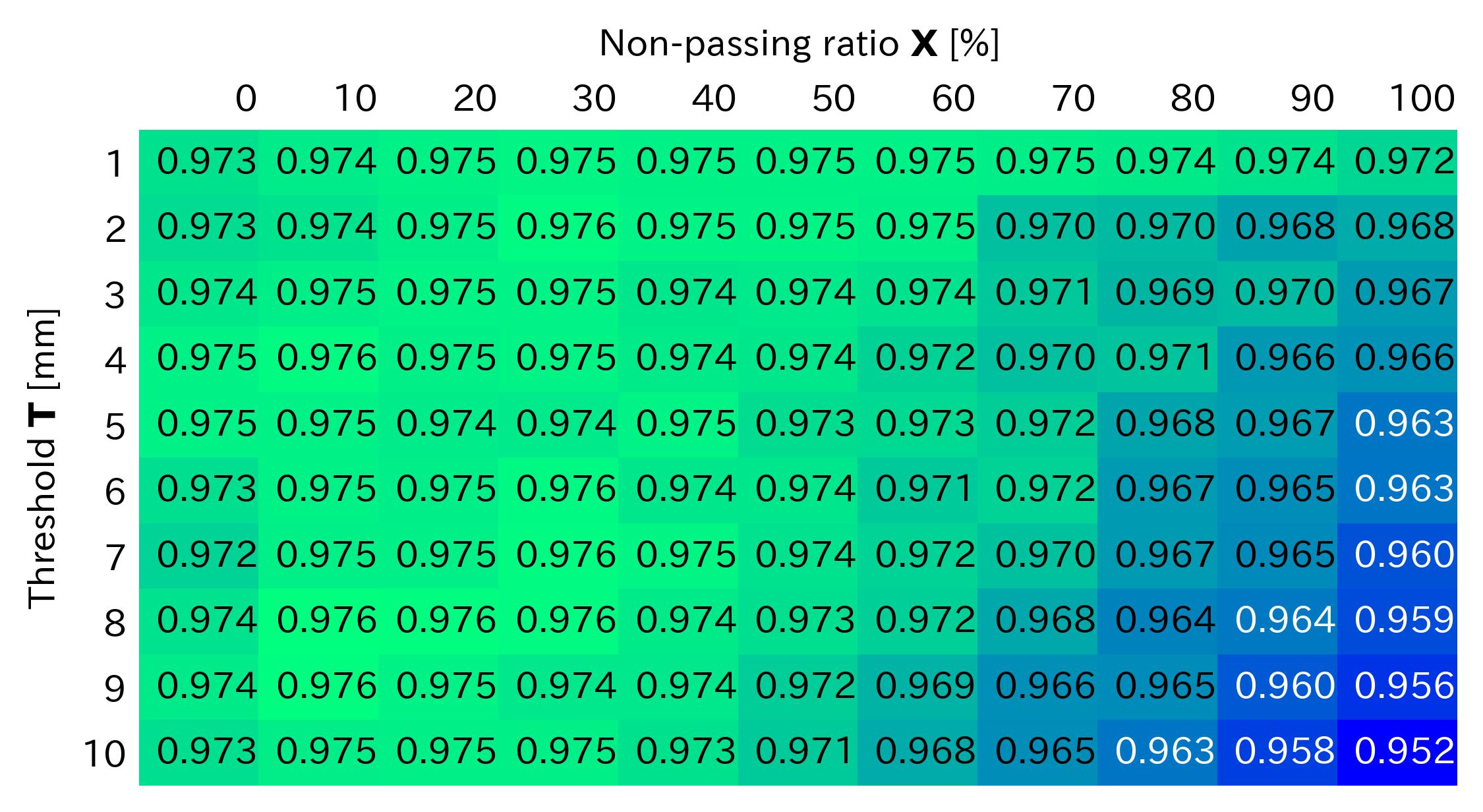}
        \vspace{-10pt}
        \caption{Cross-validation results of $R^2$ for $ER$ ($N=40$, accurate)}
        \label{fig:exp2_fitts_er_accuracy_R2_loocv}
        \Description{Leave-one-W-out cross-validated R^2 heatmap for the ER model in Experiment 2 (example N=40, “accurate”). Predictive accuracy declines as X increases and T is loosened.}
    \end{minipage}
    \vspace{-5pt}
\end{figure*}

We additionally conducted leave-one-$W$-out cross-validation (LOOCV)\hl{, in which we iteratively hold out one width level \(W\) as unseen data, train the model on the remaining levels, and then evaluate prediction accuracy on the held-out level.}
An example for $N=40$ under the ``accurate'' instruction is shown in Figure~\ref{fig:exp2_fitts_mt_accuracy_R2_loocv}.
Compared to using all $W$ levels, $R^2$ is slightly lower, but the same trend holds: model accuracy deteriorates from the upper-left to the lower-right.
Thus, beyond model fit, our screening is effective for predicting user performance under new (unseen) task conditions.
Complete LOOCV results are included in the supplementary materials.

\subsubsection{Error Rate ($ER$)}
Figures~\ref{fig:exp2_fitts_er_speed_R2_sim} and \ref{fig:exp2_fitts_er_accuracy_R2_sim} show $R^2$ for Equation~(\ref{eq:exp2_fitts_er}) when varying $T$ and $X$.
Again, $R^2$ declines from the upper-left to the lower-right of each heatmap: lenient thresholds and large non-passing proportions degrade fit.
Thus, the degradation observed in Experiment~1 is replicated.
Because Experiment~2 increased the number of $W$ levels and repetitions per task condition and controlled target sizes in mm, these results more strongly support that screening via the size-adjustment pre-task is effective.
The effect of $N$ on $ER$ mirrored the $MT$ results, as we observed that increasing $N$ raised $R^2$ overall, most prominently in the upper-left region of the heatmaps (strict $T$ with small $X$), and partially buffered the loss of fit at larger $X$.

Unlike Experiment~1, even under the ``fast'' instruction (Figure~\ref{fig:exp2_fitts_er_speed_R2_sim}), increasing $X$ reduced model fit.
As described, because smartphone settings make it harder to control tapping operations using only the dominant hand's index finger than PC settings, it may induce more noise to the data and thus amplify the negative effect of $X$.

As a cross-validation example, Figure~\ref{fig:exp2_fitts_er_accuracy_R2_loocv} shows the ``accurate'' instruction with $N=40$.
As with $MT$, $R^2$ is slightly lower than when using all $W$ levels, yet the same upper-left to lower-right degradation trend holds.
Thus, when researchers aim to predict $ER$ under unobserved conditions, our screening remains effective.

\subsection{Limitations}
By restricting devices to iPhones, Experiment~2 enabled comparison of the size-adjustment outcome to correct references.
In both the pre-task and the main task, we controlled the visual stimuli in the mm scale, which enabled more rigorous evaluations.
In the main task, increasing $W$ levels and repetitions further allowed for precise validation of the models.
However, as seen in Figures~\ref{fig:exp2_fitts_er_speed_R2_sim} and \ref{fig:exp2_fitts_er_accuracy_R2_sim}, particularly for the $ER$ model, the top-left cells (threshold $T=1$~mm, non-passing proportion $X=0\%$) were not always the best-fitting, leaving some limitations from Experiment~1 unresolved.
A plausible cause is that, in both Experiments~1 and~2, errors triggered re-aiming until success.
An inattentive participant seeking to finish quickly might adopt a strategy of tapping both quickly and accurately, making it difficult to distinguish them from conscientious participants.

\section{Experiment 3: iPhone-Based Experiment without Re-Aiming}
\subsection{Task and Design}
As in Experiment~2, Experiment~3 rigorously evaluates both the size-adjustment pre-task and the pointing task on iPhones.
The key difference is that, in Experiment~3, the next trial begins immediately even when an error occurs (i.e., no re-aiming).
This design aims to address the limitation observed in Experiments~1 and~2, where a re-aiming policy made inattentive participants difficult to distinguish from conscientious ones.
Except for the handling of errors in the pointing task, the experimental design is identical to Experiment~2, and we conduct the same simulations.

\subsection{Participants}
A total of 575 participants completed the experiment.
Each received 250~JPY (approximately 1.7~USD).
The mean task completion time was 5~min~27~s, yielding an effective hourly payment of 2{,}752~JPY (approximately 18.8~USD).
We excluded 1 participant with missing data and 55 for whom the iPhone PPI could not be identified, leaving 519 participants for analysis (250 male, 266 female, 3 other).

\subsection{Results}
\subsubsection{Size-Adjustment Task}
\begin{figure}[ht]
    \vspace{-10pt}
    \centering
    \includegraphics[width=0.85\linewidth]{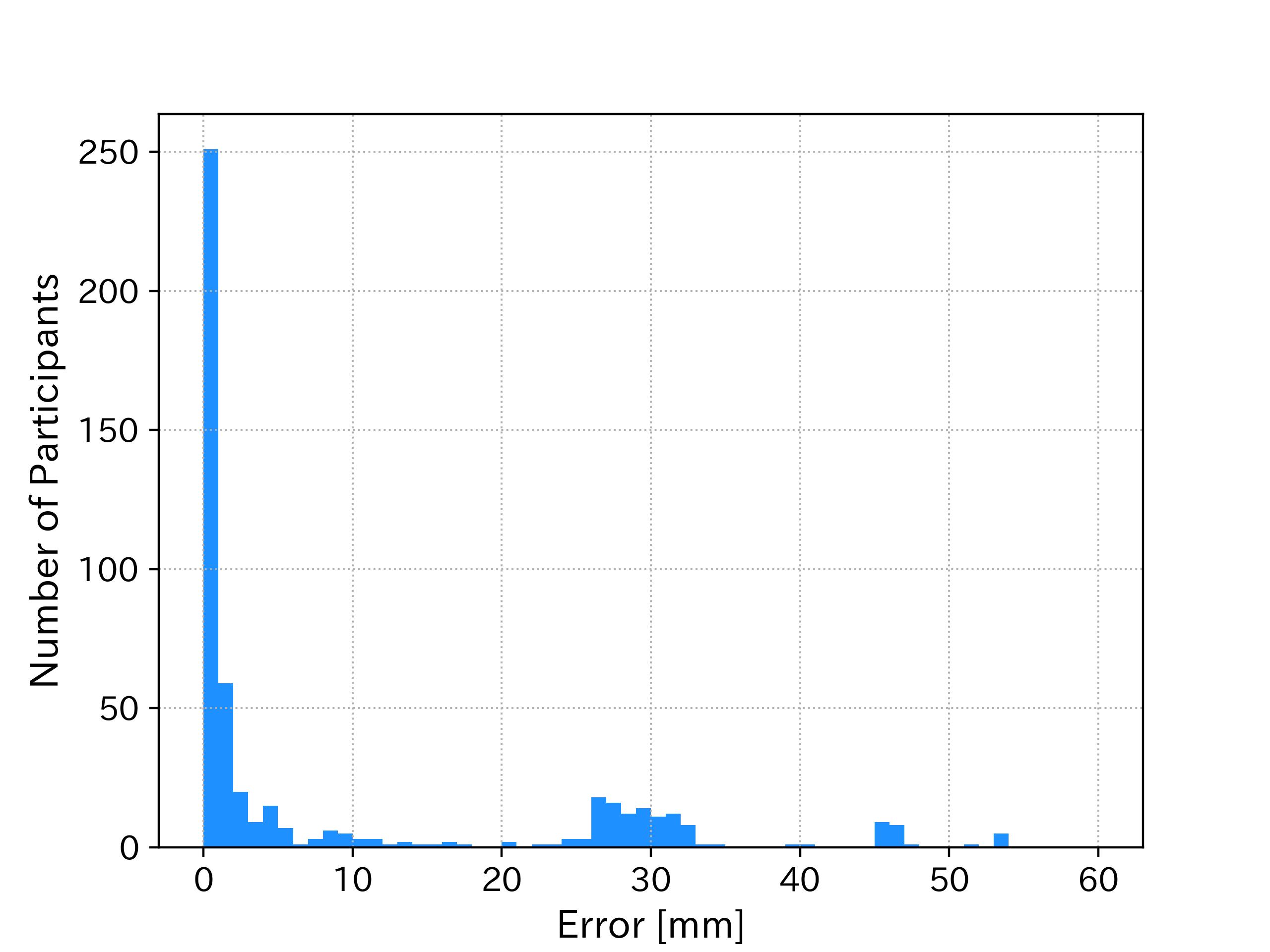}
    \vspace{-5pt}
    \caption{Distribution of size-adjustment errors}
    \label{fig:exp3_size_diff}
    \Description{Histogram of absolute size-adjustment errors (mm) in Experiment 3 (iPhone). Distribution similar to Experiment 2 with both small and large errors present.}
\end{figure}

Figure~\ref{fig:exp3_size_diff} shows the distribution of absolute errors relative to the correct size in the size-adjustment task.
Most participants exhibited small errors, specifically, 310 (60\%) were below 2~{mm}, but some showed large errors, with 143 (28\%) at or above 10~{mm}.
The operation time for the adjustment averaged 8.44~{s} (SD=7.66~{s}).
Five participants recorded 0~{s}, indicating no adjustment at all.

\begin{figure*}[t]
    \centering
    \begin{minipage}[b]{0.48\linewidth}
        \centering
        \includegraphics[width=\linewidth]{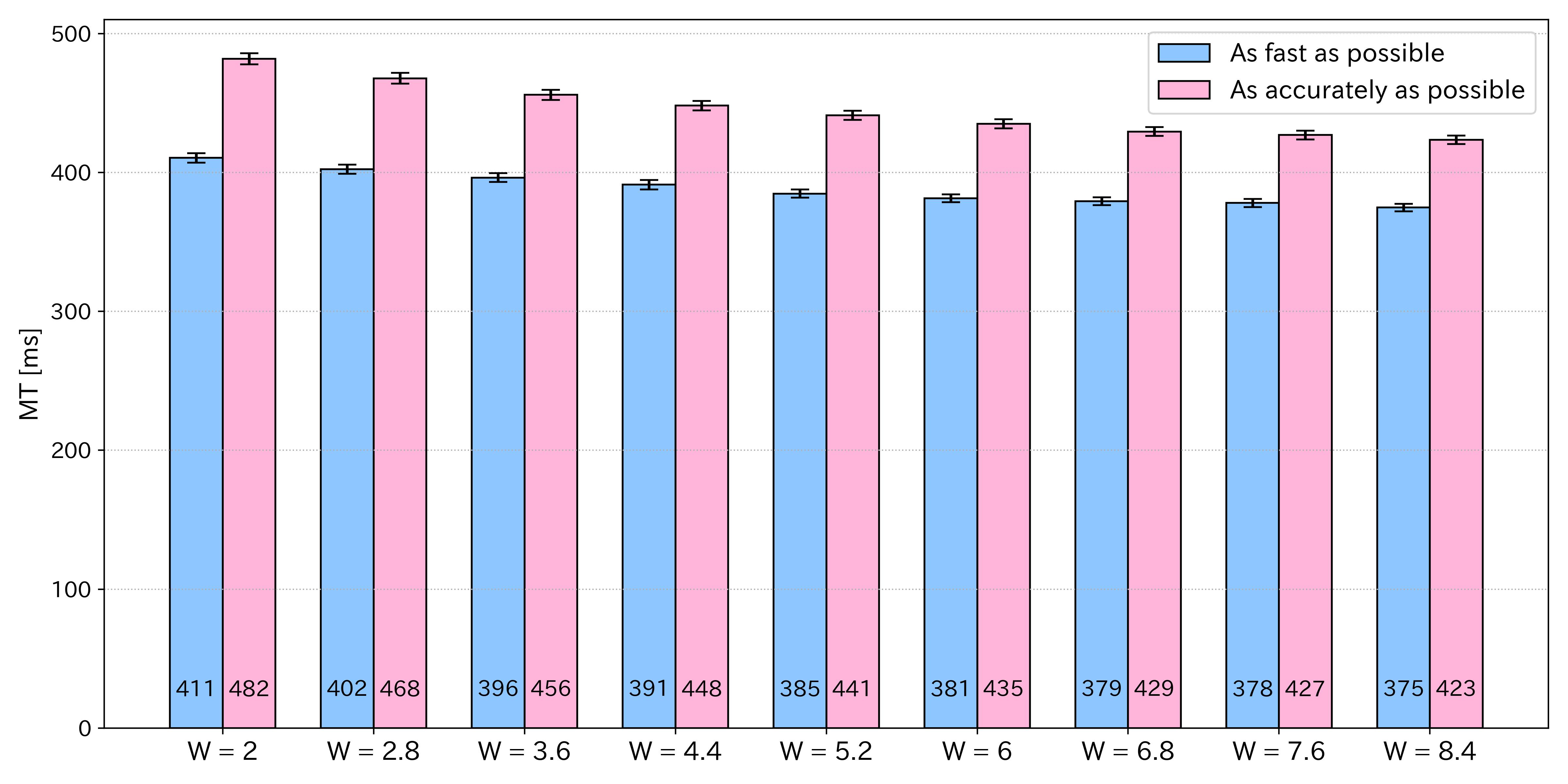}
        \vspace{-10pt}
        \caption{\hl{$MT$ results for each $W$ under the fast and accurate instruction conditions}}
        \label{fig:exp3_fitts_mtAll}
        \Description{Mean movement time (MT) with 95\% CIs by W and instruction in Experiment 3 (no re-aiming). MT slightly shorter than Experiment 2.}
    \end{minipage}
    \hfill
    \begin{minipage}[b]{0.48\linewidth}
        \centering
        \includegraphics[width=\linewidth]{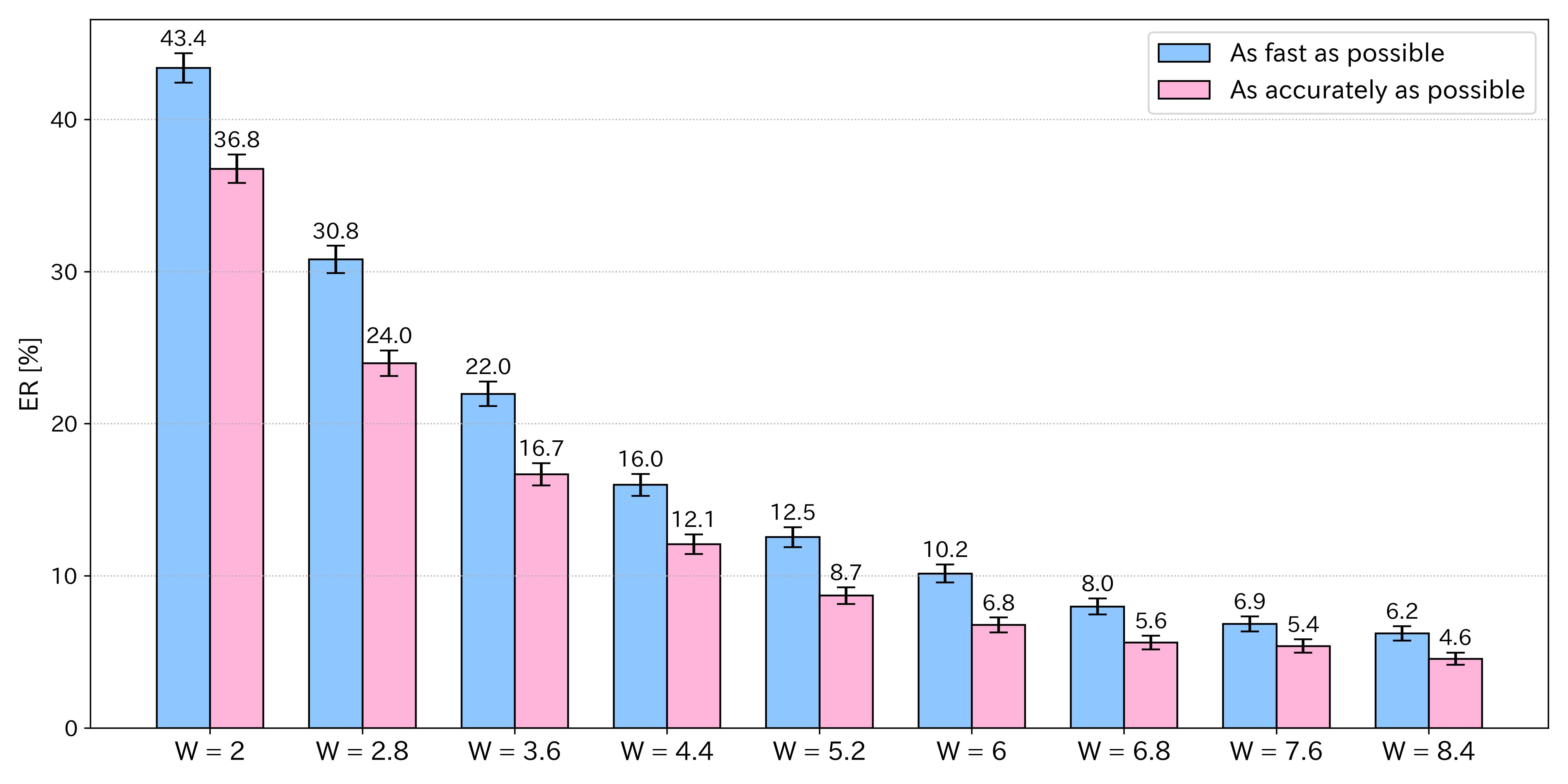}
        \vspace{-10pt}
        \caption{\hl{$ER$ results for each $W$ under the fast and accurate instruction conditions}}
        \label{fig:exp3_fitts_erAll}
        \Description{Mean error rate (ER) with 95\% CIs by W and instruction in Experiment 3 (no re-aiming). ER slightly higher than Experiment 2, especially at small W.}
    \end{minipage}
    \vspace{-5pt}
\end{figure*}

\subsubsection{Pointing Task}
Applying the same outlier handling as in Experiment~2 (Section~\ref{sec:exp2_result_fitts}), we excluded 1{,}011 trials for tap coordinates, 2{,}717 trials for $MT$, and four participants for $MT$.
In total, 181{,}796 trials (97.3\%) were retained for analysis.
Figures~\ref{fig:exp3_fitts_mtAll} and \ref{fig:exp3_fitts_erAll} present the pointing task results for Experiment~3: $MT$ by $(W \times \text{instruction})$ and $ER$ by $(W \times \text{instruction})$, respectively.
\hl{
For $MT$, we found significant main effects of instruction ($F_{1,514}=205.8$, $p<0.001$, $\eta_g^2=0.037$) and $W$ ($F_{8,4112}=252.2$, $p<0.001$, $\eta_g^2=0.011$).
The interaction effect of $W \times$ instruction was significant ($F_{8,4112}=32.79$, $p<0.001$, $\eta_g^2=0.00063$).
For $ER$, we found significant main effects of instruction ($F_{1,514}=67.46$, $p<0.001$, $\eta_g^2=0.012$) and $W$ ($F_{8,4112}=1396$, $p<0.001$, $\eta_g^2=0.27$).
The interaction effect of $W \times$ instruction was significant ($F_{8,4112}=18.61$, $p<0.001$, $\eta_g^2=0.0026$).
}
As in Experiment~2, $ER$ was more strongly affected by $W$ than $MT$\hl{; for example, under the fast condition, when $W$ increased from 2.0 to 8.4~mm, the change ratio of $ER$ was 86\% while that of $MT$ was 8.8\%.}
In addition, relative to Figure~\ref{fig:exp2_fitts_erAll}, $MT$ was slightly shorter and $ER$ slightly higher.

\subsection{Simulation}
\begin{figure}[ht]
    \centering
    \includegraphics[width=0.85\linewidth]{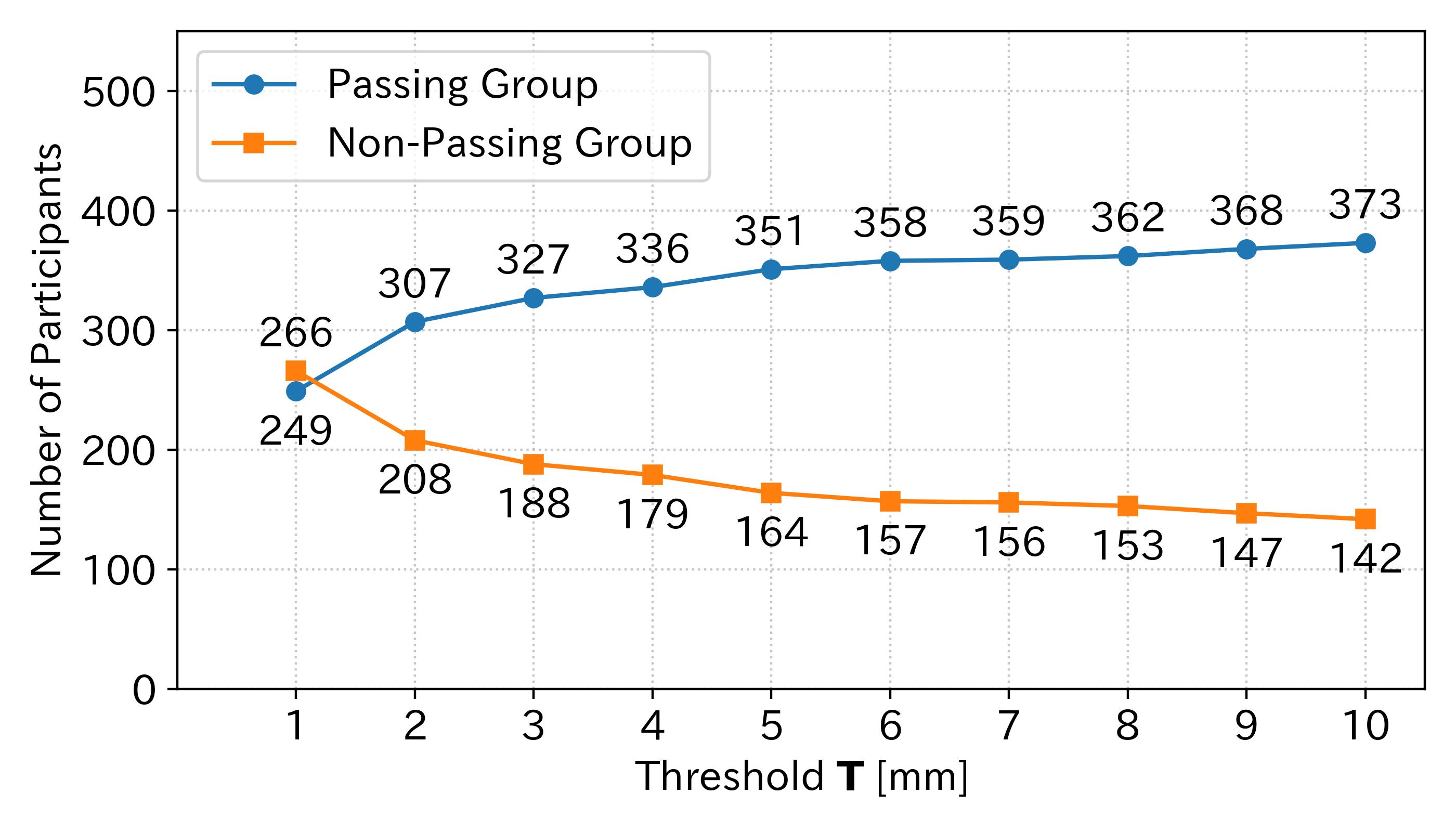}
    \caption{Numbers of passing vs.\ non-passing participants as a function of the threshold}
    \label{fig:exp3_group_N}
    \Description{Counts of passing vs. non-passing participants versus threshold T (mm) in Experiment 3. Group sizes vary systematically with T as in Experiment 2.}
\end{figure}

The simulation conditions and procedures followed Experiment~2 (Section~\ref{sec:exp2_sim}).
Figure~\ref{fig:exp3_group_N} shows the numbers of passing vs.\ non-passing participants as a function of the threshold $T$ in Experiment~3.

\begin{figure*}[ht]
    \centering
    \includegraphics[width=0.49\linewidth]{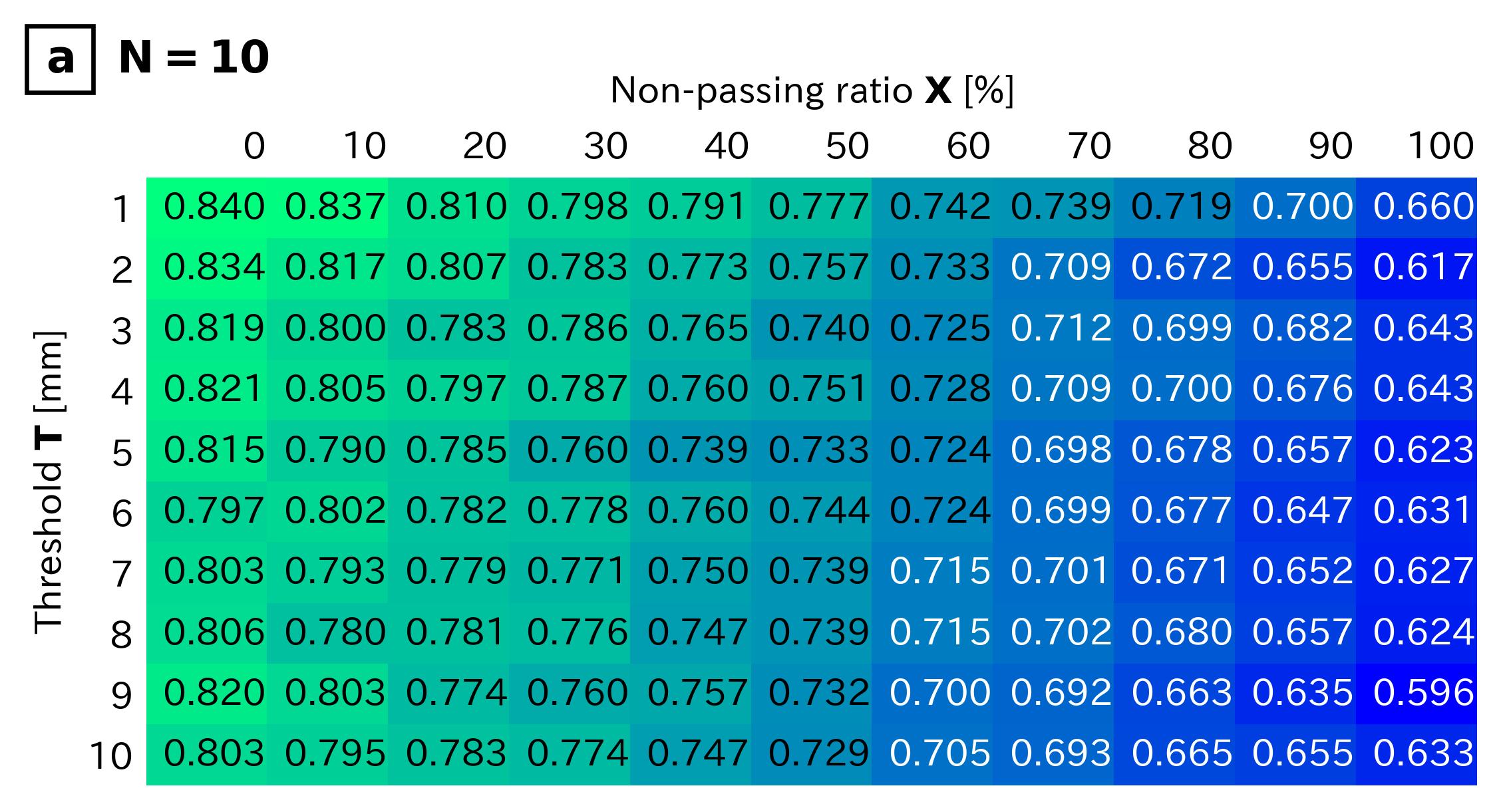}
    \includegraphics[width=0.49\linewidth]{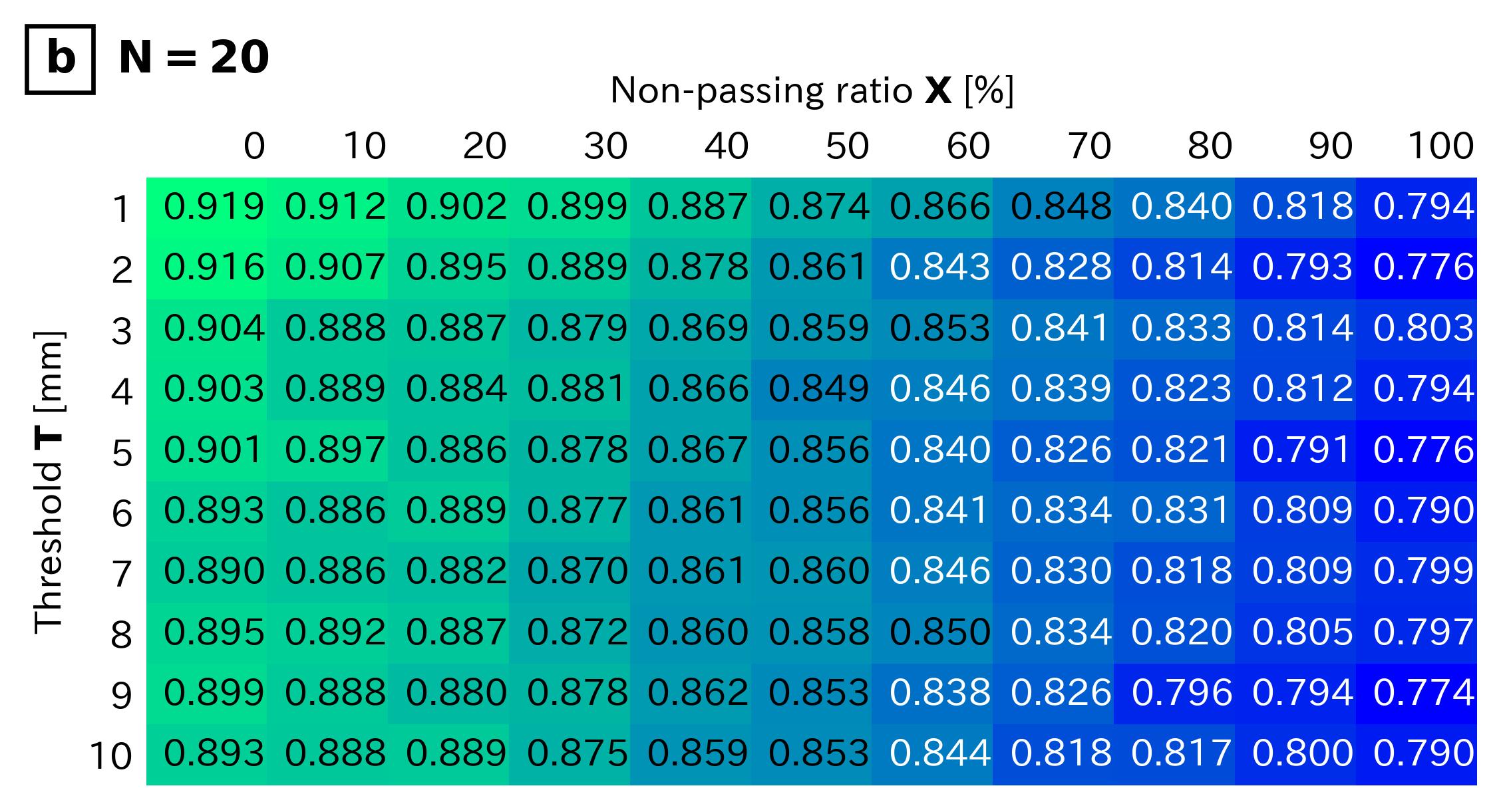}\\[1ex]
    \includegraphics[width=0.49\linewidth]{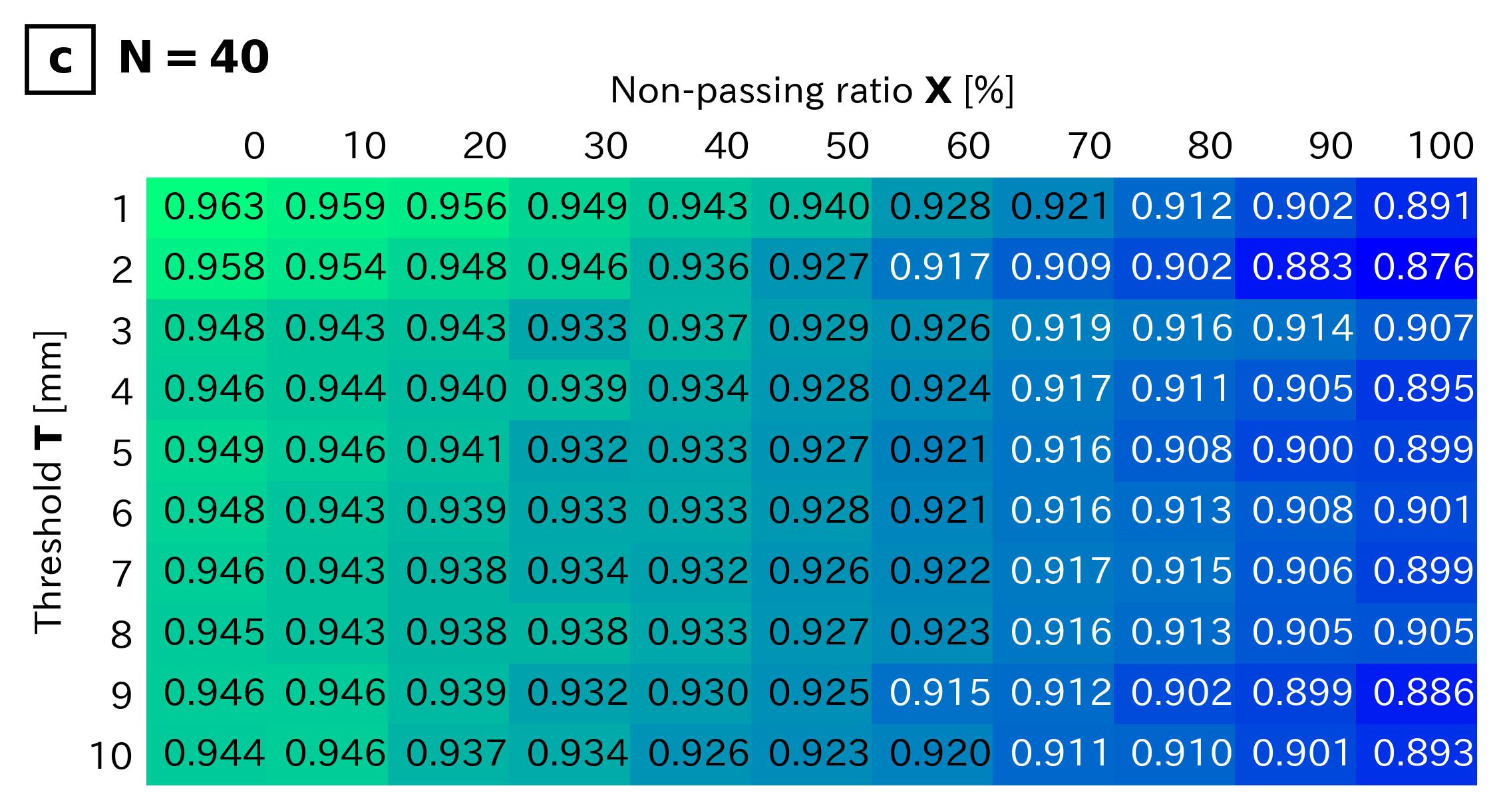}
    \includegraphics[width=0.49\linewidth]{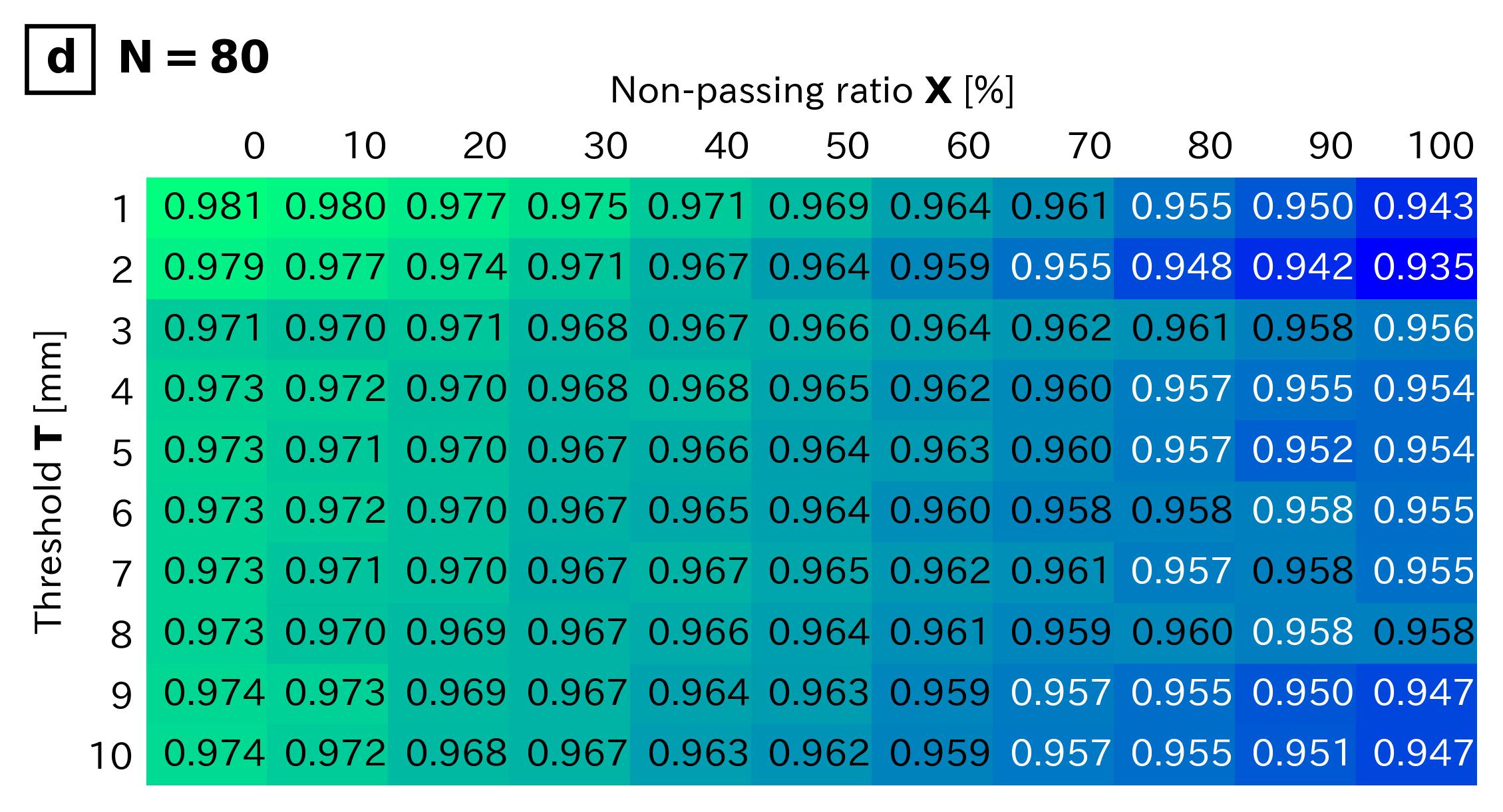}
    \caption{Experiment~3: Goodness of fit to Equation~(\ref{eq:exp1_fitts_mt}) for $MT$ (fast)}
    \label{fig:exp3_fitts_mt_speed_R2_sim}
    \Description{Four heatmaps (N=10,20,40,80) of R^2 for the MT model under “fast” instruction in Experiment 3 (no re-aiming). Degradation with larger X and looser T is more pronounced than in Experiment 2.}
\end{figure*}
\begin{figure*}[ht]
    \centering
    \includegraphics[width=0.49\linewidth]{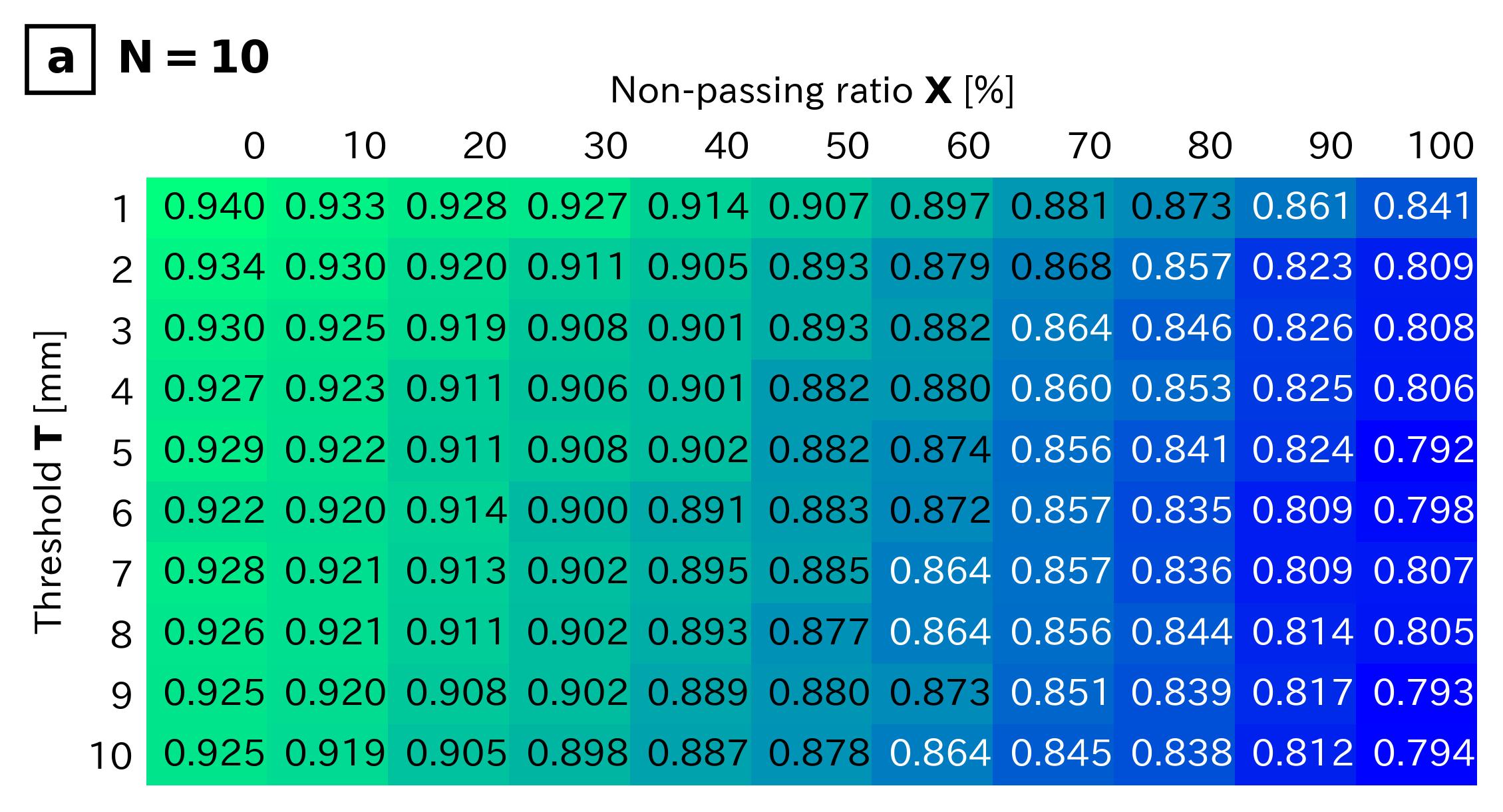}
    \includegraphics[width=0.49\linewidth]{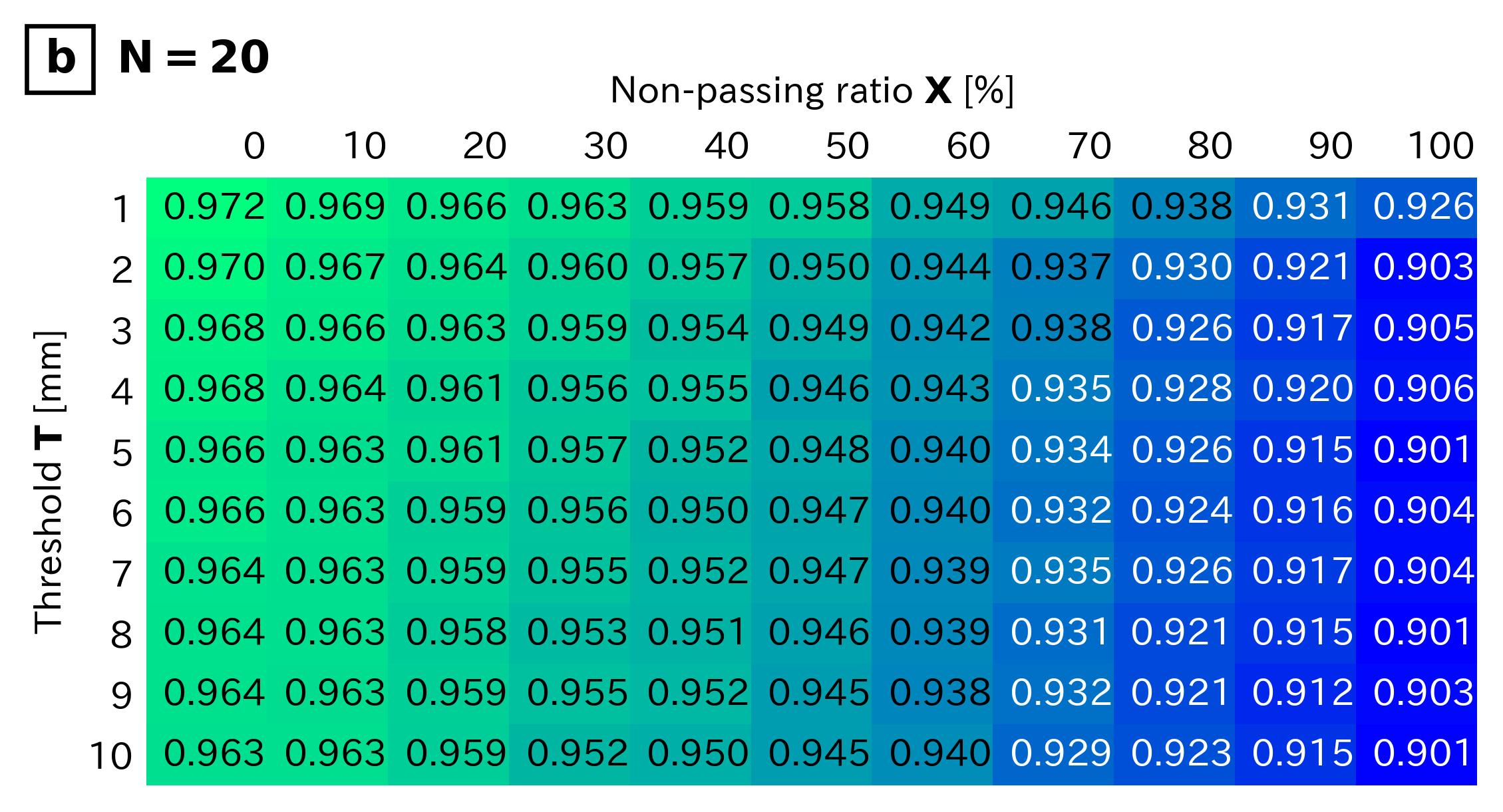}\\[1ex]
    \includegraphics[width=0.49\linewidth]{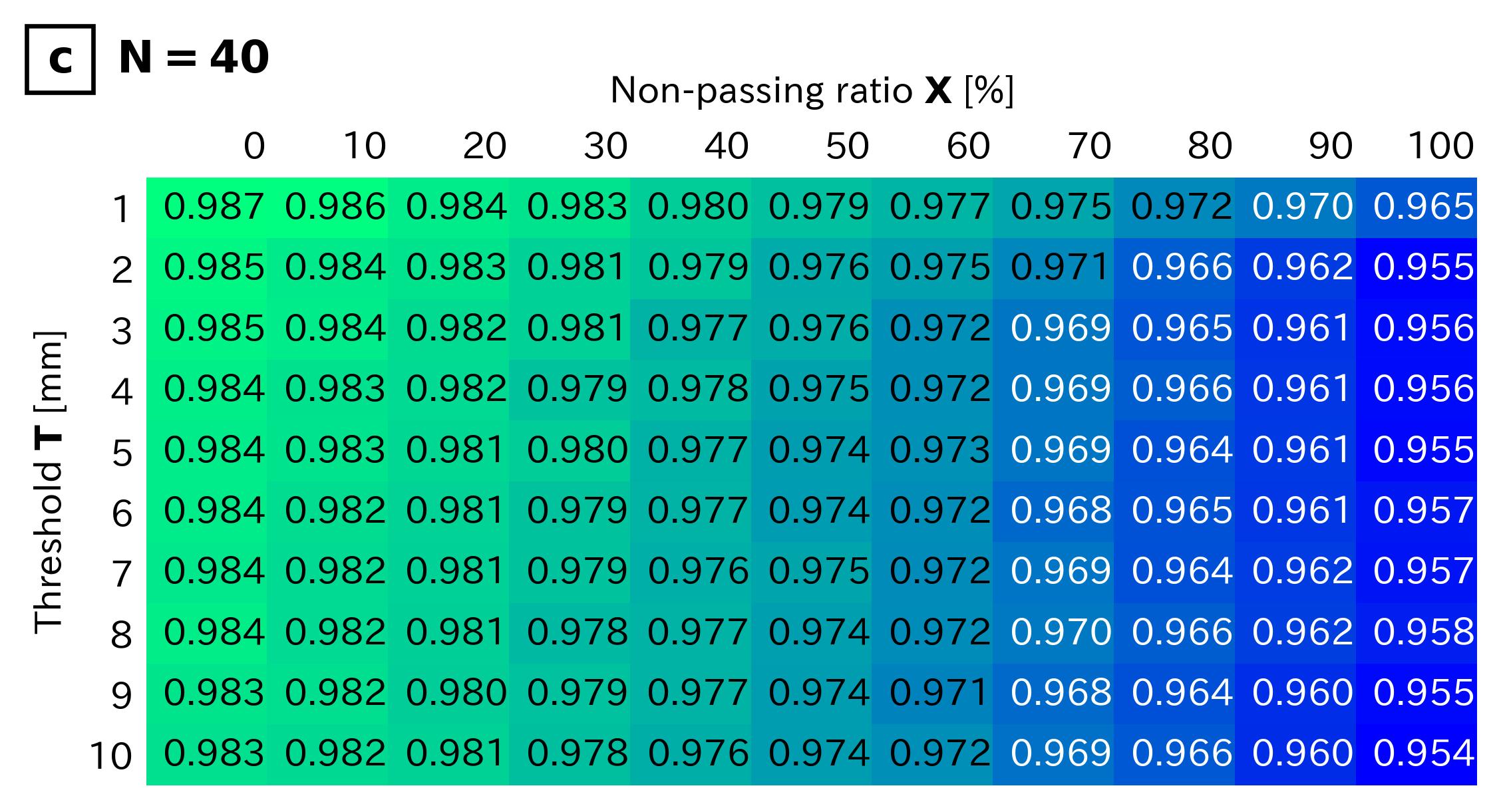}
    \includegraphics[width=0.49\linewidth]{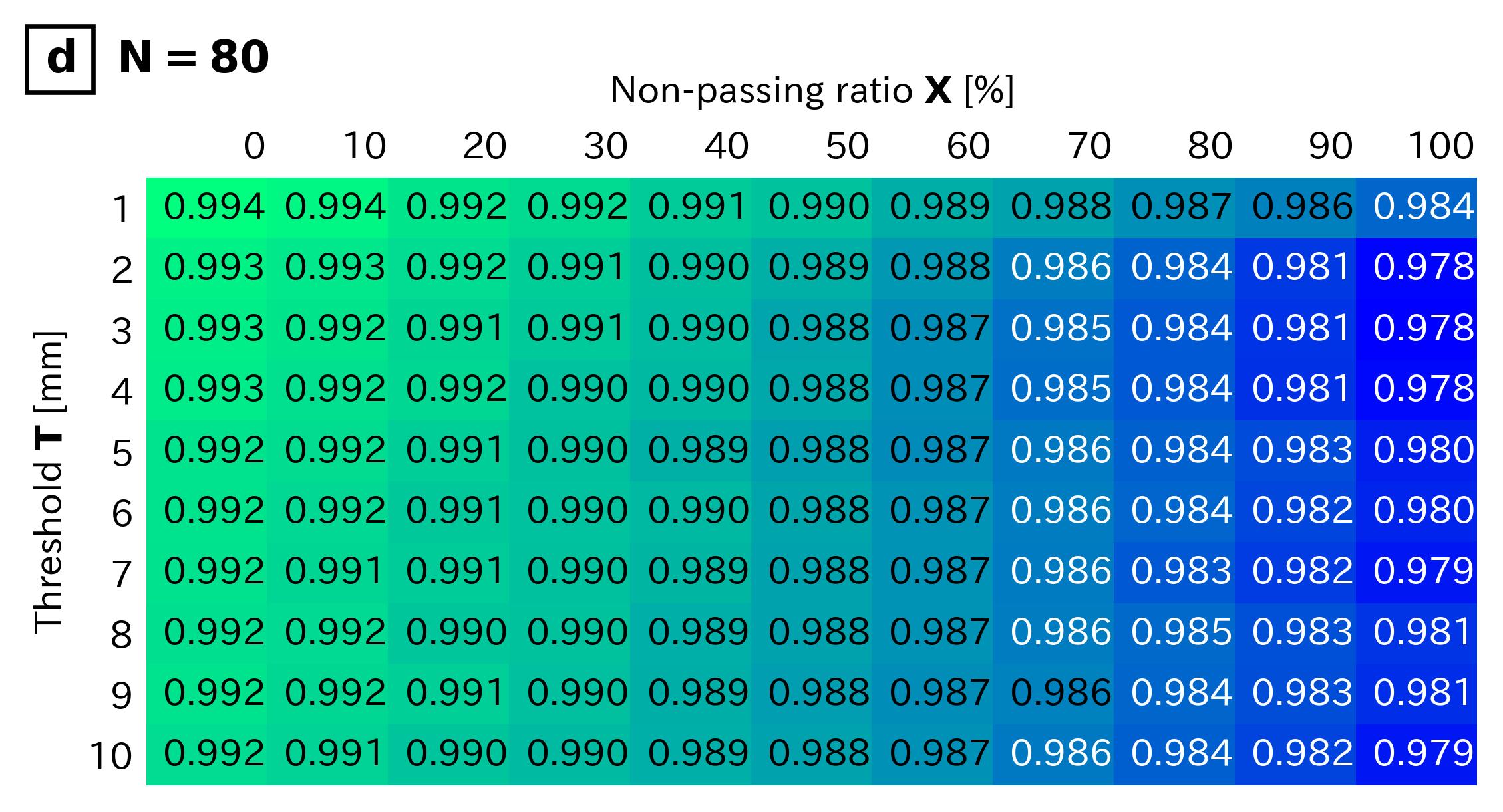}
    \caption{Experiment~3: Goodness of fit to Equation~(\ref{eq:exp1_fitts_mt}) for $MT$ (accurate)}
    \label{fig:exp3_fitts_mt_accuracy_R2_sim}
    \Description{Four heatmaps (N=10,20,40,80) of R^2 for the MT model under “accurate” instruction in Experiment 3. Stronger upper-left to lower-right drop, especially for small N.}
\end{figure*}
\begin{figure*}[ht]
    \centering
    \includegraphics[width=0.49\linewidth]{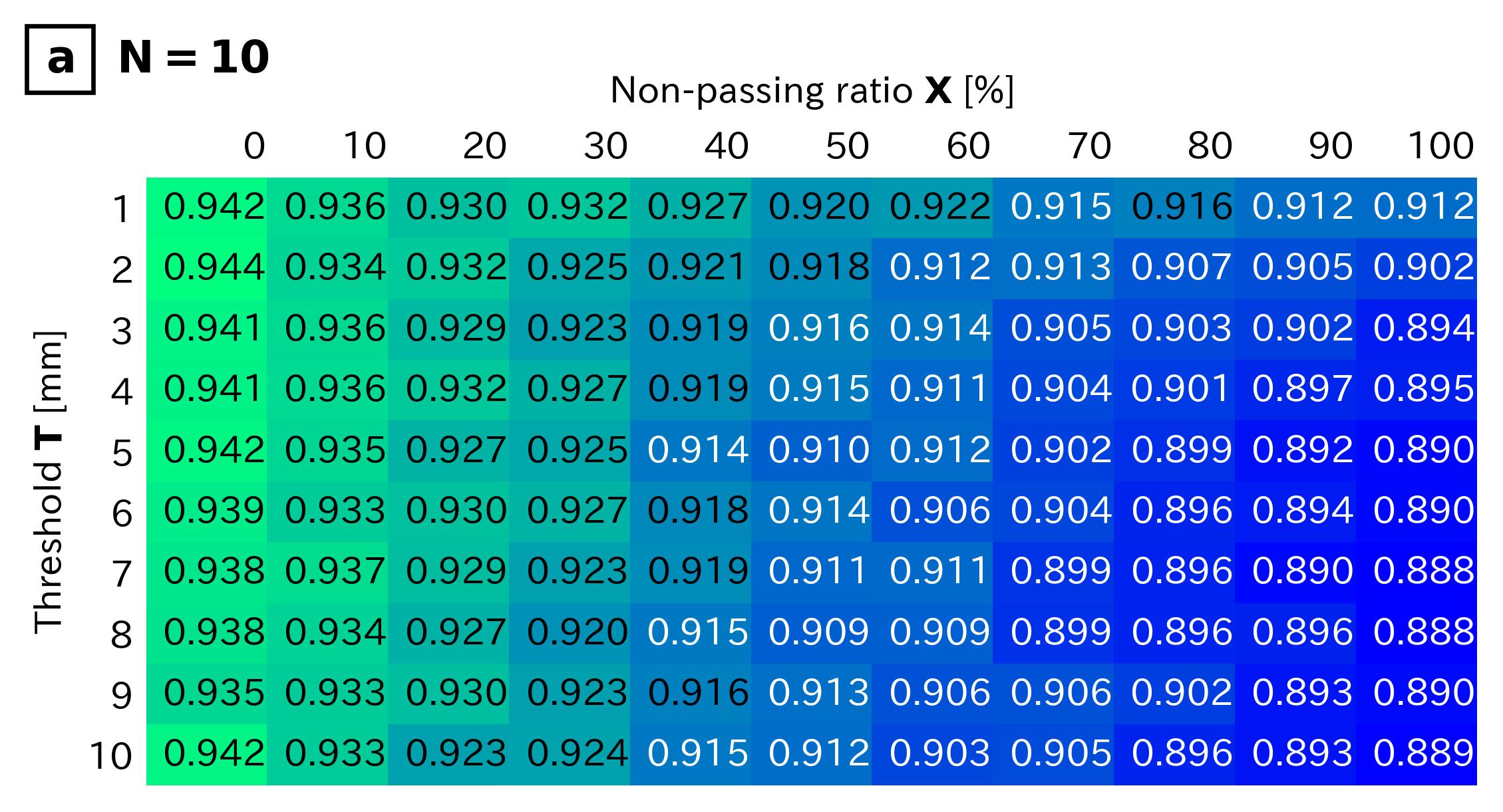}
    \includegraphics[width=0.49\linewidth]{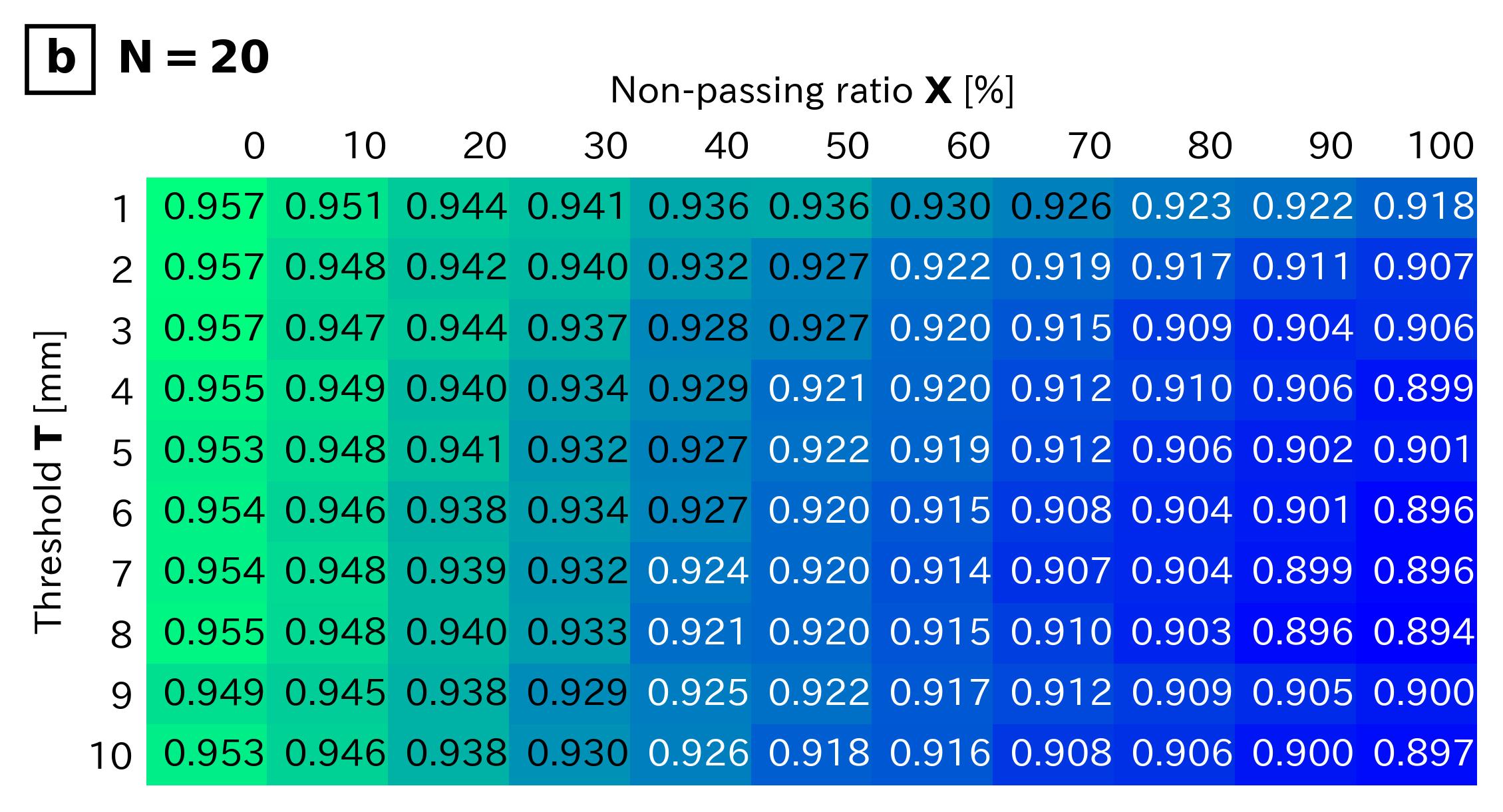}\\[1ex]
    \includegraphics[width=0.49\linewidth]{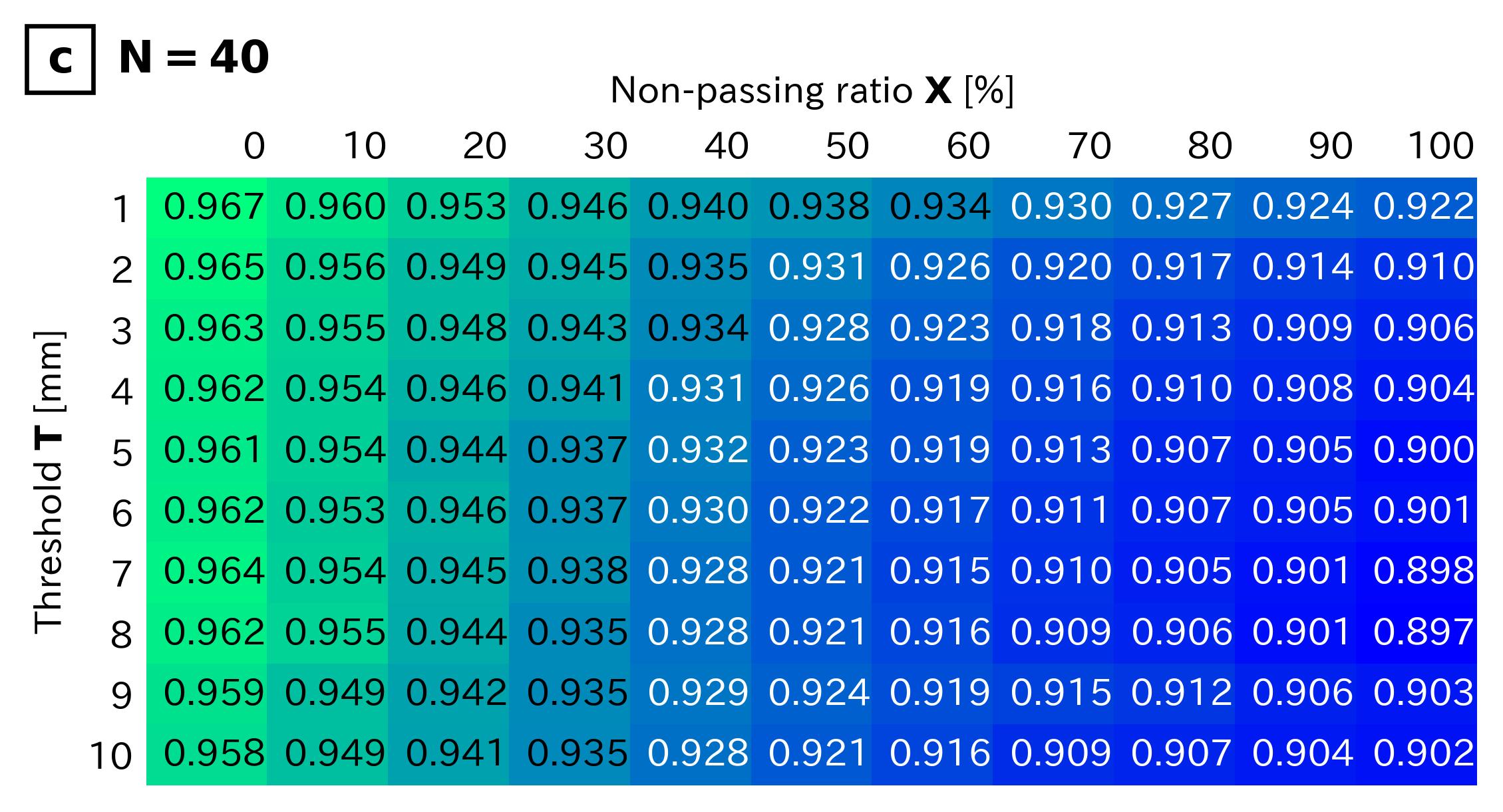}
    \includegraphics[width=0.49\linewidth]{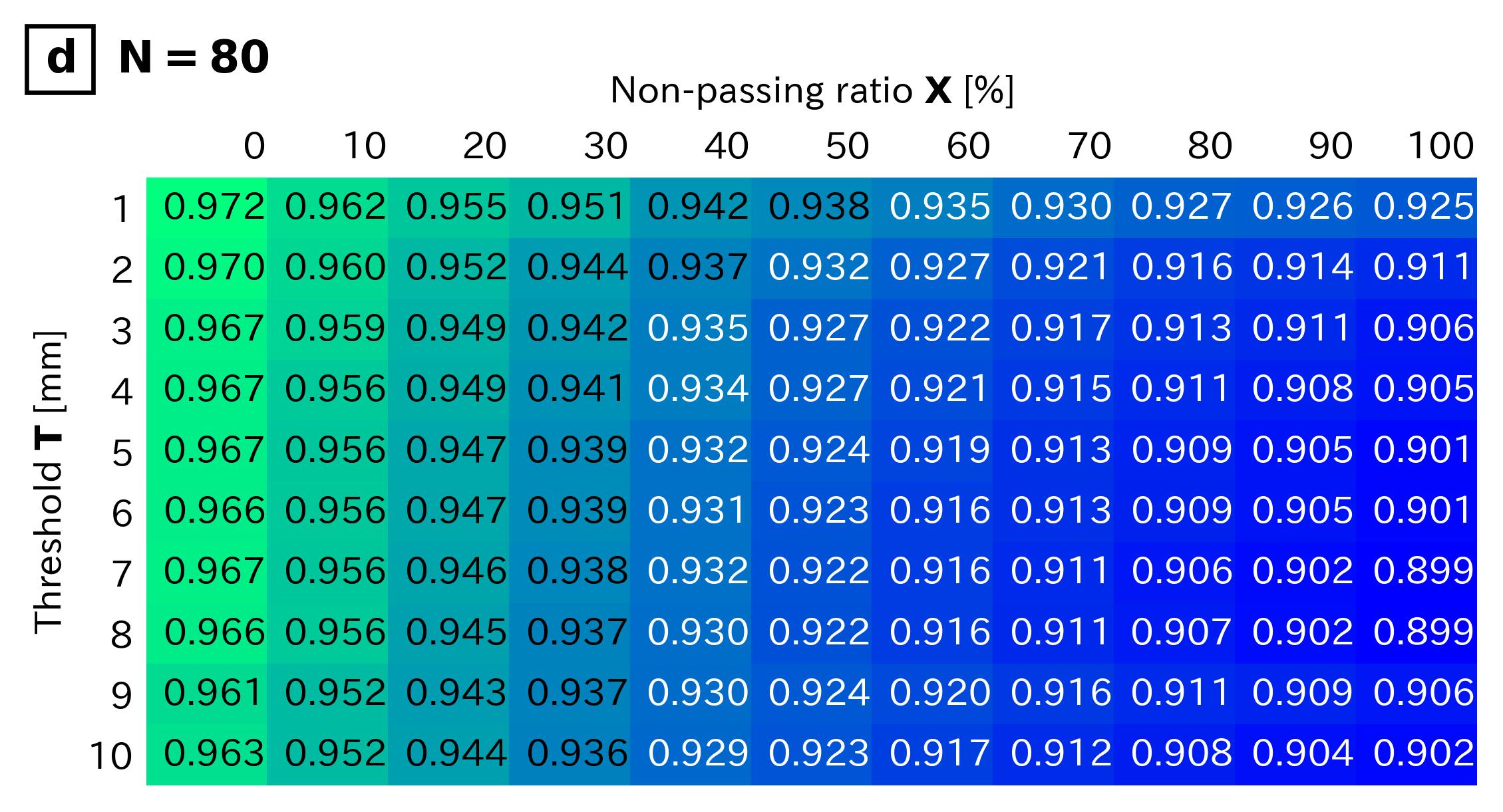}
    \caption{Experiment~3: Goodness of fit to Equation~(\ref{eq:exp2_fitts_er}) for $ER$ (fast)}
    \label{fig:exp3_fitts_er_speed_R2_sim}
    \Description{Four heatmaps (N=10,20,40,80) of R^2 for the ER model under “fast” instruction in Experiment 3 (no re-aiming). Fit decreases with larger X and looser T; effects are stronger than in Experiment 2.}
\end{figure*}
\begin{figure*}[ht]
    \centering
    \includegraphics[width=0.49\linewidth]{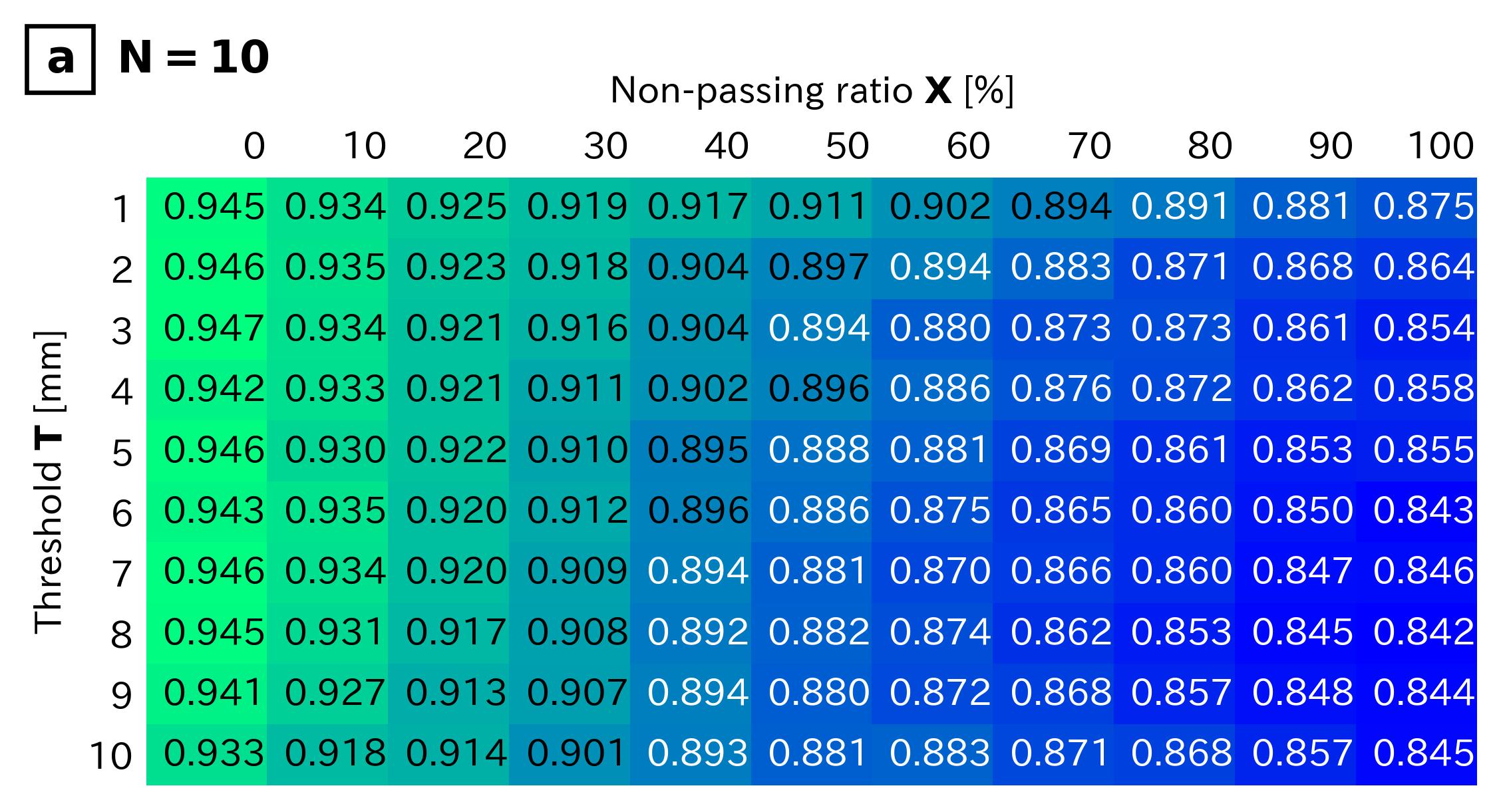}
    \includegraphics[width=0.49\linewidth]{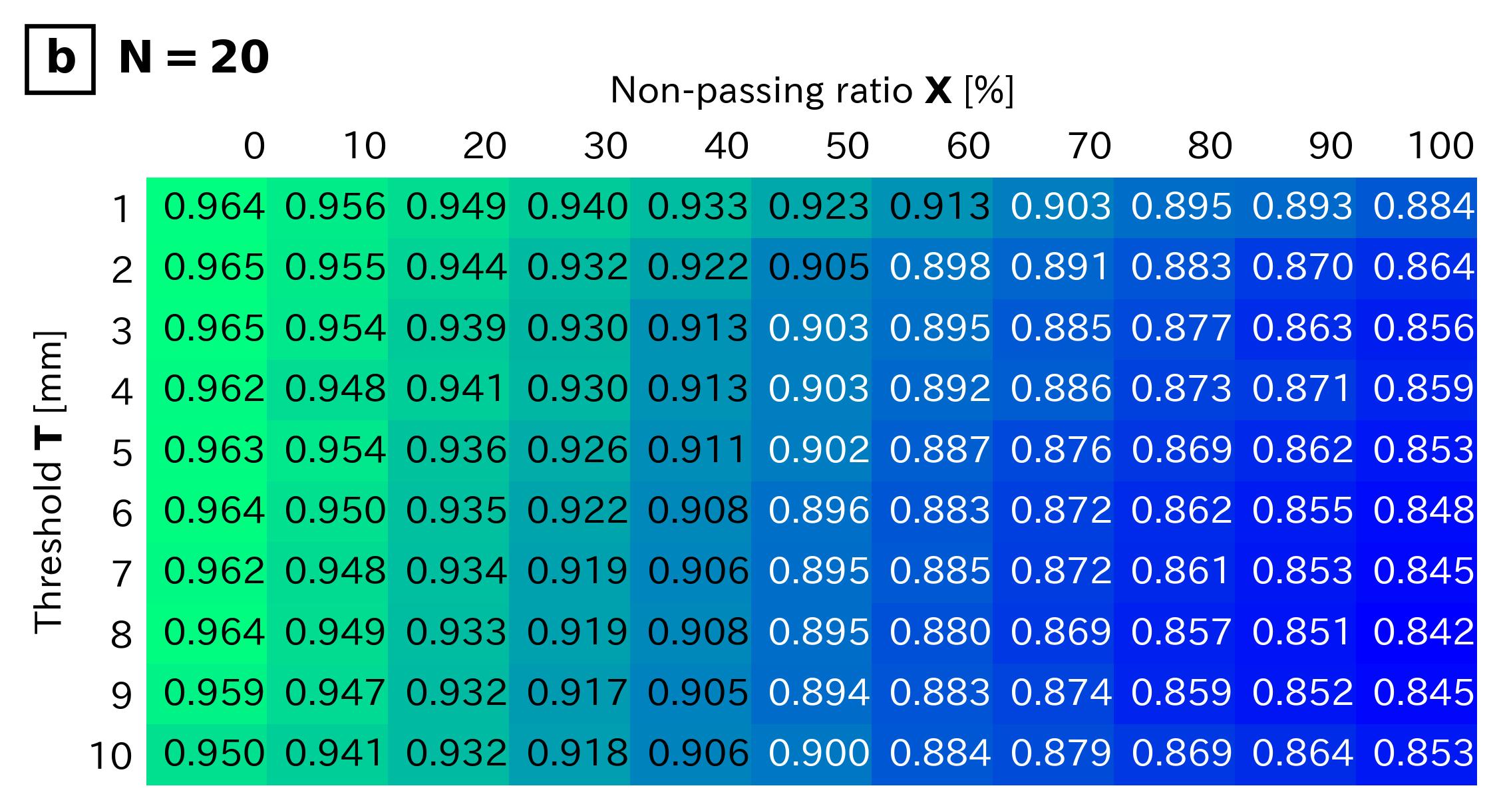}\\[1ex]
    \includegraphics[width=0.49\linewidth]{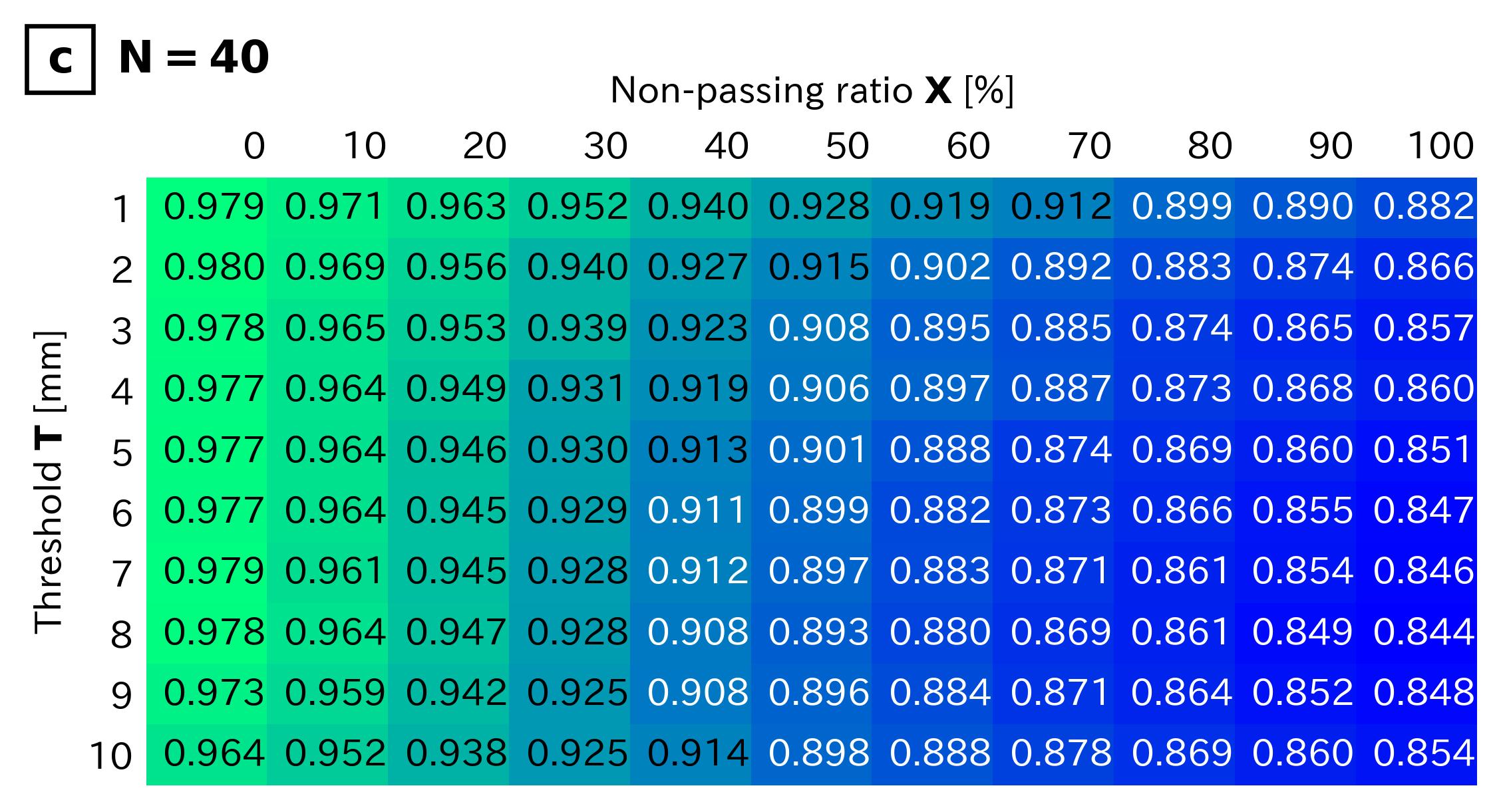}
    \includegraphics[width=0.49\linewidth]{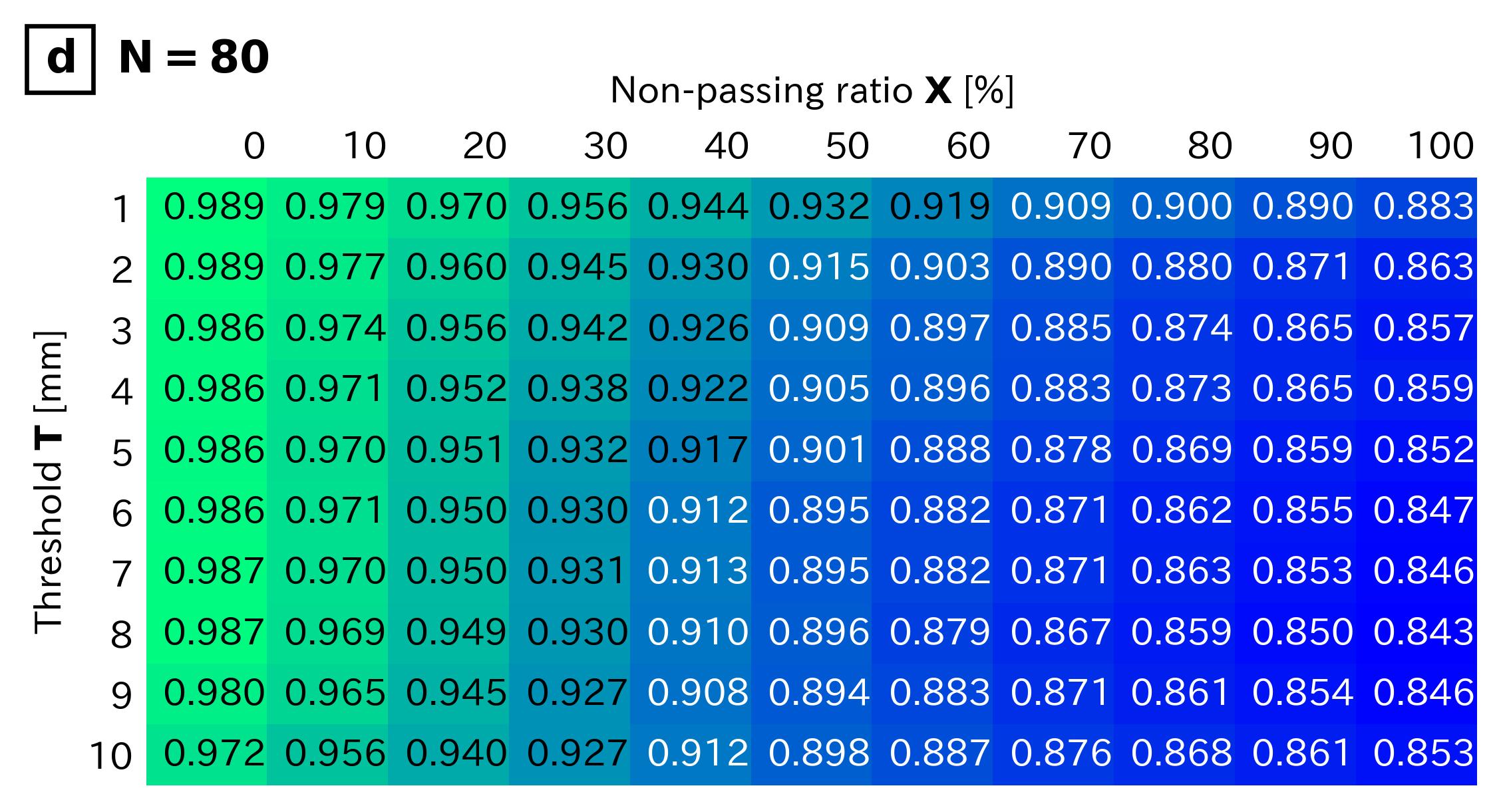}
    \caption{Experiment~3: Goodness of fit to Equation~(\ref{eq:exp2_fitts_er}) for $ER$ (accurate)}
    \label{fig:exp3_fitts_er_accuracy_R2_sim}
    \Description{Four heatmaps (N=10,20,40,80) of R^2 for the ER model under “accurate” instruction in Experiment 3. Top-left cells (strict T, X=0) often yield the best fit.}
\end{figure*}
\begin{figure*}[ht]
    \centering
    \begin{minipage}[b]{0.48\linewidth}
        \centering
        \includegraphics[width=\linewidth]{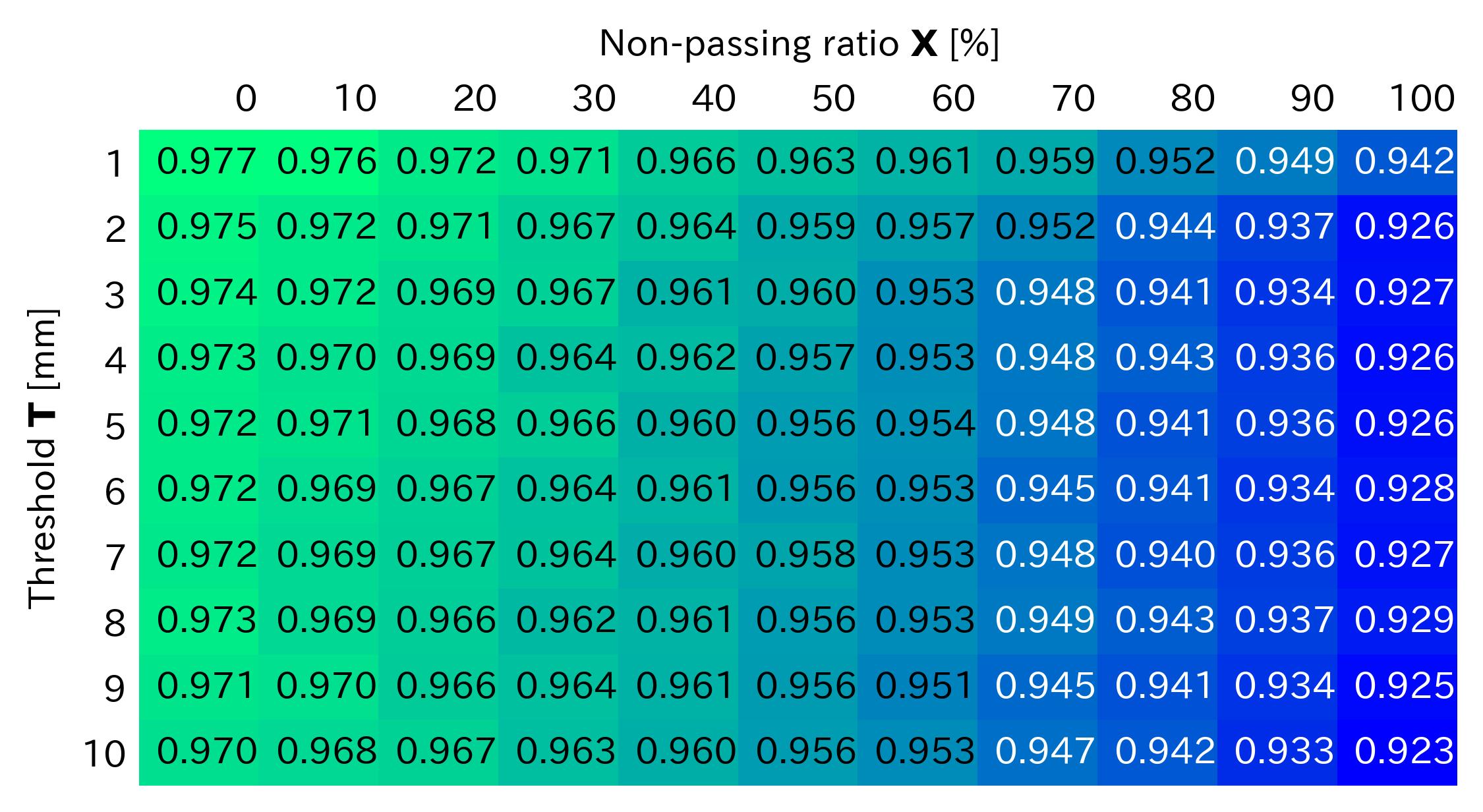}
        \vspace{-10pt}
        \caption{Cross-validation results of $R^2$ for $MT$ ($N=40$, accurate)}
        \label{fig:exp3_fitts_mt_accuracy_R2_loocv}
        \Description{Leave-one-W-out cross-validated R^2 heatmap for the MT model in Experiment 3 (example N=40, “accurate”). Predictive accuracy declines as X increases and T loosens.}
    \end{minipage}
    \hfill
    \begin{minipage}[b]{0.48\linewidth}
        \centering
        \includegraphics[width=\linewidth]{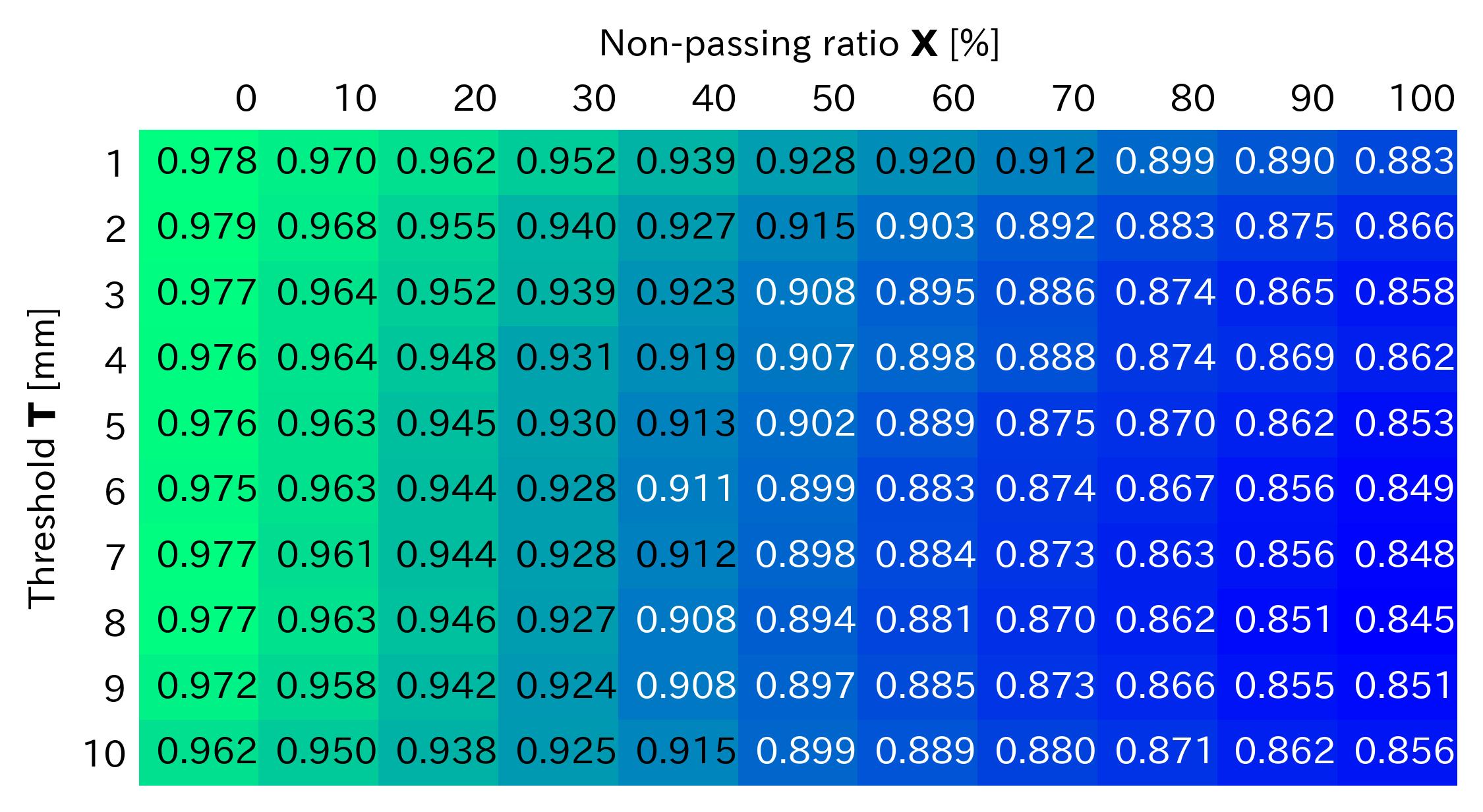}
        \vspace{-10pt}
        \caption{Cross-validation results of $R^2$ for $ER$ ($N=40$, accurate)}
        \label{fig:exp3_fitts_er_accuracy_R2_loocv}
        \Description{Leave-one-W-out cross-validated R^2 heatmap for the ER model in Experiment 3 (example N=40, “accurate”). Predictive accuracy mirrors the main-fit heatmaps, degrading as X grows and T loosens.}
    \end{minipage}
    \vspace{-5pt}
\end{figure*}

\subsection{Simulation Results}
\subsubsection{Movement Time ($MT$)}
Figures~\ref{fig:exp3_fitts_mt_speed_R2_sim} and \ref{fig:exp3_fitts_mt_accuracy_R2_sim} present $R^2$ for Equation~(\ref{eq:exp1_fitts_mt}) in Experiment~3 as $T$ and $X$ vary.
In both figures, $R^2$ decreases from the upper-left to the lower-right of each heatmap: lenient thresholds combined with larger non-passing proportions lower model fit.
Thus, the degradation observed in Experiments~1 and~2 is replicated in Experiment~3.

Compared with Experiment~2, the magnitude of degradation with increasing non-passing proportion is somewhat larger in Experiment~3.
Specifically, in Figures~\ref{fig:exp2_fitts_mt_speed_R2_sim} and \ref{fig:exp2_fitts_mt_accuracy_R2_sim}, even when fit decreased, most cells remained above 0.88 for $N=10$ or $N=20$.
By contrast, in Figures~\ref{fig:exp3_fitts_mt_speed_R2_sim} and \ref{fig:exp3_fitts_mt_accuracy_R2_sim}, values sometimes dropped to around 0.65 for $N=10$ and around 0.78 for $N=20$.
This difference should stem from the change introduced in Experiment~3, i.e., skipping re-aiming and proceeding immediately after errors.
This design mitigates the prior limitation where inattentive participants behaved similarly to conscientious (rapid and accurate) ones.
Consequently, changes in model fit are more pronounced.

As before, increasing $N$ improves $R^2$ in the upper-left regions and attenuates the effect of larger $X$.
Nevertheless, $R^2$ still varies with $X$ in Figures~\ref{fig:exp3_fitts_mt_speed_R2_sim} and \ref{fig:exp3_fitts_mt_accuracy_R2_sim}, and the declines are larger in Experiment~3.
Thus, in settings where participants can complete tasks without following instructions (as here), our screening method becomes even more valuable.

As a cross-validation example, Figure~\ref{fig:exp3_fitts_mt_accuracy_R2_loocv} shows the case $N=40$ under the ``accurate'' instruction.
Similar to the results in Experiment 2, compared to using all $W$ levels, overall predictive accuracy is slightly lower, yet the same upper-left to lower-right degradation trend appears.

\subsubsection{Error Rate ($ER$)}
Figures~\ref{fig:exp3_fitts_er_speed_R2_sim} and \ref{fig:exp3_fitts_er_accuracy_R2_sim} show $R^2$ for Equation~(\ref{eq:exp2_fitts_er}) in Experiment~3 as $T$ and $X$ vary.
Again, $R^2$ decreases from the upper-left to the lower-right in each heatmap: lenient thresholds and larger non-passing proportions reduce fit.
Thus, the degradation observed in Experiments~1 and~2 is also observed here.
The influence of $N$ follows the same trend described above.

Compared to Experiment~2, the declines in $R^2$ with increasing non-passing proportion are somewhat larger in Experiment~3.
In Figures~\ref{fig:exp2_fitts_er_speed_R2_sim} and \ref{fig:exp2_fitts_er_accuracy_R2_sim}, most cells remained above 0.93 even when fit decreased; in Figures~\ref{fig:exp3_fitts_er_speed_R2_sim} and \ref{fig:exp3_fitts_er_accuracy_R2_sim}, values sometimes dropped to around 0.85 for $N=10$ or $N=20$.
Moreover, unlike Figures~\ref{fig:exp2_fitts_er_speed_R2_sim} and \ref{fig:exp2_fitts_er_accuracy_R2_sim}, Experiment~3 shows a tendency for fit to decline even when the non-passing proportion is small.
Another difference is that, in most heatmaps, the top-left cell (threshold $T=1$, non-passing proportion $X=0$) attains the best fit.
Whereas Experiment~2 exhibited a limitation in which this cell was not always optimal (e.g., Figure~\ref{fig:exp2_fitts_er_accuracy_R2_sim}), Experiment~3 generally achieves the best fit there (Figures~\ref{fig:exp3_fitts_er_speed_R2_sim}--\ref{fig:exp3_fitts_er_accuracy_R2_sim}), and this is also true for the $MT$ results (Figures~\ref{fig:exp3_fitts_mt_speed_R2_sim}--\ref{fig:exp3_fitts_mt_accuracy_R2_sim}).

A likely reason is, again, the immediate progression after errors in Experiment~3.
Inattentive participants may accept higher error rates to act quickly, deviating from the pointing behavior assumed by the $ER$ model.
This amplifies the effects of $N$ and $X$ on model fit and yields the best performance at the top-left cell (strict threshold, no non-passing participants).

As a cross-validation example, Figure~\ref{fig:exp3_fitts_er_accuracy_R2_loocv} shows the case $N=40$ under the ``accurate'' instruction.
As with Figure~\ref{fig:exp3_fitts_mt_accuracy_R2_loocv}, overall predictive accuracy is slightly lower than when using all $W$ levels, but the same upper-left to lower-right degradation trend holds.

\section{DISCUSSION}
\subsection{RQ1: Can a simple GUI pre-task effectively screen out nonconforming participants?}
Across three experiments using the size-adjustment task as a pre-task and a pointing task as the main experiment, we found that the goodness of fit of models estimating GUI performance (movement time and error rate) changes with the proportion of non-passing participants mixed into the sample (e.g., Figure~\ref{fig:exp3_fitts_er_accuracy_R2_sim}).
Consistently, keeping the proportion $X$ of the non-passing group small (suppressing the inclusion of nonconforming participants) and setting a stricter error threshold $T$ in the size-adjustment task tended to improve model fit.
In other words, screening based on size-adjustment error contributes to higher goodness of fit for the performance models in the main experiment.

A key advantage is that nonconformity is judged without using outcomes from the main experiment, as we selected participants solely based on the pre-task UI interaction.
This elevates the quality of those who proceed to the main experiment and reduces the risk of incorrect conclusions in model evaluation (e.g., that a given model poorly estimates UI performance).
Moreover, the fact that screening can be achieved with a simple UI operation within nine seconds on average supports ease of adoption and practical utility for researchers.
Taken together, size-adjustment-based screening is an effective method to enhance data quality by improving the model fit of the main experiment.

Overall, the answer to RQ1 is: \textit{a brief GUI pre-task—op\-er\-a\-tion\-al\-ized here as size adjustment—effectively screens out nonconforming participants and thereby improves the goodness of fit of performance models in the main experiment}.

\subsection{RQ2: How do the non-passing proportion and the error threshold affect model fit and predictive accuracy in the main experiment?}
We systematically manipulated the size-adjustment error threshold $T$ and the non-passing proportion $X$ and evaluated their effects on the fit and predictive accuracy of the $MT$ and $ER$ models.
Overall, stricter $T$ (smaller allowable error) and smaller $X$ tended to yield higher $R^2$.

In the PC mouse setting (Experiment~1), the $MT$ model's $R^2$ was stably above 0.98 and only weakly sensitive to $T$ or $X$ (Figures~\ref{fig:exp1_fitts_mt_speed_R2_sim} and \ref{fig:exp1_fitts_mt_accuracy_R2_sim}).
By contrast, for the $ER$ model under the ``accurate'' instruction, the heatmap showed a clear decline in $R^2$ as $T$ was relaxed and $X$ increased (Figure~\ref{fig:exp1_fitts_er_accuracy_R2_sim}).

When restricting to smartphones (Experiment~2), declines in fit with relaxed $T$ and larger $X$ were observed not only for the $ER$ model but also for the $MT$ model (Figures~\ref{fig:exp2_fitts_mt_speed_R2_sim}, \ref{fig:exp2_fitts_mt_accuracy_R2_sim}, \ref{fig:exp2_fitts_er_speed_R2_sim}, and \ref{fig:exp2_fitts_er_accuracy_R2_sim}).
For unobserved target-width conditions, the same degradation trends were reproduced by leave-one-$W$-out cross-validation (e.g., Figures~\ref{fig:exp2_fitts_mt_accuracy_R2_loocv} and \ref{fig:exp2_fitts_er_accuracy_R2_loocv}).
Experiment~3 replicated these tendencies as tightening $T$ and suppressing $X$ improved fit (Figures~\ref{fig:exp3_fitts_mt_speed_R2_sim}, \ref{fig:exp3_fitts_mt_accuracy_R2_sim}, \ref{fig:exp3_fitts_er_speed_R2_sim}, and \ref{fig:exp3_fitts_er_accuracy_R2_sim}).

We also observed cases where the seemingly ideal setting for experimenters (the smallest $T$ with $X=0$) did not always yield the highest $R^2$.
This suggests that design factors, such as the number of $W$ levels, repetitions, and the error-handling policy (presence/absence of re-aiming), can affect evaluation resolution and warrant consideration in experimental design.
In sum, stricter $T$ and smaller $X$ improve fit and predictive accuracy, especially for the $ER$ model, and in smartphone settings the $MT$ model benefits as well.

Overall, the answer to RQ2 is: \textit{stricter thresholds $T$ and smaller non-passing proportions $X$ systematically increase model fit $R^2$ (and, where assessed, predictive accuracy), with especially strong effects for $ER$ and clear benefits for $MT$ in smartphone settings.}

\begin{figure*}[ht]
\centering
\includegraphics[width=\linewidth]{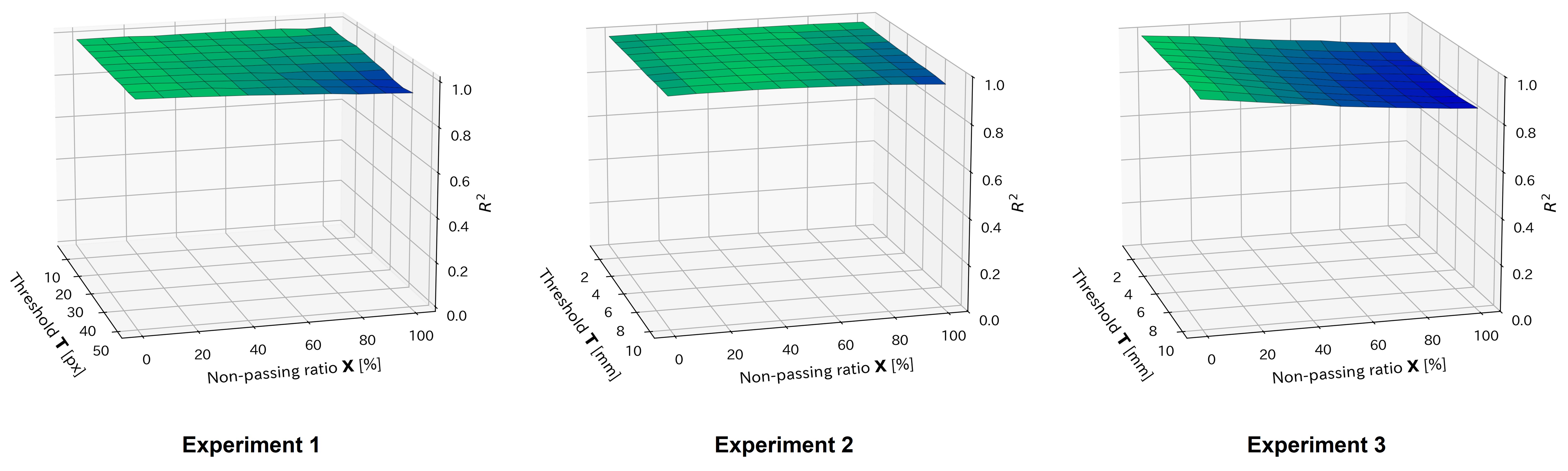}
\caption{Comparison of $ER$ model fit heatmaps across Experiments 1–3 ($N=40$, accurate)}
\label{fig:heatmap_all}
\Description{Three side-by-side 3D surfaces of ER model fit (R^2) vs. threshold T [px] and non-passing ratio X [\%] for N=40 under the ‘accurate’ instruction. Experiments 1 and 2 show only small decreases—surfaces stay near R^2≈1.0 even at larger T and X—whereas Experiment 3 shows a clear, stronger decline as T and X increase.}
\end{figure*}

\subsection{RQ3: Is the proposed screening consistently effective across devices and error-handling procedures?}
We evaluated the same pre-task-based screening method across three experiments spanning PC (mouse) and smartphone (touch).
Comparing Experiment~1 (PC) with Experiments~2 and~3 (smartphone), we observed a common pattern that reducing the non-passing proportion $X$ and tightening the error threshold $T$ increased $R^2$.
As examples of device comparisons, see Figures~\ref{fig:exp1_fitts_er_accuracy_R2_sim}, \ref{fig:exp2_fitts_er_accuracy_R2_sim}, and \ref{fig:exp3_fitts_er_accuracy_R2_sim} for the $ER$ results under the ``accurate'' instruction.
Improvements were especially pronounced for the $ER$ model and extended to the $MT$ model in smartphone settings (e.g., Figures~\ref{fig:exp2_fitts_mt_speed_R2_sim} and \ref{fig:exp3_fitts_mt_speed_R2_sim}).
To streamline figure references, Figure~\ref{fig:heatmap_all} summarizes the $ER$ results across experiments.

A direct comparison of the two smartphone procedures, Experiment~2 (re-aim until success) and Experiment~3 (proceed immediately after errors), reveals similar improvement gradients with respect to $T$ and $X$, indicating robustness across error-handling policies. 
This effect is also captured compactly in Figure~\ref{fig:heatmap_all}.
That is, across device differences and the presence/absence of re-aiming, stricter $T$ and smaller $X$ consistently yield higher fit.
In addition, from Figure~\ref{fig:heatmap_all}, we see that Experiments 1–2 (PC/iPhone with re-aiming) show only slight reductions in $R^2$ as $T$ is relaxed and the non-passing proportion $X$ grows, whereas Experiment 3 (no re-aiming) exhibits a marked, monotonic decline.
Therefore, selecting a stricter $T$ and keeping $X$ small is especially crucial when errors are not corrected via re-aiming.

Even under smartphone conditions where target width $W$ was controlled in mm via device PPI identification, we still observed fit improvements with stricter $T$ and smaller $X$, supporting generalizability in practical deployments.
At the same time, the best-looking setting (smallest $T$, $X=0$) did not always produce the highest $R^2$.
Therefore, sufficient numbers of $W$ levels and repetitions and the choice of error-handling can influence evaluation resolution and should be considered.

Overall, the answer to RQ3 is: \textit{the method is consistently effective across device differences and across re-aiming procedures}.

\subsection{Limitations and Future Work}
\label{sec:limitations}
Our screening cannot identify participants who act in a task-con\-form\-ing manner in the pre-task yet unexpectedly in the main task.
\hl{Thus, even after screening, the remaining sample may still include participants whose behavior in the main task is inattentive or only partially compliant with the instructions.}
Moreover, our approach requires setting a threshold $T$ for the size-adjustment task.
Because model fit and predictive accuracy change gradually with $T$ and with the non-passing proportion $X$, uniquely determining the best $T$ is difficult.
A stricter $T$ reduces the inclusion of nonconforming participants but induces a smaller sample size and thus leads to imprecise model evaluation.
Further, design factors such as the error-handling policy (presence/absence of re-aiming), the number of $W$ levels, and the number of repetitions may contribute to the phenomenon where the ``best-looking'' condition does not always yield the best fit.

\hl{Another limitation concerns potential bias introduced by the proposed screening method.
Participants with motor or visual impairments may find it inherently difficult to achieve small resizing errors even when they follow the instructions conscientiously.
Whether such participants should be excluded depends on the purpose of the experiment.
In our study, the goal is to evaluate the estimation accuracy of GUI performance models under standardized task conditions.
Hence, substantial motor or perceptual limitations would directly influence these models' accuracy and complicate comparisons across participants, so excluding such cases is consistent with our objectives.
However, for studies aiming to observe diverse real-world GUI behaviors or to investigate accessibility-related differences, removing these participants would suppress meaningful variability and could be undesirable.
In addition, some apparent ``nonconformity'' may stem from unclear instructions or suboptimal interface design rather than worker intent, suggesting that improvements in task design and pre-task screening should be considered jointly.}

Our findings and discussion are limited to the case where the pre-task is image resizing and the main task is pointing.
\hl{Both tasks are relatively simple GUI operations with modest cognitive and motor demands and share similar low-level control characteristics (e.g., precise positioning on a 2D display), which may limit the extent to which our findings directly generalize to more complex interaction scenarios.}
Whether the same screening idea works when one or both tasks differ requires further investigation.
Future work will thus test the screening approach with other combinations of pre-tasks and main tasks.
\hl{One conceptually interesting configuration would be to invert the roles of pre-task and main task, i.e., using a pointing task as the pre-task and a resizing task as the main task.
If participants who yield low $R^2$ in pointing performance also tend to exhibit large resizing errors, this would support the idea that operational carefulness transfers across tasks.
At the same time, prior work indicates that pointing-based performance models, especially error-rate models, can produce low $R^2$ even for conscientious participants when the number of trials per condition is small~\cite{yamanaka2021utility}, and that people with motor-control difficulties may show high error rates despite carefully following instructions~\cite{Findlater20dataset}.
Therefore, interpreting a poor model fit in reversed configurations as straightforward indicators of ``inattention'' would require caution.}
\hl{We also examined using the number of re-aiming attempts in the main task as a direct indicator of careless behavior, but we did not observe a clear correlation with pre-task errors (Pearson's $r < 0.2$; see the supplementary materials).}

Across our three experiments, we consistently observed that stricter $T$ and smaller $X$ increased $R^2$, and that appropriate screening was possible using only pre-task UI interaction data, without using outcomes from the main experiment to judge nonconformity.
These results suggest that effective screening may be achievable even when the pre-task differs from the main task.
Thus, researchers may not need to change the pre-task every time the main task changes.
Rather, screening should transfer to other GUI-based main experiments, such as those to evaluate key-typing~\cite{Soukoreff03metric,Zhang19text} and path-steering performance~\cite{Accot97,Kasahara24chi}.

Alternative pre-tasks are also conceivable.
Potential pre-tasks would be brief, require simple GUI-based interactions, and share operational characteristics with the main task.
We will explore such pre-tasks for more effective screening.
Another possible approach would be to employ a gold-standard task~\cite{Kazai11gold} or attention check~\cite{Oppenheimer09imc} as a pre-task, thereby restricting the main task to workers who pass it.
Still, we demonstrated that the results of pre-tasks, especially measures of operational carefulness (size-adjustment error), are in fact distributed continuously rather than dichotomously.
\hx{Because gold tasks and attention checks are often used to make an explicit pass/fail decision of conformity (even though their outcomes can also be treated as continuous measures such as accuracy or response time), the advantage of our proposed method is that the pre-task is designed to yield a task-relevant continuous error signal and thus account for this gradient.}
\hl{Future work should empirically compare the screening capabilities of these candidate pre-task mechanisms to clarify their respective advantages and appropriate use cases in GUI experimentation.}

\subsection{\hl{Positioning within HCI and Suitable Downstream (Main) Tasks}}
\label{sec:main_task}
From a broader HCI perspective, our work contributes to the meth\-od\-o\-log\-i\-cal and infrastructural foundations that support empirical research on GUI interaction.
As crowdsourcing becomes increasingly common for collecting large-scale behavioral data in tasks involving pointing, selection, movement, or text entry, ensuring that the collected performance data are reliable is essential.
Our screening method complements prior infrastructural contributions in HCI, such as refined statistical analysis procedures~\cite{wobbrock2011aligned,Elkin20ART} and standardized GUI performance metrics~\cite{Kasahara24chi,Wobbrock11dim}, by providing a practical mechanism for improving data quality \emph{before} conducting the downstream main experiment.
Rather than proposing a new interaction technique, this work strengthens the basis on which such techniques, as well as comparative evaluations of existing models, can be reliably assessed in crowdsourced settings.

\hl{The proposed screening method is also broadly applicable to downstream GUI-interaction tasks beyond the pointing paradigm used in our experiments.
Relevant examples include goal crossing~\cite{Accot02,yamanaka2020necessary}, path steering~\cite{Accot97,Kulikov05}, dragging~\cite{mackenzie1991comparison,Inkpen01drag}, typing~\cite{banovic2019limits,Soukoreff95limit}, and object selection techniques~\cite{yamanaka2022effectiveness,Bjerre17}.
These tasks all rely on quantitative performance measurements (e.g., speed, accuracy, error patterns) and are sensitive to noisy or inattentive behavior in crowdsourced environments.
By screening out participants whose pre-task behavior indicates insufficient task compliance, our method can improve the reliability of collected data, thereby supporting the validity of experimental results in model fitting, performance comparisons, and replications.}

\section{CONCLUSION}
This work introduced a pre-experiment screening based on a size-adjustment task and validated its effectiveness across three experiments spanning PCs (mouse) and smartphones (touch).
For RQ1, we showed that selecting participants solely from pre-task UI outcomes (size-adjustment error) consistently improved the goodness of fit of performance models in the main experiment.
In particular, the smaller the inclusion of the non-passing group and the stricter the threshold $T$, the higher the fit.

For RQ2, manipulating the threshold $T$ and the ratio of inclusion of nonconforming participants $X$ had systematic effects on model fit $R^2$ (and, in Experiments 2 and 3, on predictive accuracy via LOOCV), demonstrating that stricter $T$ and smaller $X$ are especially effective for the $ER$ model, with improvements extending to the $MT$ model in smartphone settings.
For RQ3, we observed the same pattern of improvements across device differences (PC/smartphone) and error-handling procedures (re-aim until success vs.\ proceed immediately after errors), confirming the robustness of the proposed method.

These results indicate that a brief pre-task finished in nine seconds on average can enhance participant quality \textit{before} the main experiment and reduce the risk of erroneous conclusions in model evaluation.
In practice, a conservative (stricter) choice of $T$ and keeping $X$ small should be the default policy, while also ensuring sufficient sample size $N$, number of target-width levels $W$, and repetitions.
On smartphones, the effect was confirmed even when stimulus size was controlled in millimeters via device PPI identification, supporting generalizability to real deployments.
At the same time, uniquely determining $T$ is nontrivial: there are trade-offs with $X$ and participant recruitment, and residual influences from error-handling and trial design remain.
Developing guidance tailored to study goals and constraints, and optimizing protocols accordingly, are important directions.

\bibliographystyle{ACM-Reference-Format}
\bibliography{main}

\end{document}